\documentclass[natbib]{svjour3}
\usepackage{mathptmx}           %postscript
\usepackage{lscape}
\usepackage{graphicx}
\usepackage{colortbl}
\usepackage{rotating}
\usepackage{color}       %dvips option
\usepackage{ulem}
\usepackage{booktabs}
\usepackage{ctable}

\newcommand{\pderiv}[2]{\frac{\partial #1}{\partial #2}} %partial derivative
\newcommand{\deriv}[2]{\frac{d #1}{d #2}} %ordinary derivative
\newcommand{\tensor}[1]{\underline{\underline{#1}}}
\newcommand{\tensorc}[1]{\underline{\underline{\underline{#1}}}}

\newcommand{\threedot}{\stackrel{.}{:}}

\begin{document}
\title{A review on solar wind modeling: kinetic and fluid aspects}
\author{Marius M. Echim \and Joseph Lemaire \and \O ystein Lie-Svendsen }
\institute{Marius M. Echim  \at Belgian Institute for Space Aeronomie,
  Avenue Circulaire 3,
1180 Brussels, Belgium, \email{marius.echim@aeronomie.be} and
Institute for Space Physics, M\u{a}gurele, Rom\^{a}nia
\and Joseph Lemaire \at Belgian Institute for Space
Aeronomie, Avenue Circulaire 3,
1180 Brussels, Belgium
\and \O ystein Lie-Svendsen \at Norwegian Defense Research
  Establishment (FFI), P.O. Box 25, NO--2027 Kjeller, Norway\\
  \email{Oystein.Lie-Svendsen@ffi.no} and
  Institute for Theoretical Astrophysics, Univ.\ of Oslo, Norway}
%\author{\O ystein Lie-Svendsen}
%\institute{\O ystein Lie-Svendsen \at Norwegian Defense Research
%  Establishment (FFI), P.O. Box 25, NO--2027 Kjeller, Norway\\
%  \email{Oystein.Lie-Svendsen@ffi.no} \and
%  Institute for Theoretical Astrophysics, Univ.\ of Oslo, Norway}
%%\author{Marius M. Echim}
%\institute{Marius E chim \at Belgian Institute for Space Aeronomie, Avenue Circulaire 3,
%1180 Brussels, Belgium \and
%  Institute for Space Sciences, Magurele, Romania}

\maketitle

\begin{abstract}
The paper reviews the main advantages and limitations of the kinetic exospheric
and fluid models of the solar wind (SW). The general theoretical background is also outlined:
the Boltzmann and Fokker-Planck equations, the Liouville and Vlasov equations,
the plasma transport equations derived from an "equation of change". 
The paper also provides a brief history of the solar
wind modeling. It discusses the hydrostatic model imagined by Chapman, the first supersonic
hydrodynamic models published by Parker and the first generation subsonic kinetic
model proposed by Chamberlain. It is shown that a correct estimation of the electric field
as in the second generation kinetic exospheric models developed by Lemaire and Scherer, provides
a supersonic expansion of the corona, reconciling the hydrodynamic and the kinetic
approach. The modern developments are also reviewed emphasizing the characteristics of
several generations of kinetic exospheric
and multi-fluid models. The third generation kinetic exospheric models
considers kappa velocity distribution function (VDF) instead of a Maxwellian at the
exobase and in addition they treat a non-monotonic variation of the electric potential with the
radial distance; the fourth generation exospheric models include Coulomb collisions based on the 
Fokker--Planck collision term. Multi-fluid models of the solar wind provide a coarse grained description
of the system and reproduce with success the spatio-temporal variation of SW macroscopic
properties (density, bulk velocity). The main categories of multi-fluid SW models
are reviewed: the 5-moment, or Euler, models, originally proposed by Parker to describe
the supersonic SW expansion; the 8-moment and 16-moment fluid models, the gyrotropic
approach with improved collision terms as well as the gyrotropic models based on observed
VDFs. The outstanding problem of collisions, including the long range Coulomb encounters,
is also discussed, both in the kinetic and multi-fluid context. Although for decades
the two approaches have been seen as opposed, in this paper we emphasize their complementary.
The review of the kinetic and fluid models of the solar wind contributes also to a
better evaluation of the open questions still existent in SW modeling
and suggests possible future developments.
\end{abstract}

\section{Introduction}
\label{sec:Introduction}
The existence of a corpuscular solar outflow with 
velocities of the order of $10^2$ km/s has been deducted by
\citet{Biermann53} based on observations of the cometary
tails. The   term ``solar wind'' was coined by Eugene Parker 
(\citeyear{Parker58a}) to describe the supersonic 
expansion of the solar corona. 
The concept received a theoretical description
the same year \citep{Parker58}.
%Since then the supersonic expansion of the solar corona continues
%to be the  subject of sustained scientific investigation. 
The solar wind (SW) models describe the physical mechanisms that accelerate
the plasma from the solar corona to supersonic velocities,  and
account for the SW properties at 1 Astronomical Unit (AU).
Indeed, in the Earth's neighborhood
the solar wind is quasi-continuously sampled  in-situ  by numerous spacecraft 
from almost the beginning of the
space era. One of the reasons for monitoring  the solar wind is that
  its sudden variations have a large
  impact on the Earth's middle and upper atmosphere, 
   {ionosphere and magnetosphere. In this paper we
review the kinetic and fluid models of the solar wind, with emphasis on their
historical evolution  and the main theoretical concepts stemming from
the classical transport theory. We are aware that an exhaustive
review of solar wind modeling is an outstanding task, beyond the scope
of our study. Therefore we
privilege the historical approach, aiming  to provide the reader with
the evolution of ideas and theoretical concepts
and approaches from which emerge kinetic and
fluid  models of the solar wind.}
% Therefore a physical description of the
%solar wind acceleration and properties is needed.

%Therefore fluid equations provide a coarse-grained description 
%of the dynamical state that still can 
%at spatio-temporal scales of the order 
%of ?c, ?c, RL, and/or TL.

The main difference between the kinetic and fluid description
results from the fundamental physical variables on which
is based each representation. 
The velocity distribution function (VDF) of each species, 
$f_s(\vec r, \vec v, t)$, provides a
 {rich} information on the
plasma dynamics; it is the key-variable of the kinetic
theory both in plasma and neutral gases. 
It gives at time $t$ the number of particles of species $s$ within a 
6-dimensional volume of the phase space, $d\vec r$ $d\vec v$,
centered in  ($\vec r$, $\vec v$) at the time t.
The computation of the VDF of each species from a
kinetic equation is sometimes very difficult. The fluid description,
based on the lower order moments of $f_s$,
%On one hand the kinetic models describe the evolution of
%the velocity distribution function (VDF) and of its moments, for each
%species $s$ of the gas/plasma; on the other hand the coarse-grained
%fluid models describe the spatio-temporal variation of several
%macroscopic variables 
i.e. the density $n_s(\vec r,t)$, the bulk
velocity $\vec u_s(\vec r, t)$, the temperature $T_s(\vec r, t)$, the
pressure tensor, $\tensor{p_s}$, is simpler and often sufficient for
interpretation of  experimental results.
%  In various practical situations the
%fluid modeling can provide a satisfactory description of the dynamical
%state even though {collective
%processes  (like micro-instabilities) are disregarded and the physics 
%at the microscopic, particle level is truncated. This may be the case
%in most space weather applications.} 

Nevertheless,  a thorough theoretical description of the
physics governing the dynamics of an ensemble of ionized 
molecules moving almost without collisions in the gravitational and
electromagnetic field
must necessarily take into account the smallest scales 
and  the corresponding velocity distribution functions. 
{Therefore,  the kinetic formulation should not be overlooked but comprehensively developed.}
%A critical review of the the kinetic and fluid modeling 
%can be found in the paper by \citet{Grad1958}. 
In this paper we discuss the merits and limitations of
the fluid and the kinetic  approach, 
applied to the problem of the supersonic
expansion of the solar corona. The complementarity of the
two has been also discussed recently by 
\citet{Parker2010}.
%these fundamental aspects will be developed in the next sections 
%where we will discuss fluid and kinetic models for space plasmas, 
%with emphasis on the solar wind. 
The models discussed in the following chapters are adapted
to plasma and electromagnetic/gravitational field parameters 
typical for our Sun; their application can be extended to the case of 
stellar  winds \citep{Meyer-Vernet2007}.
%The scope of this paper is not to be an exhaustive review of all
%solar wind modeling efforts. Several excellent  books and 
%papers have been already published during the last decades.
%Nor do we intend to provide a detailed analysis
%on how the results of models compare with data. 
%Solar wind models based on wave-particle interactions  
%are not discussed, a comprehensive 
%review on this subject has been published recently by \citet{Marsch2006}.

The validity of any physical description or model of a dynamical system is
determined by the length and time scales of the physical processes 
controlling  the state of the system at the smallest scales. 
In a dense neutral gas or dense non-magnetized plasma the latter are
determined by the binary collision between molecules ($e^-$, $H^+$, etc.). 
In plasma the smallest spatio-temporal scales depend 
on the density, temperature, as well as on the 
properties of the embedded electromagnetic field.
%A satisfactory physical description must
%"cover" the entire dynamical range of spatio-temporal scales linked 
%to the fundamental processes to be modeled. 
%The kinetic theory treats the individual dynamics of molecules; 
In  neutral gases and dense plasmas  the basic 
length scale is the mean-free-path, 
$\lambda_c$ -  the distance traveled by molecules between two collisions;
 the basic time scale is  the collision time, $\tau_c$ - the time
spanned between two encounters. 
In plasmas an additional  fundamental scale is the Debye length,  $\lambda_D$, the screening
distance for the Coulomb force.
When the plasma is magnetized  the   (Larmor)
radius of gyration, $R_L$  and the
gyration period $T_L$ are also 
fundamental scales.   For a broad range of densities, thermal energies and
ambient electromagnetic field, the spatial and temporal 
kinetic scales, typical for the dynamics of individual particles, can
be very small compared to the conventional, fluid macroscopic scales,
but sometimes their range overlap, giving rise to complex,
multi-scale processes. 
In this case the validity of the fluid approach
becomes questionable.

%The solar wind is an  example of an astrophysical
%multi-scale plasma process. The coarse-grained properties 
%of the solar wind (density, average velocity, temperature) 
%are described by fluid models; they have  predicted a supersonic
%expansion of the solar corona even before its discovery
%by interplanetary spacecraft.  In addition to a good 
%description of the macroscopic structure of the solar wind,
%kinetic models provide insight about the  physical processes
%that accelerate electrons and ions to supersonic velocities,
%as well as about  the energy distribution and transport.
%%In this paper we outline some of the fundamental features of
%%steady state kinetic models and multi-fluid models of the solar wind.
%%We emphasize their key-features and 
%%trace their common ``roots'' into the general space plasma physics theory.

 {Several reviews on solar wind modeling have been published
  in the past. Two of the most recent are the  works of}
\citet{Marsch2006} and the book  by   \citet{Meyer-Vernet2007},
 { where the basic concepts of solar wind modeling are
  illustrated and discussed from various perspectives.}
\citet{Marsch2006}  {emphasizes the role of wave-particle interactions and 
their kinetics for solar wind acceleration and dynamics;
  \citet{Meyer-Vernet2007} gives a physical perspective on the
  kinetics of solar wind models.}
 {In the following sections we provide the reader 
with a historical perspective on the solar wind modeling for over
half a century, emphasizing the classical transport theory,
a domain where the authors of this review contributed the most.}
A briefer account of this history can be found in 
\citet{LemairePierrard2001}. 

%In this way we can assess what still remains to be achieved
%by including into the models additional physical processes.

The paper is organized as follows: in section
\ref{fundamental_equations} we outline the theoretical
background by discussing the fundamental equations to be considered in
solar wind modeling.  {It shows
that the kinetic and fluid approaches have the
same theoretical root: the Boltzmann equation for
gases and plasmas.}
Section \ref{sec:history} gives an overview
of the early solar wind models, and put them into an historical
perspective. In section \ref{sec:kineticSW} we discuss 
the exospheric solar wind models, emphasizing
their recent evolution during the last decades. 
The merits and limitations of modern kinetic models are pointed out.
In section \ref{sec:fluid-models} we provide a technical review
of the multi-fluid models of the solar wind. We discuss 
their fundamental properties and results. The list of
mathematical definitions and symbols used throughout the paper
can be found in Appendix \ref{definitions}.
Appendix \ref{Boltzmann_solutions} includes a discussion on the 
Chapman-Enskog and Grad solutions
of the Boltzmann equation. It outlines the limitations of these
two approaches when the velocity distributions departs
significantly from displaced Maxwellians as it is the case
for the solar wind plasma at larger heliocentric distances. 
%from normal solutions to Grad expansion
%and Newton iteration. The aim of the Appendix
%\ref{Boltzmann_solutions} is first  to
%provide the reader  with the theoretical background and the 
%mathematical formulation of the current descriptions
%of solar wind models. In
%Appendix \ref{Boltzmann_solutions}  the classical 
%contributions in the field of kinetic theory are recalled with 
%a list of relevant references.
%We describe in the Appendix \ref{Boltzmann_solutions}
%some classes of solutions for the Boltzmann equation.
%In Appendix B we also discuss key-aspects  of the mathematical
%formulation. These 
%aspects are included in our review hoping they will stimulate
%young theoreticians to achieve further advances in this open field
%of investigation.}
%ready to join this open field of investigation.%

\section{Fundamental kinetic and hydrodynamic equations}
\label{fundamental_equations}
%Any mathematical description of a natural phenomenon such
%as the supersonic expansion of the solar corona,
%results from a set of fundamental equations. 
The equations relevant for the solar wind modeling are
 the fundamental equations of plasma physics. 
%They shall be briefly reviewed in the following paragraphs.
In sections \ref{sec:kineticSW} and \ref{sec:fluid-models}, we discuss particular solutions 
of these equations, applied to the supersonic  coronal expansion. 
% {In Appendix In Appendix \ref{Boltzmann_solutions} we briefly outline the
%mathematical
%framework of some of the most prominent methods to solve the
%Boltzmann equation.}

\subsection{The Boltzmann equation}
\label{Boltzmann_section}

A formal derivation of the Boltzmann equation from Liouville's theorem
of statistical mechanics can be found, for instance, 
in the paper by \citet{Grad1958} or in the
monographs by \citeauthor{Jancel+Kahan} (\citeyear{Jancel+Kahan}, pages
262--264) or \citeauthor{Uhlenbeck+Ford} (\citeyear{Uhlenbeck+Ford}, pages
118--138). A more intuitive derivation, close to the original
heuristic argument of Boltzmann, can be found
in the monographs by \citeauthor{Chapman1970}
(\citeyear{Chapman1970}, pages 46--68),  \citeauthor{Montgomery+Tidman} 
(\citeyear{Montgomery+Tidman}, pages 4--8) and in several recent
textbooks like those of  \citet{Gombosi1994} and 
\citet{SchunkNagy2004}.

Let $f_s(\vec v, \vec r,t) d\vec r\, d\vec v$ denote the number of
particles of species $s$ inside a phase space volume $d\vec r\, d\vec
v$ located at $(\vec r, \vec v)$, where $\vec r$ denotes position
and $\vec v$ velocity.  The time evolution of the phase
space density $f_s$ is then given by the continuity equation
\begin{equation}
  \label{eq:boltzmanncont}
  \pderiv{f_s}{t} = - \nabla_r \cdot ( \vec v f_s) - \nabla_v \cdot
  (\vec a_s f_s) + \sum_{t=1}^{N}\left(\frac{\delta f_{st}}{\delta t}\right)_\mathrm{coll}
\end{equation}
where the operators $\nabla_r$ and $\nabla_v$ denote derivatives with
respect to spatial and velocity coordinates and $\vec a_s=\vec F_s/m_s$ is the
acceleration, where  $\vec F_s$ is the total external force acting 
on the particle and $m_s$ is the particle mass.  

The first term on the right-hand side expresses the
change in the number of particles in the phase space volume because
particles enter and leave (at the velocity $\vec v$) the volume
$d\vec r$. The second term expresses the analogous change because
particles enter and leave the volume $d\vec v$ in velocity space.
%Indeed the acceleration $\vec a_s$ expresses the rate at which particles change
%their velocity.  
The term denoted $\delta f_{st}/\delta t_\mathrm{coll}$
corresponds to the rate of change of the VDF $f_s(\vec r, \vec v, t)$ due to 
collisions between species $s$ and $t$. It includes self-collisions.
%In case of a multi-species system, this term comprises 
%collisions between particles of the same kind but also between
%different species of particles.
In the most  general case the VDF can also change due to other processes,
e.g. ionisation and recombination. These processes 
may be important particularly in the solar corona but they are
not discussed here. Note that (\ref{eq:boltzmanncont}) is 
nothing but the expression of the conservation of particles, 
and as such bears a close resemblance, both
in form and content, to the standard continuity equation
(\ref{eq:cont}) that we shall encounter later on in the fluid
description.

%Expressing the acceleration as $\vec a_s = \vec F_s/m_s$, where $F_s$
%is the total external force acting on the particle and $m_s$ the
%atomic mass of the particle, and 
Assuming that the force $\vec F_s$ is either 
independent of $\vec v$,  or is the Lorentz force, 
the Boltzmann equation (\ref{eq:boltzmanncont}) may be
cast in a concise form:
\begin{equation}
  \label{eq:boltzmann}
\frac{\mathcal{D}f_s}{\mathcal{D}t}= \sum_{t=1}^{N}{J_{st}(f_s,f_t)}
%  \pderiv{f_s}{t} + \vec v \cdot \nabla_r f_s + \frac{\vec F_s}{m_s} \cdot
%  \nabla_v f_s = \frac{\delta f_s}{\delta t}_\mathrm{coll}.
\end{equation}
%\begin{equation}
%\mathrm{\frac{\partial f_s} {\partial t} +
%  \mbox{\boldmath$v$}\cdot\frac{\partial f_s}
%       {\partial\mbox{\boldmath$r$} } } = \mathcal{J}
%\label{Boltzmann_neutral}
%\end{equation}
%In equation (\ref{eq:boltzmann_epsilon}) $\vec F_s$ denotes the external force,
%$J_s$ is the collision integral and 
%$\epsilon$ is a small parameter associated  to the
%smallest spatial scale (can be the mean-free-path) or the smallest
%time period (the collision time) or both.  In the reminder of the
%Appendix  we denote the left hand side of equation
%(\ref{eq:boltzmann_epsilon}) by $\dot{f}$.
where we used the notation $\mathcal{D}/\mathcal{D}t=\partial/\partial
t + \vec v \cdot \vec \nabla_{\vec r} + \vec F_s/m_s$ and
$J_{st}(f_s,f_t)$ denotes the right-hand side of the Boltzmann equation, or the
Boltzmann collision integral, that is discussed in some detail in the following
paragraphs.

In practical situations it is useful to define
 the ``peculiar'' velocity,
$\vec c_s \equiv \vec v - \vec u_s$,  
 in addition to the actual velocity $\vec v$
 \citep{Grad1958,Schunk1977};
$\vec u_s$ is the mean velocity of species $s$.
 After this change of variable and
considering that the force term, $\vec F_s = \vec F_{sG} + \vec F_{sEM}$,
is the sum of gravitational, $\vec F_{sG}=m_s \vec g$ and electromagnetic, 
$\vec F_{sEM}= e_s\vec  E + e_s \vec v \times \vec B$ forces,
the Boltzmann equation (\ref{eq:boltzmann}) takes the form
%(\citeauthor{Jancel+Kahan}, \citeyear{Jancel+Kahan}, pg. 409):
%\begin{equation}
%  \label{eq:boltzmann_actual_v}
%  \frac{Df_s}{Dt} + \vec u_s \cdot \nabla_r f_s + 
%  \left( \frac{\vec F_s}{m_s} -\frac{D\vec u_s}{Dt} \right) \cdot
%  \nabla_{\vec c_s} f_s - 
%  \left( \vec \nabla_{\vec c_s} f_s \right) \vec c_s :
%  \vec\nabla_r \vec u_s = 
%  \frac{\delta f_s}{\delta t}_\mathrm{coll}
%\end{equation}
\citep{Schunk1977}:
\begin{eqnarray}
  \pderiv{f_s}{t} + (\vec c_s +\vec u_s)\cdot \vec \nabla f_s
  -\frac{D_s \vec u_s}{Dt}\cdot \vec \nabla_{\vec c_s}f_s - \vec
  c_s\cdot \vec \nabla \vec u_s \cdot \vec \nabla_{\vec c_s}f_s +
  \nonumber & & \\
\left[ \vec g + \frac{e_s}{m_s}\left( \vec E + \vec u_s \times \vec
  B\right)  \right]\cdot \vec \nabla_{\vec c_s}f_s
  +\frac{e_s}{m_s}\left( \vec c_s \times \vec B  \right) \cdot \vec
  \nabla_{\vec c_s}f_s &=&  \frac{\delta f_s}{\delta t}_\mathrm{coll},
 \label{eq:boltzmann_actual_v}
\end{eqnarray}
where the convective derivative is defined by:
\[
 \frac{D}{Dt}\equiv \pderiv{}{t} + \vec u_s \cdot \vec
 \nabla_{\vec r}.
\]
In this formulation the $m$-order moment of the velocity distribution function is
defined by:
\begin{equation}
\label{eq:moment_VDF}
n(\vec r, t)\mathcal{\vec M}_s^{(m)}(\vec r, t) = \int \int \int {{\vec c}_s^m f_s(\vec c_s, \vec r, t)}\mathrm{d} \vec c_s
\end{equation}
where $n(\vec r, t)$ is the zero-order moment ($m=0$) giving the number
density. 
%A similar expression corresponds to formulation (\ref{eq:boltzmann}).
The first order moment ($m=1$) corresponds to  the average velocity, the second-order
moment determines the pressure tensor, etc.  For convenience, a list
of definitions of the moments of order $m \leq 3$  and of their corresponding
plasma macroscopic quantities is given in Appendix \ref{definitions}.

A key problem for solving the Boltzmann equation is the treatment of 
the collision term,  $\delta f_s/\delta t_\mathrm{coll}$ 
in (\ref{eq:boltzmann}) and (\ref{eq:boltzmann_actual_v}).  
%An excellent discussion can be found
%in the seminal paper by \citet{Grad58}. 
Classical review papers and textbooks (e.g.
the paper by \citeauthor{Grad1958} \citeyear{Grad1958},  or the monographs
by \citeauthor{Chapman1970} \citeyear{Chapman1970}, pg 56-66, or
\citeauthor{Jancel+Kahan} \citeyear{Jancel+Kahan}, \citeyear{Jancel+Kahan}, pg. 264-267),
give a formal derivation of the collision term, based on the hypothesis of
molecular chaos and elastic binary collisions. 
The  following expression are found:
\begin{equation}
  \label{eq:boltzmann_J}
%  \frac{\delta f_s}{\delta t}_\mathrm{coll} &=& \int_{0}^{2\pi}
%  \int_{0}^{\pi} \int_{-\infty}^{+\infty} \int_{-\infty}^{+\infty} 
% \int_{-\infty}^{+\infty} \left( f'_s f'_t -
%  f_s f_t\right)  \left| \vec g_{st}\right|  b \hspace{0.075cm}
%  db  \hspace{0.075cm} d\epsilon  \hspace{0.075cm} dv_{j x}
%  \hspace{0.075cm}  dv_{j y}  \hspace{0.075cm}  dv_{j z}\nonumber \\
% \frac{\delta f_s}{\delta t}_\mathrm{coll}  
J_{st}(f_s)=  \int_{0}^{2\pi} 
  \int_{0}^{\pi} \int_{-\infty}^{+\infty} \int_{-\infty}^{+\infty} 
 \int_{-\infty}^{+\infty} \left( f'_s f'_t -
  f_s f_t\right)  \left| \vec g_{st}\right|  \sigma_{st}(g_{st}, \theta)  \hspace{0.075cm} 
  \sin \theta d\theta  \hspace{0.075cm}  d\phi   \hspace{0.075cm}
  d\vec v_{t} 
\end{equation}
%In (\ref{eq:boltzmann_J})
%$\theta$ is the polar angle, $b$ is the impact parameter (the distance at
%which the particles would pass one from another
%if there were no collisions), 
where $\vec g_{st}= \vec v_s - \vec v_t$ 
is the relative velocity between particle $s$ and
$t$; $\theta$ is the  zenital/deviation angle of  $\vec g_{st}$, and $\phi$ is
the azimuthal angle;
%after collision with $bdbd\epsilon=\sigma(g_{ij},\theta)d\Omega$; 
$\sigma_{st}$ is the differential collision cross section;
%and $\Omega$ is the solid angle;  
primed and non-primed quantities are 
considered  after and respectively before
collisions between particles $s$  and $t$ 
($s\equiv t$ in case of self-collisions).

The functional relation between $\sigma_{st}$, $|\vec g_{st}|$  and $\theta$
is determined by the potential energy of the system of colliding particles and by
the collision kinematics. 
When the colliding molecules/particles can be
approximated by rigid spheres, 
 the differential collision cross-section, $\sigma_{st}(|\vec g_{st}|, \theta)$ is a constant.
%A graphical representation of the kinematics of an elastic binary
%collisions may  be found in the monograph by Chapman and Cowling
%(\citeyear{Chapman1970}, pg. 56-62).
%Some of the main difficulties results from the 
%nonlinearity of  $J_s$.

Note that in the lower corona and the inner solar wind 
% besides elastic Coulomb collisions, other 
additional processes take place, like  ionisation, recombination, 
charge exchange and radiation; they  are not included in the
collision term (\ref{eq:boltzmann_J}) and are therefore neglected
in models of the more distant solar wind.
 {Wave--particle interactions cannot be neglected
when the energy density of the electromagnetic waves
(MHD, ion cyclotron, whistlers, etc.) is comparable to the kinetic energy of the 
plasma particles. The effects of wave-particle interactions is 
addressed in the review by \citet{Marsch2006} and in monographes of the
solar corona} (e.g., \citeauthor{Aschwanden2009}
\citeyear{Aschwanden2009}).

\subsection{The Fokker-Planck equation}
\label{Fokker-Planck}

%A direct application of the Boltzmann equation 
In a plasma, in addition to binary encounters, the charged particles
interact/collide through the long range Coulomb potential.
The differential cross section of Coulomb collisions between  
particles with charge $e_s$ and $e_t$ is given by:
\begin{equation}
\sigma_{st}(g,\theta)=\frac{e_s^2 e_t^2(m_s+m_t)}{4g^4 m_sm_t
  \sin^4(\theta/2)}
\label{Coulomb_crosssec}
\end{equation}
Therefore the  Boltzmann  collision term described by (\ref{eq:boltzmann_J}) 
diverges in the case of Coulomb interactions for small impact
parameters, $\theta \to 0$, and the hypothesis of an infinitely small
time of the binary interaction is no more justified.
%the Boltzmann
%collision term has to be refined for partially ionised gases and plasmas.
%and Vlasov equations
%(\ref{eq:Vlasov})--(\ref{eq:Maxwell}) with charge and current
%densities given by (\ref{eq:moment_VDF}), represent two plasma
%descriptions in which the long-range and respectively
%short-range/binary collisions are neglected.  In the plasma universe
%one can find many examples for which one or the other of these two
%approaches can be applied.  Nevertheless, the effect of collisions
%between charged particles in multi-scale space plasmas is even more
%complex since a particle interacts simultaneously with all other
%particles within the Debye sphere and, by definition of a plasma,
%there are many particles inside the Debye sphere.  
One can approximate, however, that each long range Coulomb
collision leads to a tiny deflection of the particle trajectory,
described by  the Boltzmann integral (\ref{eq:boltzmann_J}) 
and then simply add the effect of all collisions
\citep{Rosenbluth1957}. In order to avoid the divergence
for small deflection angles the  integral (\ref{eq:boltzmann_J}) is cut-off
at some angle $\theta_{min}$ that depends on the Debye length. In
a proton-electron plasma this relationship is 
given by \citep{Montgomery+Tidman}:
\begin{equation}
\sin\frac{\theta_{min}}{2} \approx \frac{e^2}{mg^2L_D}
\label{theta_min}
\end{equation}
where the Debye length:
\[
L_D=\sqrt{\frac{kT}{8\pi n_0 e^2}}
\]
is defined as the distance over which the Coulomb field of a test-charge is
screened-off by collective effects of the plasma charges. 
%The Debye screening is a fundamental property of plasmas in general. 
%Hence one can use the the Boltzmann collision term is still valid, but in order to
%evaluate the term analytically or numerically it must be cast in a
%form taking into account the dominance of small-angle collisions.
%Assuming that the probability $\psi(\vec v, \Delta \vec v)$ for
%increasing  a particle's velocity from $\vec v$ to $\vec v + \Delta \vec v$
%in a time $\Delta t$ does not depend on time (the hypothesis of
%a Markovian process) one can then express the
%rate of change of the velocity distribution function due to multiple
%collisions as below (\citeauthor{Rosenbluth1957},
%\citeyear{Rosenbluth1957}; 
%see also \citeauthor{Clemmow+Dougherty}, \citeyear{Clemmow+Dougherty}, p.\ 423--431):

Following this line of thought, the distribution function $f_s$ in the 
Boltzmann integral can be expanded in terms of
$\Delta \vec v_s=\frac{m_t}{m_s+m_t}\Delta \vec g$, up to the
second order. The collision term at the right hand side of the
Boltzmann equation can be rewritten as:
\begin{equation}
\label{eq:Fokker_Planck_Rosenbluth}
J_{st} = \pderiv{}{v_i}\left[ -A_i^{s t} f_s +
  \frac{1}{2}\pderiv{}{v_j}\left( B_{ij}^{s t} f_s  \right)\right]
%\left(\frac{\delta f}{\delta t}\right)_{coll} = \Gamma \left\{
% - \frac{\partial}{\partial \vec  v_1}\cdot \left[ f(\vec v_1) \pderiv{H(\vec v_1)}{\vec v_1} \right]+
%\frac{1}{2}\frac{\partial^2}{\partial \vec v_1 \partial \vec  v_1}:
%\left[f(\vec v_1) \frac{\partial^2G(\vec v_1)}{\partial \vec
%  v_1 \partial \vec v_1} \right] \right\}
\end{equation}
%\begin{equation}
%\label{eq:Fokker_Planck}
%\left(\frac{\delta f}{\delta t}\right)_{coll} = -\frac{\partial}{\partial \vec
%  v}\cdot\left(f\left<\Delta \vec v\right>\right) +\frac{1}{2}\frac{\partial^2}{\partial
%  \vec v \partial \vec v}:\left( f\left< \Delta \vec v \Delta \vec v \right>\right)
%\end{equation}
%where:
%\[
%\left\{ \begin{array}{c} \left< \Delta \vec v\right> \\
%\left< \Delta \vec v \Delta \vec v\right> \end{array} \right\}
%=\frac{1}{\Delta t}\int{\psi\left(\vec v, \Delta \vec v\right) 
%\left\{ \begin{array}{c} \left< \Delta \vec v\right> \\
%\left< \Delta \vec v \Delta \vec v\right> \end{array} \right\} }
%d\left(\Delta \vec v\right)
%\]
%\comment{ols: Because Fokker-Planck can be derived from Boltzmann, it
%  is not necessary to show~(\ref{eq:Fokker_Planck}) --- we could jump
%  straight to the Rosenbluth potentials?}
where tensor notation is used with summation convention over
identical indices; $i$, $j$
correspond to Cartesian components, $s$, $t$ indicate
species. 
The term (\ref{eq:Fokker_Planck_Rosenbluth}) is known as the Fokker-Planck
collision term;  $A_i^{st}$ is the coefficient of
dynamical friction and corresponds to slowing-down effects; the term
$B_{ij}^{st}$ is the coefficient of diffusion in velocity space.
By assuming an inverse square law for the
inter-particle force and a small Debye length,
%Fokker-Planck transport coefficients,   
$A_i^{st}$  and $B_{ij}^{st}$,
from (\ref{eq:Fokker_Planck_Rosenbluth}) can be written (\citeauthor{Rosenbluth1957}, 
\citeyear{Rosenbluth1957}; \citeauthor{Clemmow+Dougherty},
\citeyear{Clemmow+Dougherty}, p.\ 423--431):
%\begin{equation}
%\label{Fokker_Planck_Rosenbluth}
%\left(\frac{\delta f_s}{\delta t}\right)_{coll} =
%\pderiv{}{v_i}\left[ -A_i^{s t} f_s +
%  \frac{1}{2}\pderiv{}{v_j}\left( B_{ij}^{s t} f_s  \right)\right]
%%\left(\frac{\delta f}{\delta t}\right)_{coll} = \Gamma \left\{
%% - \frac{\partial}{\partial \vec  v_1}\cdot \left[ f(\vec v_1) \pderiv{H(\vec v_1)}{\vec v_1} \right]+
%%\frac{1}{2}\frac{\partial^2}{\partial \vec v_1 \partial \vec  v_1}:
%%\left[f(\vec v_1) \frac{\partial^2G(\vec v_1)}{\partial \vec
%%  v_1 \partial \vec v_1} \right] \right\}
%\end{equation}
\begin{eqnarray}
\label{Rosenbluth_FP_coefficients_A}
A_{i}^{s t} & = & \frac{1}{4 \pi}\left( \frac{e_s
  e_t}{\epsilon_0 m_s}  \right)^2 \frac{m_s +
  m_t}{m_t} \ln \Lambda \pderiv{}{v_i} \int{\frac{f_t
  (\vec v')}{\left|\vec v - \vec v'  \right|} d^3 v'} \\
\label{Rosenbluth_FP_coefficients_B}
B_{ij}^{s t} & = & \frac{1}{4 \pi}\left( \frac{e_s
  e_t}{\epsilon_0 m_s}  \right)^2 \ln \Lambda 
 \frac{\partial^2}{\partial v_i \partial v_j} 
 \int{{\left|\vec v - \vec v'  \right|}f_t(\vec v')  d^3 v'} 
\end{eqnarray}
where primes denote post-collision quantities and 
\[
\Lambda = \frac{12 \pi \left( \epsilon_0 kT \right)^{3/2}}{n^{1/2} e^3}
\]
is the plasma parameter, with $\epsilon_0$ the electric permittivity of
vacuum, $k$ the Boltzmann constant, $T$ the plasma temperature,
$n$ the plasma density, and $e$ is the elementary charge.
%The Fokker-Planck equation can be,
%however, derived directly from the principles
%of the statistical mechanics. 
The form of the collisional integral given in 
(\ref{eq:Fokker_Planck_Rosenbluth})--(\ref{Rosenbluth_FP_coefficients_B}) 
is precisely the same as the one derived by \citet{Landau1936} 
for the kinetic equation of plasma state,  from statistical mechanics
arguments; therefore it is sometimes also
called the Landau collision term. 

%An additional level of complexity and an increased degree of
%collective effects included in the collision term is achieved 
%in the Balescu-Lenard description of plasma kinetics 
%\citep{Balescu1960,Lenard1960}. 
%Since no solar wind model based on the Balescu-Lenard approach
%has been yet proposed we do not develop
%any further this approach. We  
%mention it here as a possible development
%for a next generation of  solar/stellar winds models.  \comment{ols: I
%don't think it is relevant at all to the solar wind.  I believe those
%collective effects are only important at very high densities (when
%correlations among particles increase), while the corona and solar
%wind is an extremely \emph{low} density environment.  Suggest we
%remove this paragraph.}

\subsection{The Liouville and Vlasov equations}
\label{Vlasov}

When the mean-free path of plasma particles is
large compared to the characteristic spatial dimension of the system
itself, and the time between collisions is larger
than the characteristic time scale, and due to the long-range Coulomb
interactions between  charged particles,
the many-body interactions cannot be neglected anymore in the
description of plasma dynamics.
%In other words the orbit of a particle is deflected due to the
%interaction with many neighbors.
%This situation is encountered for fully ionised, collisionless
%plasmas, like, for instance, the solar wind.
The effect of collective encounters is not appropriately evaluated by the
Boltzmann integral (\ref{eq:boltzmann_J}) describing binary, elastic collisions.
%based on the one-particle velocity distribution
%function, $f_s(\vec r, \vec v, t)$. 
The Liouville theorem gives then the appropriate theoretical framework for 
the description of plasma dynamics.

%\sout{In contrast to the Boltzmann equation,}
The fundamental concept of the statistical approach is the
phase space probability density, $D_N(\vec X_1, \vec X_2, \ldots, \vec
X_N,t)$, where $\vec X_i\equiv(\vec r_i, \vec v_i)$ and $N$ is the
total number of particles. 
$D_N$ denotes the probability that particle 1 is located in [$\vec r_1,
  \vec r_1 + d\vec r$] and its velocity pertains to the interval
 [$\vec v_1, \vec v_1 +   d\vec v$], while particle 2 is  located in [$\vec r_2,
  \vec r_2 + d\vec r_2$] and its velocity pertains to [$\vec v_2, \vec v_2 +
  d\vec v$], etc. Thus $D_N$ is defined in the $6N$ dimensional
space of the positions and velocities of the
 $N$ plasma constituents (electrons, protons, etc.). 
The Liouville theorem states that 
{the statistical} ensemble is represented  in
  the $6N$ dimensional space by a  ``cloud'' whose volume does not
change with time, like in incompressible flows:
\begin{equation}
\pderiv{D_N}{t}+\sum_{j=1}^{N}\left\{ \pderiv{D_N}{\vec r_j}\cdot 
\dot{\vec r}_j + \pderiv{D_N}{\vec v_j}\cdot \dot{\vec v}_j  \right\} = 0
\label{Liouville}
\end{equation}
where $\dot{\vec r}_j$ and $\dot{\vec v}_j$ denote time derivatives.
Note that equation (\ref{Liouville}) gives only a formal expression of
the Liouville theorem, in its most general form, also known as a
master equation.  General solutions of (\ref{Liouville}) have not been
found yet. A discussion of the physical content of 
$D_N$ and  particular forms and solutions
of the master equation (\ref{Liouville}) for different
interaction potentials between particles can be found in 
\citet{Balescu}. 
%A description on how the Boltzmann equation (\ref{eq:boltzmann})
%is derived from Liouville's theorem (\ref{Liouville}) is given by
% \citet{Grad1958}.

The key-elements of the statistical theory of plasma physics are
described in classical monographs (e.g., \citeauthor{Montgomery+Tidman}, 
\citeyear{Montgomery+Tidman}, chapters 4-6; \citeauthor{Balescu},
\citeyear{Balescu}, pg. 26-55).
In the BBGKY (Born-Bogoliubov-Green-Kirkwood-Yvon) approach 
the strategy to solve (\ref{Liouville})  is based on a
hierarchy of equations derived  by
integration of  (\ref{Liouville})  in subdomains
of the  phase space. The BBGKY hierarchy of equations 
couples the lower order, or reduced, probability distribution
functions to the higher order ones (see \citeauthor{Bogoliubov1962},
\citeyear{Bogoliubov1962}; also \citeauthor{Montgomery+Tidman},  
 \citeyear{Montgomery+Tidman}, pg. 41-50). 
%The lower order
%probability distribution functions are obtained by integrating
%$D_N(\vec X_1, \vec X_2, \ldots, \vec X_N)$ in subdomains of the $6N$
%dimensional space. 
The reduced one-particle distribution function, $f_1$,
is computed by integration in the $6(N-1)$ dimensional sub-domain
defined by the coordinates and velocities of the other $N-1$ particles:
\begin{equation}
f_1 (\vec X_1)= \int\int \ldots \int{D_N(\vec X_1, \vec X_2, \ldots, \vec
  X_N)d\vec X_2, d\vec X_3, \ldots, d\vec X_N}
\label{VDF_BBGKY}
\end{equation}
$f_1$ defined by (\ref{VDF_BBGKY}) must be identified with the 
velocity distribution function described by the
Boltzmann equation, (\ref{eq:boltzmanncont}), (\ref{eq:boltzmann}) 
or (\ref{eq:boltzmann_actual_v}).
The BBGKY chain of equations relates $f_1(\vec X_1)$ with 
$f_2(\vec X_1, \vec X_2)$, $f_3(\vec X_1, \vec X_2,  \vec X_3)$ and all the 
higher order reduced probability density. The first equation of this
chain can be written in a simplified form, when there is no
magnetic field and the factor
$G=1/(n_0L_D^3) \to 0$ (see, e.g.,
\citeauthor{Montgomery+Tidman}, \citeyear{Montgomery+Tidman}, pg. 48-49):
\begin{equation}
\pderiv{f_1}{t} + \vec v_1 \cdot \pderiv{f}{\vec r_1} -
\frac{n_0}{m_s}\left[ \int \pderiv{\phi_{12}}{\vec r_1} f_1(\vec r_2,
  \vec v_2) d\vec r_2 d\vec v_2  \right]\cdot \pderiv{f_1}{\vec v_1} = 0
\label{Vlasov_BBGKY}
\end{equation}
with $\phi_{12}$  the two-particle 
interaction potential and $n_0$ the mean particle density.
When (\ref{Vlasov_BBGKY}) is valid, all the higher order corrections
of the BBGKY hierarchy are equal to zero.
Equation (\ref{Vlasov_BBGKY}) is also known as
the Vlasov equation. 
%The previous paragraphs
%outline  the statistical  mechanics 
%``roots'' of the Vlasov approach.

Since the one-particle distribution function $f_1$ from
(\ref{VDF_BBGKY}) and (\ref{Vlasov_BBGKY}) has the same physical
content as the velocity distribution function from the Boltzmann
equation (\ref{eq:boltzmann}), the Vlasov equation is also called the 
Boltzmann equation without collisions.  Note also that 
\citet{Balescu}  developed an alternative way 
to solve the Liouville equation (\ref{Liouville}) based on  
\citeauthor{Prigogine}'s (\citeyear{Prigogine}) 
method of diagrams for non-equilibrium statistical mechanics. 
Balescu's  approach is based on a
direct solution for $D_N(\vec X_1, \vec X_2, \ldots, \vec
X_N,t)$ from (\ref{Liouville}). 
When $G\to 0$ the Prigogine-Balescu solution
leads to the same Vlasov equation (\ref{Vlasov_BBGKY}).
In the following we adopt the kinetic notation introduced in section 
\ref{Boltzmann_section} and write the Vlasov equation in the standard
form, taking into account a gravitational and magnetic field:
\begin{equation}
  \frac{\partial f_s} {\partial t} +
  \left[ \frac{q_s}{m_s}
    \left( \vec E  + \vec v \times  \vec B  \right) + m_s \vec a_g \right]
  \cdot \nabla_v f_s +
  \vec v \cdot \nabla_r f_s = 0
\label{eq:Vlasov}
\end{equation}
where $\vec a_g$ is the gravitational acceleration.
In (\ref{eq:Vlasov}) 
%the collective effects
%are included in the force term. Indeed, 
the electric ($\vec E$) and magnetic ($\vec B$) fields are  {derived from} the 
charge ($\rho$) and current density ($\vec j$); thus
equation (\ref{eq:Vlasov}) must be coupled to Maxwell's equations:
\begin{eqnarray}
\vec \nabla \cdot \vec B & = & 0 \label{divB}\\
\vec \nabla \times \vec E + \pderiv{\vec B}{t}  & = & 0 \\
\epsilon_0 \vec \nabla \cdot \vec E & = & \rho_c^{ext} + \sum_s{e_s n_s} \\
\frac{1}{\mu_0} \vec \nabla \times \vec B - \epsilon_0  \pderiv{\vec
  E}{t} & = & {\vec j}_{ext} +  \sum_s{\vec j_s}
\label{eq:Maxwell}
\end{eqnarray}
where $e_s$ is the charge of  species $s$.
The external charge and current densities
\footnote{The external charges and 
  currents were generally ignored in geophysical and astrophysical 
applications of Vlasov-Maxwell equations, (\ref{eq:Vlasov}) to (\ref{eq:Maxwell}). 
Therefore, the electric  field  $\vec E$ is 
restricted to the field generated by
local charges or induced inside the plasma by the  motion (convection)  across 
magnetic field lines. 
However, electrostatic fields are also produced by 
polarization induced by, e.g. the gravitational force or centrifugal effects, 
as well as by thermoelectric effects. 

Similarly, the magnetic field intensity $\vec B$,
%in equations   (\ref{eq:Vlasov}) to (\ref{eq:Maxwell}) 
were often restricted to the internal diamagnetic contribution of 
the local current density. The contribution of external
  (non-local or distant) electric currents or magnets 
were often overlooked in ideal MHD applications. However, it should be pointed out,
that adding this external curl-free component of B does not affect/modify 
eqs. (\ref{divB}) to (\ref{eq:Maxwell}),  nevertheless
it can certainly changes the solution of eq. (\ref{eq:Vlasov}) 
and consequently the velocity distribution function $f_s$.},
$\rho_c^{ext}$ and ${\vec j}_{ext}$, have been separated
from $n_s$ and $\vec j_s$, the internal contribution of the plasma itself, 
given by the zero and first order moments of $f_s$
(see (\ref{eq:moment_VDF}) and Appendix \ref{definitions}).
A Vlasov-Maxwell equilibrium is defined by solutions to 
the coupled system of equations
(\ref{eq:Vlasov})--(\ref{eq:Maxwell}) with the
definitions (\ref{eq:moment_VDF}).
%Some authors \citep{Clemmow+Dougherty} call the entire set of equations the
%``Vlasov equations'' to stress the coupling between the equation
%of state for the velocity distribution function, its first order
%moments and Maxwell's equations. 
%It is important to recall that 
%here that 
%where the effects of long range, collective interactions are included
%in the electric field of eq. (\ref{eq:Vlasov}).
% describing collective
%effects inside and outside the Debye sphere.

%\begin{eqnarray}
%\rho_s = e_s 
%\label{density:Maxwell}
%\end{eqnarray}
%In (\ref{eq:Vlasov}) the electric and magnetic fields are determined by 
% external as well as by own internal plasma charge and current
% density. The latter are determined by the moments of the velocity
%distribution function given by (\ref{eq:Vlasov}).

 {In summary, the chain of equations described in this section
illustrate the microscopic description of plasmas.
The kinetic approach considers} the micro-physics of particle dynamics, 
for each plasma component species. Therefore, the kinetic theory
has the following  advantages:
%identify some of the advantages of thkinetic models of the solar wind,
%that will be illustrated in the following sections:
%Although not limited at the list mentioned below,
%kinetic models strong points come from :
\begin{itemize}
\item Kinetic models describe the velocity distribution for
  each component species separately and treat self-consistently the
  coupling between the particle dynamics, external forces and the electromagnetic field;
\item Collisions and wave--particle interaction can be included; 
  relevant solutions have thus been developed based on the Fokker-Planck equation;
%\comment{ols: However, proton
%    self-collisions, which are essential, are extremely difficult to
%    include consistently in a kinetic model.}
\item The entire set of moments equations  is
  satisfied by the solutions of the kinetic solution;
\item The moments of steady state velocity distributions are
 analytical functions determined by the
  electromagnetic field and the  VDF parameters at the boundaries of
 the integration domain;
%\item Many particles species can be included and treated independently;
\item The kinetic solutions are self-consistently  coupled to Maxwell's equations;
%\item  The problem of  wave-particles interactions can also be
%  addressed in the framework of the kinetic theory; this aspect is not
%considered here.
\end{itemize}
The advantages and the limitations of solar wind models based on the
kinetic theory are discussed in section \ref{sec:kineticSW}.

\subsection{Plasma transport equations}
\label{Transport_eqs}

There are physical situations, including the case of the solar wind,
where the details of the velocity distribution functions are not
measured directly. Only the spatial and temporal 
variation of the lower order moments can then be evaluated,
the density, $n_s(\vec r, t)$, the temperature, 
 $T_s(\vec r, t)$, the bulk velocity,  $\vec u_s(\vec r, t)$,
the pressure,  $P(\vec r, t)$, etc.
%practical situations, including some applications of the solar wind
%models, the level of details entailed by the kinetic equations
%may exceed what is actually needed.
%Indeed, when one is interested only on an average, 
%coarse grained description of the system one looks for 
%a description of the spatio-temporal evolution of
%macroscopic quantities like the density, temperature, bulk velocity,
%pressure, heat flux. 
These  observables can be also determined from the solutions
of the Boltzmann, Vlasov or Fokker-Planck equations. Their 
temporal and spatial evolution is derived by integrating the kinetic equations
in the velocity space. One then solves the so-called ``equations of
change'':
\begin{equation}
\label{eq:Change}
\int\int\int\Phi_s(\vec v)\frac{\mathcal{D}f_s}{\mathcal{D}t}d\vec v=
\sum_{t=1}^{N}\int \int \int {\Phi_s(\vec v)J_{st}(f_s,f_t)} d\vec v
\end{equation}
where $\Phi_s(\vec v)$ is a generic notation standing for various 
powers of the velocity.  {From (\ref{eq:Change})  one retrieves the partial derivative
equations describing the spatio-temporal evolution of the moments
of the VDF, also known as the plasma transport equations.
%Note also that the hydrodynamic description is obtained from
%(\ref{eq:Change}) with the ``summational invariants'':
%$\Phi(\vec v)=1$,  $\Phi(\vec v)=m_s\vec v$,  $\Phi(\vec v)=\frac{1}{2}m_s\vec v^2$.
If one replaces in (\ref{eq:Change}) $\Phi_s=m_s$, $\Phi_s=m_s c_{si}$,
and  $\Phi_s=\frac{1}{2}m_s c_s^2$, where $c_{si}$ are the components
of the peculiar velocity $\vec c_s= \vec v - \vec u_s$ 
(with $\vec u_s$ the average velocity  of species $s$ defined in 
Appendix \ref{definitions}) and after integration over
velocity one  obtains the continuity, momentum and energy
transport equations for species $s$  \citep{Schunk1977}:}
%\begin{eqnarray}
%koi & = &
%\end{eqnarray}
\begin{eqnarray}
  \label{eq:cont}
  \pderiv{n_s}{t} + \nabla \cdot (n_s \vec u_s) &=& 0\\
  m_s n_s \frac{D_s \vec u_s}{Dt} + \vec \nabla \cdot
  \tensor{p_s} - n_sm_s\vec g -n_s e_s( \vec E + \vec u_s \times \vec B) &=& \frac{\delta \vec
    M_s}{\delta t}_{coll}\label{eq:mom} \\
  \frac{D_s}{Dt} \left( \frac{3}{2} n_skT_s \right) + \frac{5}{2}
  n_skT_s\left(\nabla \cdot \vec u_s\right) +\vec \nabla \cdot \vec  q_s 
+  \tensor{p_s}:\vec \nabla \vec u_s   &=&  
  \frac{\delta E_s}{\delta t}_{coll}  \label{eq:emom}
\end{eqnarray}
where the right hand sides denote the collision terms, and 
$n_s$ is the number density, $\vec u_s$ is the average velocity,
$\tensor {p_s}$ is the pressure tensor, $k$ is the Boltzmann constant,
$T_s$ is the temperature (see Appendix \ref{definitions} for definitions of these
macroscopic variables).
%\comment{(ols: Added next two sentences.)}  
In (\ref{eq:emom}) - (\ref{eq:mom}) the
production and loss of particles are  disregarded, which is important 
in the transition region and
below.  If these terms are included the right-hand side of (\ref{eq:cont})
becomes nonzero, and additional terms are added to (\ref{eq:mom})--(\ref{eq:emom}).
%the momentum and energy equations acquire
%additional terms as well.  
In (\ref{eq:cont})--(\ref{eq:emom}) the
force term is of electromagnetic and gravitational origin.
  Since $m_s$, $m_s v_i$, and $\frac{1}{2}m_s v^2$ 
%(also called \emph{summational invariants}) 
are conserved in binary, elastic
collisions, the corresponding moments of the Boltzmann collision
operator (\ref{eq:boltzmann_J}) vanish when collisions between
particles of the same species are considered. Therefore the right
hand-side term in eqs. (\ref{eq:mom}) and (\ref{eq:emom}) quantify
only effects of collisions between different species, $s$ and $t$.

When  equations (\ref{eq:cont})--(\ref{eq:emom}) are
summed over all species $s$ one obtains, after some
algebra, the equations of conservation of mass, momentum and energy 
for the plasma (\citeauthor{Montgomery+Tidman},
\citeyear{Montgomery+Tidman}, p.\ 198):
\begin{eqnarray}
  \label{eq:cont_plasma}
  \pderiv{\rho_m}{t} + \nabla \cdot (\rho_m \vec U) &=& 0\\
 \rho_m \frac{D\vec U}{Dt} + \vec \nabla \cdot
 \tensor{P}  -\rho_m\vec g - \rho_c\left(\vec E - \vec J
 \times \vec B \right) &=& 0  \label{eq:momentum_plasma}\\
 \frac{3}{2} nK\frac{DT}{Dt} - \frac{3}{2}kT\nabla \cdot
 \sum_s{n_s\vec U_s} +\vec \nabla \cdot \vec q +
 \tensor{P}:\vec \nabla \vec U -(\vec J - \rho_c
 \vec U)\cdot\left[ \vec E + \vec U \times \vec B \right]  &=& 0
 \label{eq:e5_plasma}
\end{eqnarray}
%\comment{ols: I assume $d/dt$ above should be replaced by $D/Dt$ (the
%  convective derivative)?}
where $\rho_m$ is the plasma mass density, $\vec U$ is the plasma bulk
velocity, $\vec J$ is the net current density, $\tensor P$ is the
plasma pressure tensor, $\vec q$ is the plasma heat flux (see Appendix
A for the definitions of these plasma parameters).
%The definitions of the plasma macroscopic variables occurring in
%(\ref{eq:cont_plasma})--(\ref{eq:e5_plasma}) are given in Appendix 1.

Both systems of equations, (\ref{eq:cont})--(\ref{eq:emom}) and
(\ref{eq:cont_plasma})--(\ref{eq:e5_plasma}) are not closed; they
contain more unknowns than equations. 
Indeed the heat flux $\vec q_s$, $\vec q$
and pressure tensor $\tensor{p_s}$, $\tensor{P}$ are still functions of the velocity
distribution function. They cannot be determined from the first three
transport equations given above, without making additional assumptions
and approximations.  In Appendix  \ref{Boltzmann_solutions} we discuss
two classical  methods for
closing this chain of equations (Chapman--Enskog, Grad).
One possible simplifying assumption is to
consider that the VDF of all species is an
isotropic Maxwellian (in the frame of reference comoving with
the velocity $\vec u_s=\vec U$); 
in this case $\vec q_s=0$ and the pressure tensor $\tensor{p_s}$ is
isotropic. Under these assumptions the equations
(\ref{eq:cont})--(\ref{eq:emom}) become the one-fluid Euler or
  5-moment equations:
\begin{eqnarray}
\pderiv{n}{t} + \vec \nabla_{\vec r}\left(n\vec U\right) & = & 0
\label{Euler:continuity} \\
\pderiv{\vec U}{t} + \vec U \cdot \vec \nabla_{\vec r} \vec U & = &
\frac{\vec F}{m} - \frac{1}{mn}\vec \nabla_{\vec r}\left( nkT \right)  
\label{Euler:momentum} \\
\left(\pderiv{}{t} + \vec U \cdot \vec \nabla_{\vec r}  \right)\left(
Tn^{-2/3}  \right) &=& 0
\label{Euler:energy}
\end{eqnarray}
The first hydrodynamic models of the solar wind were
based on the one-fluid Euler approximation that will be discussed in more details
in sections \ref{Parker_model} and \ref{sec:fluid-models}.

 {The next approximation considers the 
  electrons and protons as separate fluids with different
  temperatures and isotropic pressure tensors for each species, but
  the same bulk velocity, $\vec u_s = \vec U$. 
%  This is the main assumption of the two-fluid class of
%  models derived from (\ref{eq:cont})--(\ref{eq:emom}) that also 
%  takes into account that the electron to proton mass ratio is much
%  smaller than one, $m_e/m_p \ll 1$, and gives the following 
%  set of two-fluid equations:
%\begin{eqnarray}
%\pderiv{\vec U}{t} + \vec U \cdot \vec \nabla_{\vec r} \vec U & = &
%\frac{\vec F}{m_p} - \frac{1}{m_pn}\vec \nabla_{\vec r}\left[ nk(T_e +T_p)\right]  
%\label{2fluid:momentum} \\
%\left(\pderiv{}{t} + \vec U \cdot \vec \nabla_{\vec r}  \right)\left(
%Tn^{-2/3}  \right) &=& 0
%\label{2fluid:energy}
%\end{eqnarray}}
% {The equation of continuity for the two-fluid description is the same 
%with (\ref{Euler:continuity}), given by the Euler approach. These
%equations are at the basis of a series of early SW fluid models 
(e.g.  \citeauthor{HartleSturrock68}, \citeyear{HartleSturrock68}).
The corresponding two-fluid equations and their relationship to the Euler set
are discussed in sections \ref{sec:5-moment-appr} and
\ref{sec:8-moment}.}
%The kinetic and fluid models of the solar wind discussed in the next
%chapters may be considered direct applications of various
%approximations and solutions of the equations discussed above.  

 {The general theoretical framework outlined in this
  section reveals the complementarity between kinetic and fluid approaches.}
The kinetic exospheric models are applications of the Vlasov and/or
Fokker-Planck equations and are discussed in section~\ref{sec:kineticSW}.
Multi-fluid models  {may be also viewed as}  
applications of the Boltzmann and/or
Fokker-Planck,  {based on various expansions of the velocity distribution
functions and corresponding to different approximations of the transport equations.
The latter are reviewed  in section}~\ref{sec:fluid-models}. 
%Simplified forms of these transport
%equations have been used to describe the distribution of the 
%plasma density in the solar corona and later on in the solar wind.
%We recall first a brief history of these early approaches used to
%model the solar corona and solar wind.http://www.evz.ro/index.html
% described  by \citet{Grad58}.
%in fact direct applications of the methods discussed by 

%In the Appendix we give a brief review of the main classes of solutions
%of the Boltzmann equation. 
%We outline the general theoretical background http://www.evz.ro/index.htmlhttp://www.evz.ro/index.html
%and the relationship between kinetic and fluid approximations of 
%space plasma physics. 

%In the reminder of this section we will mainly discuss methods to solve the Boltzmann equation (1) and will follow Grad (1959).  The basic 

%\section{A Half a Century of solar wind modeling}
%\label{sec:kineticSW}
%Several reviews of the solar wind have been published in the past. 
%The book titled "Basics of the Solar Wind" by Nicole Meyer-Vernet
%(2007) is seminal work. 
%A comprehensive article on "Kinetic Physics of the 
%Solar Corona and Solar Wind" by Marsch (2006) is available on Internet. 
%The following sections will recall in detail in a historical
%perspective solar wind modeling efforts over half a century. 
%A briefer account of this history can be found in the article of 
%\citet{LemairePierrard2001}. 

%\subsection{A physical description of the solar corona}
\section{A historical account of  modeling the solar corona and the solar wind}
\label{sec:history}
%\comment{ols: Changed heading of this subsection (it is more a
%  historical account than a physical description).}

During all solar eclipses a bright halo (the solar corona) can be seen
around the Sun when it is masked by the Moon.  The coronal luminosity
extends then sometimes up to 10 solar radii ($10~R_S$).  This
brightness in the visible spectrum is mainly due to scattering of the
photospheric light by free electrons of the fully ionized coronal gas
(Thompson scattering).
 
It is established since 1942 \citep{Edlen1942} that this extended
solar coronal plasma is heated to more than one million degrees. The
high temperature was established from the discovery of the high degree
of ionisation of the atoms emitting the yellow, green and red coronal
emission lines. It was later confirmed by the broad Doppler width
of these spectral lines.  The heating mechanism is still not fully
understood (see, e.g., \citeauthor{Gomez1990}, \citeyear{Gomez1990};
\citeauthor{Heyvaerts1990}, \citeyear{Heyvaerts1990};
\citeauthor{Hollweg1991}, \citeyear{Hollweg1991};
\citeauthor{Zirker1993},  \citeyear{Zirker1993}; 
\citeauthor{Narain_et_Ulmschneider1996}, \citeyear{Narain_et_Ulmschneider1996}).  

The coronal plasma is constituted of fully ionized
Hydrogen, Helium (5--10\%), as well as smaller traces of highly
ionized heavier atoms.  Up to 1958, it was considered that the density
and kinetic pressure of the coronal plasma is maintained in
hydrostatic equilibrium by the Sun's gravitational field.
This scenario would correspond then to $\vec u_s=0$ in 
equations (\ref{eq:cont})-(\ref{eq:emom}).

Due to this high temperature ($T_e \approx 10^6$~K), the plasma thermal conductivity
is extremely large. Therefore the coronal temperature was assumed to 
be almost uniform, i.e., independent of heliographic altitude. 
The radial distribution of the coronal brightness was generally fitted 
to theoretical electron density distributions considered to be 
in isothermal hydrostatic equilibrium  \citep{vandeHulst53}. 

\subsection{Chapman's hydrostatic models of the solar corona}

The first non-isothermal model of the corona was proposed in the late 50's
by \citet{Chapman57}.  Arguing that thermal conductivity of the coronal
plasma is large, but not infinitely large and proportional to
$T_e^{5/2}$, 
where $T_e$ is the free electron temperature  {(equal to $T_p$, the
proton temperature)}, 
\citet{Chapman57} obtained  a distribution for $T_e(r)$ as a
function of $r$, the radial distance; 
he postulated that heat conduction is the only  mechanism to 
transport heat away from the base of the corona into interplanetary
space.  By requiring that $T_e=0$ at infinity 
(i.e. tending to that of the interstellar medium), 
he determined that  $T_e(r)$ should decrease as 
$r^{-2/7}$ from over  $T_e=10^6$~K  at the base
of the corona, to zero for $r \rightarrow \infty$. 
Assuming $T_e(r_0)$ , the coronal temperature at $r_0
  =  1.06~R_S$ , is equal to $10^6$~K, 
the electron temperature at 1~AU would still have a high value 
($2 \times 10^5$~K) in Chapman's conductive model. 
This surprising result lead him to argue that 
the terrestrial thermosphere could be heated from above by thermal 
conduction, i.e., by heat being conducted down from interplanetary
space into the upper atmosphere of the Earth \citep{Chapman57,Chapman59}.
\footnote{This interesting suggestion of Chapman has unfortunately 
never been carried out further, and remains widely unknown or
overlooked.  We believe that this hypothesis warrants more 
attention from those modeling the temperature distribution 
in the upper terrestrial atmosphere. Indeed, this downward heat 
flow may explain the positive temperature gradient observed 
above the mesosphere, and the persistence of high 
temperatures up to very high altitudes in the thermosphere.}

Applying the theoretical $r^{-2/7}$ temperature distribution
\citet{Chapman57} calculated the distribution of the electron pressure
($p_e = n_e k T_e$) and density ($n_e$) in the corona assuming it is
in hydrostatic equilibrium. He found that his theoretical electron
(and ion) density had a minimum value at 0.81~AU\@.  Using a density
of $2\times 10^8$ electrons/$\mathrm{cm^3}$ at the base of the corona
--- obtained from eclipse observations --- he found that the minimum
value of the electron density should be equal to
$340~\mathrm{cm}^{-3}$ at $r = 174~R_S$; beyond this heliocentric
distance in Chapman's hydrostatic model, $n_e(r)$ increases
indefinitely with altitude; at 1~AU, $n_e = n_p= 342~\mathrm{cm}^{-3}$
with the same density of protons and electrons due to the
quasi-neutrality of the interplanetary plasma.  The properties at the
base of the solar corona and at 1~AU corresponding to Chapman's
conductive and hydrostatic model are listed in Table~\ref{tab:table1} (second column;
model $\{0\}$).
%\comment{ols: Moved table 1 here, since it is referenced here the
%  first time.}
%\begin{landscape}
%\begin{table}[htb]

%Table 1
%Comparison between measurements and \textit{hydrodynamic models} of the solar 
%wind of the number density, bulk velocity, parallel and perpendicular 
%temperatures, temperature anisotropies, energy flux and heat
%conduction flux. During quiet solar wind conditions, the observations 
%taken from Hundhausen (1968) at 1~AU For the high speed solar wind, 
%the observations are made by Helios-1 \& 2 \citep{Maksimovic95}. %

%Table 1 + legend
%------------------------------------------------
Such a theoretical density profile (i.e. a density increasing versus
altitude) implies that the assumed hydrostatic equilibrium is 
convectively unstable. 
%In retrospect, it seems disconcerting 
%that Chapman did not infer from this surprising theoretical result 
%that the solar corona cannot be in hydrostatic equilibrium, 
%but must be expanding as proposed by \citet{Parker58} one year later.
In one of his subsequent paper, \citet{Chapman61} argued that the
interplanetary medium might become turbulent beyond 0.81~AU
($174~R_S$) but he did not envisage the more realistic hypothesis that
the corona might not be in global hydrostatic equilibrium, but
expanding continuously as observed later on by in-situ measurements in
the interplanetary medium.

%\comment{ols: I think subsection on the Parker model should move here,
%not only because it historically came first, but also because the text
%below makes references to the Parker model.}

\subsection{The first hydrodynamic model of the solar wind}
\label{Parker_model}

Parker's argument against hydrostatic equilibrium 
and in favor of a continuous  expansion of the solar corona is based 
on a hydrodynamic description.
\citeauthor{Parker65} (\citeyear{Parker65} 
 p. 669) argued that  ``if an atmosphere is sufficiently dense that 
the mean free path is small compared to the scale height and/or the 
radial distance $r$ from the parent body, then the ordinary
hydrodynamic equations are appropriate''. 
And according to him ``this is the situation in the corona''.
On the other hand ``if the atmosphere is so tenuous that 
the mean free path is long, the universal presence of weak 
magnetic fields and attendant instabilities maintain 
approximate statistical isotropy in the thermal motions, 
and it turns out that again the large-scale, 
low frequency variations of the gas are describable, 
in at least an approximate way, by conventional hydrodynamic 
equations with isotropic pressure. 
This appears to be the situation in the expanding corona 
at the orbit of Earth'' (\citeauthor{Parker65},
\citeyear{Parker65} p. 669).
%These arguments were then given without firm prove, 
%but have been invoqued continuously for decades within the community of MHD plasma physics.

In his  {pioneer} paper, Parker  (\citeyear{Parker58})
demonstrated that hydrostatic models of the corona 
predict kinetic pressures and densities exceeding those existing in
the interstellar medium. In his well known monograph, 
Parker (1963) estimated the interstellar density to  be 
``1 particle/cm$^3$ and 
the kinetic pressure to $1.4$ $10^{-14}$  dyne/cm$^2$''
(\citeauthor{Parker63}, \citeyear{Parker63},  p. 42).  
These values are much smaller than those predicted at infinity by
hydrostatic equilibrium for any coronal temperature decreasing with $r$ 
slower than $1/r$, as it is the case for instance in 
Chapman's conductive model ($T\approx r^{-2/7}$), 
or in an isothermal corona.  When the high coronal temperature extends 
deep into the the interplanetary medium, declining asymptotically 
slower than $1/r$,  ``the atmosphere will extend to infinity 
with a finite pressure'' which is much larger then that of 
the interstellar medium. ``Since there is nothing at infinity to 
contain such a pressure, the atmosphere would expand, 
rather then being static'' (\citeauthor{Parker63}, \citeyear{Parker63}, p. 43, par. 1).

This reasoning based on a mechanical pressure balance equilibrium
argument lead \citet{Parker58} to propose 
that the plasma is driven out of the corona, 
gaining momentum as a result of a large kinetic pressure gradient: 
i.e. due to the large pressure imbalance at the base 
of hot corona and in the cold interstellar medium. 
In other words, the excessively large 
kinetic pressure due to the high coronal temperature 
generates a pressure gradient in the 
momentum equation, accelerating the coronal plasma 
from slow (subsonic) expansion to a supersonic flow 
regime beyond some critical altitude where the radial bulk speed becomes equal to the sound speed. 

It must be emphasized that the essential part of the supersonic outflowing
is the high temperature of the corona extending far out in the solar
gravitation field due to the high thermal conductivity of the coronal
plasma and ad-hoc in-situ heating. 
Maintaining the high temperature in the escaping gas is 
responsible for the ultimate supersonic velocity (Parker, 2009,
private communication).

Parker's first solar wind model was based on the
assumption that  the coronal plasma is a fully ionized ideal  gas,
that the uniform temperature is 
the same for the electrons and protons, and that
 the radial expansion is not intermittent but
stationary.
The following one-dimensional equation was derived by
Parker from the general Euler equations 
(see Eqs. \ref{Euler:continuity} -- \ref{Euler:energy}):
\begin{equation}
 \frac{r}{U}\frac{dU}{dr} = \frac{r^3}{\left(2KT\right)/m-U^2} \left[ \frac{d}{dr}
\left(\frac{2KT}{mr^2} \right) + \left(\frac{GM_S}{r_0}\right) \frac{1}{r^4}\right]
\label{Parker_equation}
\end{equation}
where the bulk speed, $U$, and the temperature, $T$, depend
on the radial distance, $r$;  $\sqrt{\frac{GM_S}{r_0}}$ is the
escape velocity at solar altitude $r_0$.
 The implicit assumptions for solving equation (\ref{Parker_equation})
are that (i) the corona (and solar wind) are isothermal; (ii) the pressure 
tensors are diagonal and identical for the electrons 
and protons and (iii) the average velocities of the electrons
 and protons are equal to $\vec U$,  the bulk (mass averaged) 
velocity of the plasma as a whole, implying no electric current
since the plasma is quasi-neutral.
%(iii) this bulk velocity is radial and oriented outwards; 
%(iv) the magnetic field lines are Archimedian spirals whose 
%pitch is controlled by the rate of rotation of the Sun; 
%(v) the expansion is stationary (steady state). 
The modeling of plasma flow was restricted to the equatorial region.

Furthermore, due to the quasi-neutrality of plasmas, 
the electron number ($n_e$) density is necessarily 
equal to the total ionic charge densities 
($\sum_{i \ne e} Z_i n_i = n_e$) and the net
current must also be null.  
Under these additional restrictive conditions,
the different particle species are not expected 
to diffuse with respect to each other; 
the plasma can be modeled as a whole, 
almost like  a neutral gas, with a unique bulk (mass averaged) 
speed $U$, a unique characteristic temperature $T$, and a mass density $\rho$.
In Parker's SW model, the plasma distribution is spherically
symmetric, and $\vec U$ has only one component in the radial direction
($U_r$). 

\begin{figure}
\center
\includegraphics{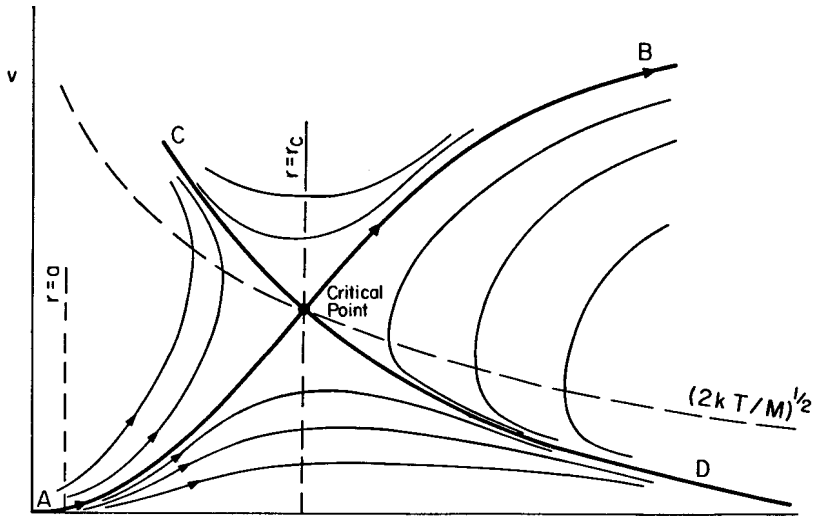}
\caption{The figure shows the variation of the solar
wind bulk velocity with the radial distance as inferred
from several classes of solutions  
for the Euler equation  (\ref{Euler:momentum}) in the form
(\ref{Parker_equation}) derived by Parker. The position of the
critical point is derived from Eq. (\ref{Parker_critical_point});
%describing Parker's model of the solar wind; 
the critical solution, AB, is the only one that
gives the supersonic acceleration of the solar wind, \citep{Parker65} \label{Fig1Parker}}
\end{figure}

 The equation (\ref{Parker_equation})
has a family of solutions with one critical point $r_c$, defined by the
solution of the equation \citep{Parker65}:
\begin{equation}
r^3\frac{d}{dr}\left( \frac{2KT}{mr^2} \right) +
\left(\frac{GM_S}{r_0r}\right) = 0
\label{Parker_critical_point}
\end{equation}
The critical point  is located somewhere between $4~R_S$ 
and $8~R_S$.  It is at this heliocentric distance that the bulk 
velocity ($U_r$) changes from subsonic values to supersonic ones
(see  Fig.~\ref{Fig1Parker}). 
%An illustration of this family of
%solution is shown in figure \ref{Fig1Parker}.
The critical point is unique if $T$ decreases with $r$ less rapidly
than $1/r$ \citep{Parker65}. Only the critical solution, passing through the
critical point, achieves the supersonic acceleration of the solar wind;
it is the only one that satisfies 
obvious boundary conditions at $r_0$, the base of the corona 
(where the density and kinetic pressure are large, and bulk velocity
small), and at infinity (where the mass density and kinetic 
pressure are small).
%The horizontal axis is the radial distance.  
The SW bulk velocity ($v$ in Parker's original paper or $U$ in 
this work) varies  from very small 
subsonic velocities at the base of the corona 
(branch ``A'' below the critical point in figure \ref{Fig1Parker}) to supersonic speeds of more 
than 300~km/s along the branch ``B''. 
Typical subsonic solutions are also illustrated by the lines 
displayed in figure  \ref{Fig1Parker} below the critical point;
in this family of hydrodynamic solutions, the expansion velocity 
increases from small values at the base of the corona ($r =
  r_0 = a$), but $u_r$ never reaches the velocity of sound; 
eventually $u_r$ decreases to zero as 
$r \to  \infty$.  
In all subsonic solutions, the plasma 
density and kinetic pressure decrease asymptotically 
to constant values, which are much too large compared to 
the corresponding value in the interstellar medium, 
just as in the hydrostatic solutions \mbox{($U = 0$)}. 
The lines asymptotic to the branch ``C'' correspond to mathematical
solutions of the stationary Euler differential equations but with no
physical relevance. They correspond to shock like acceleration with
density decreasing to zero at altitudes below that of the critical
point.  When $U(r_0)$, the bulk velocity at the reference level $r_0$,
exceeds the value corresponding to Parker's critical solution, the
hydrodynamic equations have not stationary solution extending all the
way through to $r \to \infty$; in these instances only time-dependent
hydrodynamic solutions can exist over the whole range of altitudes.
% The radial distributions of density and flow speed 
%are determined by integrating analytically or 
%numerically the Euler continuity and momentum equations,
%(\ref{Euler:continuity}) and (\ref{Euler:momentum}). 
%Note in this context that the steady state Euler's equations are ordinary 
%differential equations corresponding to the simplest 
%subset of the hierarchy of moment equations that are
%derived from Boltzmann's equation (see Sect. \ref{Boltzmann_section}, \ref{Transport_eqs}).  

%\comment{ols: Deleted footnote
%  here since this is all discussed in Sect.~\ref{sec:fluid-models}.}
% \footnote{Of course, 
% many other different approximations of the moment equations 
% or transport equations have been derived, applied and integrated 
% over the past century; the simplest and most popular approximation 
% is the Euler set of transport equations (i.e. two non-linear 
% ordinary differential equations where n and $\mathrm{u_r}$ are 
% the dependent variable and r the independent one).  
% It is this Euler approximation that has first been integrated 
% to determine the density and bulk velocity in a stationary model 
% of the solar wind. Subsequently, much more sophisticated set and 
% approximations of the hydrodynamic equations have been used to model 
% the radial (1D) distributions of the moments of the SW protons, 
% and electrons velocity distribution functions.  
% Some of them will be reviewed and discussed later on in this review.}

The existence of the  supersonic plasma flow described by 
Parker's hydrodynamic solar wind model is supported by 
Biermann's (\citeyear{Biermann53}) observations of plasma flow in 
tails of comets and  {also 
by in-situ observations of Mariner 2} \citep{NeugebauerSnyder, SnyderNeugebauer}. 
Missions like IMP 1 \& 2, Vela 2 \& 3 confirmed the permanence of the
supersonic solar wind.
Continuous in-situ observations reveal that the solar wind plasma 
is not stationary but that its average bulk velocity changes 
irregularly between 300~km/s and 450~km/s in what was called 
by \citet{Hundhausen68} and \citet{Hundhausen70} the
 ``slow'' (or ``quiet'') solar wind streams. 
By increasing the coronal temperature from $1.5 \times 10^6$~K to over 
$3 \times10^6$~K, Parker's hydrodynamic solutions are able to 
account for such a range of proton bulk velocities.
 {Later on, however, it has been recognized that a coronal temperature 
exceeding $3\times 10^6$~K would be needed for Parker's hydrodynamic 
model to account for bulk velocities of 600--900~km/s (see fig. 6.1 in
\citeauthor{Parker63}, \citeyear{Parker63}, p.75), which are 
sometimes observed at 1~AU in fast speed streams of the solar
wind.} 

 {Parker's  {pioneer} theoretical model of the 
interplanetary plasma inspired other subsequent hydrodynamic models of the solar wind.  
It influenced the theoretical modeling of other plasma flows such as
plasma escaping out of the topside terrestrial ionosphere above 
the polar caps, along open magnetotail field lines
\citep{BanksHolzer68, BanksHolzer69}, that is the polar wind.  
The history of the polar wind models and their development 
parallel to that of the solar wind models have recently been 
reviewed by} \citet{Lemaire2007}.
%Since such coronal temperatures occur to be excessive, 
%the application of the single-fluid hydrodynamic equations to model 
%fast speed solar wind streams was eventually challenged by the end of
%the 70's.

%\textcolor{red}{E. Parker: transient velocities occasionally of
%  the order of 1000 km/s and more represent the magnetic eruptions
%  called coronal mass ejection.} \comment{[He also suggests to drop last
%  sentence of the above paragraph and to correct the value of
%the temperature required to accelerate the wind to 600-900 km/s
%by replacing $10^7$ k with  $2\times 10^6$ K.]}

\subsection{Convective instability in Chapman's hydrostatic corona}

%The unexpected positive density gradient in Chapman's conductive and
%hydrostatic model is a puzzling theoretical prediction.  The origin of
%this unstable density profile was discovered by 
\citet{Lemaire68} showed that the temperature gradient in Chapman's conductive models,
$dT_e/dr$, becomes super-adiabatic at 0.22~AU ($44~R_S$), i.e. steeper
than the adiabatic temperature lapse rate  
 {(see also \citeauthor{Lemaire2010}, \citeyear{Lemaire2010})}.
% A similar conclusion was
%again recovered two decades later by \citet{LuHamilton91}.
Hence, Chapman's hydrostatic model is 
necessarily convectively unstable at an altitude below 
the minimum of the density in the hydrostatic model 
of the solar corona.  This means that the mechanism of heat conduction 
alone is not efficient to carry away all the energy 
deposited at the base of the corona, and transport it out 
into interplanetary space. 

Furthermore, \citet{Lemaire68}, showed
that, beyond the radial distance of 0.22~AU ($44~R_S$), turbulent
convection (within a plasma in average hydrostatic equilibrium, as it
is the case in the Sun's convection zone) is also insufficient to
evacuate the coronal energy flux toward outer space. Lemaire 
concluded that a continuous expansion of the coronal plasma
(i.e. a laminar hydrodynamic advection/expansion) is
required to keep the actual temperature gradient smaller than the
adiabatic temperature lapse rate at all altitudes in the corona and in
the interplanetary medium.  
%In addition to the mechanism of thermal conduction, 
%continuous outward convection (i.e. a continuous 
%near stationary expansion/explosion 
%of the corona) is thus required to remove more efficiently 
%the heat deposited in the inner corona, and to evacuate it toward
%interplanetary medium.
\footnote{It should be reminded that thermal 
heat flux is also transported inward by conduction from the 
corona to the chromosphere; this downward draining of energy 
is much larger than that evacuated by conduction upwards 
from the region where the coronal temperature has a peak value 
toward interplanetary space.  
This inward energy flux was first evaluated by \citet{Alfven41} 
and later by \citet{WooleyAllen50}. }
%The latter estimated also that the radiative loss of energy 
%from the corona to the outer space is twice as large as 
%the outward conductive flux near the Sun. 
%At low altitudes in the solar corona, where metal ions 
%are most abundant and the coronal density greatest 
%and transport by radiation is most important. 
%However, beyond a few solar radii, radiative losses of 
%energy by electromagnetic waves is much smaller than the 
%%conductive loss, and Chapman's original hypothesis 
%was non unreasonable, at least in the framework of 
%his hydrostatic equilibrium hypothesis.}

This new inference led \citet{Lemaire68} to conclude that Parker's hydrodynamic expansion of
the solar corona is needed from a thermodynamic point of view.
He argued that only a steady state "explosion" is able to evacuate 
the  energy out of the corona. Nor heat
conduction, nor the \citet{Bohm-Vitense54} turbulent convection
(i.e. hot elements raising and cooled ones falling within an
atmosphere in average hydrostatic equilibrium as in the solar
convection region) proved to be efficient enough. 
%Note that, unlike in the solar convection region, radiative transport is of minor
%importance in the solar corona.

It can be seen that Lemaire's \citeyear{Lemaire68} argument is
different but complementary  to the mechanical  one 
proposed by \citet{Parker58} to support the existence 
of a radial supersonic expansion of the solar corona. 
Parker's theoretical argument in favor of a continuous 
radial expansion of the corona is based on pressure 
imbalance conditions and the boundary conditions for
the momentum transport equation, as was pointed  out 
in the previous section.

\subsection{A brief historical perspective on the hydrodynamic models of the Solar wind}
 
In a monograph \citep{Parker63} and in subsequent
publications, \citet{Parker65, Parker67, Parker69} generalized 
his first isothermal hydrodynamic model of the solar wind.  
He replaced the hypothesis of a uniform coronal temperature, 
by the assumption that the temperature decreases with radial distances according 
to a polytropic relationship between the plasma temperature 
and its density. 
 
It would be too long to review all the alternative hydrodynamic 
SW models that flourished during four decades. 
Following the early theoretical generalization of the first SW model, 
 more sophisticated hydrodynamic models have been formulated.
Various formulations and approximations for the energy transport 
equation were coupled to the hydrodynamic continuity  and momentum
equations, Eqs (\ref{eq:cont_plasma})--(\ref{eq:momentum_plasma}), 
to compute steady state radial distributions of the solar 
wind temperature as function of the radial distance
\citep{NobleScarf63, Parker64, WhangChang65, 
Weber70, CupermanHarten70a, Durney71, Durney72}. 
In these latter one-fluid models, the effect of thermal conductivity 
was taken into account to determine the radial distribution of the 
plasma temperature, as Chapman did for the conductive model 
discussed above. In hydrodynamic models heat transport by conduction is the dominant 
mechanism close to the base of the corona, while at radial distances close 
and beyond the critical sonic point
the energy flux is predominantly carried outwards by 
advection, i.e., by the solar wind bulk motion. 
The energy is then primarily carried away by 
the supersonic expansion (or "stationary explosion") of the coronal plasma. 

The rather small effects of viscosity (due to Coulomb collisions) have 
been considered in Navier-Stokes approximations of the hydrodynamic 
momentum equations \citep{ScarfNoble65, WhangLiuChang66, Konyukov69, Eisler69, Dahlberg70}.  
Time-dependent hydrodynamic models of the solar wind expansion have
also been developed.  The first non-stationary model of the solar
corona was a self-similar, polytropic radial expansion by
\citet{Lemaire66a}.  
%Time-dependent computational codes have mostly
%been used to bypass mathematical singularities associated with
%critical points of the set of ordinary differential
%equations characterizing stationary solutions of the
%radial expansion of the corona.

Instead of single-fluid models of the solar wind, two-fluid
and multi-fluid models have been developed.  The effects of non-radial
magnetic fields, of adiabatic cooling for both electrons and protons,
the acceleration by Alfv\'en waves, electron and proton heating have
been investigated by a number of authors \citep{SturrockHartle66,
  WeberDavis67, HartleSturrock68, Urch69, CupermanHarten70b,
  CupermanHarten71, HartleBarnes70, BarnesHartle71, Whang71b, Wolff71,
  Hansteen+Leer, EsserHabbal95, Habbal95,
  TuMarsch97,TuMarsch2001,Tam+Chang, Olsen+Leer1999, XLi1999,
  Lie-Svendsen+Leer+Hansteen2001, Kim2004, Janse+Lie-Svendsen+Leer}.

These SW models are listed in the successive columns of
Tables~\ref{tab:table1} and~\ref{tab:table2}
in a more or less chronological order.
% but their ID numbers (between
%$\{\}$) have no special meaning.  
The solar wind parameters at 1~AU as
well as the boundary conditions at the base of the corona for all
these different types of stationary solar wind models are compiled in
Table 1 for the slow/quiet solar wind, and in Table 2 for the fast
solar wind.  The average solar wind
parameters observed  at 1~AU are given in the yellow shaded
columns.
% (col. 5 and 6 in Table 1; col. 2-5 in Table 2; ).  
The synsthetic characteristics of each model are summarized:
i.e. the one-fluid hydrodynamic (1F-H) models,
two-fluid hydrodynamic (2F-H) models, three-fluid hydrodynamic (3F-H),
three component hybrid/semi-kinetic (3Hb), two fluid sixteen moment
(2F-16M) models, exospheric (E) models.

Coupled multi-fluid models are more difficult to
integrate numerically.
% since they are much more sophisticated from a
%mathematical point of view. 
Furthermore, there is then an increased number of free
parameters and boundary conditions to be imposed to the
set of differential equations; this leads to a wider variety of
solutions and therefore of a greater chance to fit the 
SW measurements at 1~AU.

 As a consequence of the nonlinerarity of the set of 
transport equations, relatively small changes of the boundary conditions in the
corona can produce large amplitude changes in the values of the higher order
moments (parallel and perpendicular temperatures, stress tensors,
energy and heat fluxes) at 1~AU. Note that in-situ
 heating of the SW plasma influences the terminal flow speed of
the wind at 1~AU.
% perhaps overwhelming the effect of the boundary conditions deep in the corona. 
The temperature anisotropy at 1~AU is also influenced
heavily by in-situ heating or momentum transfer mechanisms (i.e. heat
deposition, pitch-angle scaterring, wave-particle interactions,
momentum transfer by MHD waves).
%\comment{[Last paragraph was rephrased according to Oysteyn's comment: 
%I don't agree with the last
%  statement ``Relatively small changes of the boundary conditions in the
%corona can produce large amplitude changes...''.  
%The temperature evolution is for instance to a large
%  extent determined by what happens in situ in the supersonic wind,
%  not by the boundary condition.  And in models with heating (which is
%necessary) the heating largely determines the terminal flow speed of
%the wind, not some boundary condition deep in the corona.]}

Multi-fluid models consider more than one critical point (where the bulk speed becomes
transonic and where $dU/dr$ may have two different values as illustrated in Fig.~\ref{Fig1Parker}). 
The existence of these additional mathematical singularities 
makes the search of  relevant solutions much more 
difficult and uncertain, if not questionable.
Indeed, when the mathematical singularities are located beyond the exobase, where the 
Knudsen number becomes larger than unity, the physical significance
of these singularities becomes uncertain.

The effect of non-radial (non-spherically symmetric) 
expansions has been extensively modeled.  Several radial
distributions for the extended coronal heating source and 
accelerating mechanism have been proposed with the hope to reach 
hydrodynamic models with very high SW bulk velocity, as
observed in fast speed streams \citep{Whang71a, HartleBarnes70, Hansteen+Leer}.  
Since some of these hydrodynamic multi-fluid 
solar wind models assume that the kinetic pressure tensors are
isotropic, these models cannot predict the observed 
anisotropies of the electrons and protons temperatures at 1~AU\@. 

One-fluid models predicting the same temperatures 
for the protons and electrons are in principle less 
adequate than two-fluid models where the boundary 
conditions and source terms have been adjusted so that the 
electron temperature at 1~AU is higher than the proton temperature, 
as consistently observed in the slow solar wind.  {Two-fluid models
have been developed by}   \citet{HartleSturrock68}  {accounting for
different temperature profiles of electrons and protons;} 
\citet{LeerAxford1972}   {developed a two-fluid model that
allows an anisotropic proton temperature producing more reasonable
temperatures at the Earth's orbit}.  {Other two-fluid 
models have been developed by} \citet{CupermanHarten70a,
  HartleBarnes70,  CupermanHarten71, Habbal95, TuMarsch97,
  TuMarsch2001, EsserHabbal95, Kim2004}.
The gyrotropic two-fluid model of \citet{XLi1999} and \citet{Janse+Lie-Svendsen+Leer} based on
the 16-moment transport equations with an ``improved treatment'' of heat
conduction and/or proton heating (see Sect.~\ref{sec:bimaxw}) achieves
a rather satisfactory fit to fast SW observations.  Some data
about these two models, including SW properties at 1 AU,
are presented in Table~\ref{tab:table2} and
Fig.~\ref{fig:fig4}b, see models $\{26\}$ and $\{28\}$.

The results predicted at 1~AU by the most representative of 
these models are synthetically reported in Tables 1 and 2 
respectively for the slow/quiet solar wind and fast speed stream observations.  
%Some of these results are also displayed in Figs. 4a and 4b.
Various aspects of these solar wind models have already been discussed in
previous reviews by \citet{Marsch94,Marsch2006}. 
%\textcolor{red}{$\ldots$ remove if moved to Section 5 $\ldots$ !!!!
%The most recent versions of such fluid/hydrodynamic models will 
%be presented in section \ref{sec:fluid-models}. } 

In order to reproduce the high speeds of the fast solar wind, 
the fluid models must either assume a very high
proton temperature in the corona, of the order of $10^7$~K, or extended
heating beyond the critical point.  Simple energy conservation
arguments can be used to show that this is not a shortcoming of the
fluid models, but a necessary requirement for any solar wind model,
kinetic or fluid, unless an extremely large outward heat flux can
somehow be formed in the corona.
% \comment{rephrased according to   Oystein's comment.}

%\comment{ols: And here too I want changes\ldots} 
Furthermore, in-situ observations indicate that fast solar 
wind speed streams do not originate in the coronal region 
where the temperature is highest, but
on the contrary, from polar coronal holes where the coronal
{electron} temperature is significantly lower than in the
equatorial regions, the source of the slow SW\@.  
In the ``old''
hydrodynamic solar wind models, which were driven mainly by electron heating, this
was a serious problem as the low electron temperature 
should produce a slower solar wind from
coronal holes.  However, more recent observations of polar coronal
holes by the UVCS instrument on the SOHO satellite
\citep{Kohl+etal1997} have resolved this problem.
They  showed that hydrogen, and even more so heavier ions, are much hotter than
electrons in coronal holes. 
This would imply that electrons play a smaller role in the fast
solar wind acceleration and that it is driven mostly by proton heating.
%\sout{experimental findings generated serious disappointment among the
%  modeler community; they undermined the unreserved credit given to
%  hydrodynamic SW models since 1958 and questioned the applicability
%  of the hydrodynamic transport equations to model correctly the
%  plasma expansion in the distant heliosphere.}
%\comment{rephrased according to Oystein's comment}

%\sout{Note that two decades before this  crisis  of solar wind models, 
%hydrodynamic SW models had already been criticized}
As early as 1960, it was claimed that beyond a 
certain heliospheric distance, the coronal plasma is becoming 
almost collisionless.  As a consequence, it was expected that the VDF
can strongly deviate  from a Maxwellian.  Hence higher-order 
moments of the VDF can become large since they depend
 critically on the departure from a drifting Maxwellian.  
As shown in Sect.~\ref{sec:fluid-models}, such
higher order fluid models can become laborious.
The diffusion,  viscosity and   heat conductivity coefficients 
used in the hydrodynamic equations deviate from the 
standard expressions derived for a simple  Maxwellian VDF, 
$f^{(0)}(\vec v, \vec r,t)$. 
 Within the Chapman-Enskog or 
Grad theories of non-uniform gases, $f^0$
is a zero-order approximation  of the actual 
VDF, $f(\vec v, \vec r,t)$. 
Furthermore, in the energy transport equation 
the divergence of the third order, $Q_{ijl}$, 
and fourth order moments, $R_{ijlk}$, (see definitions in Appendix A) 
have been truncated in order 
to close the hierarchy of moment equations. 
Since such a truncation is dictated merely by the convenience to limit
the mathematical complexity of the formulation, it lacks a 
sound physical argument.
These particular  methods of truncating the moment equations 
are not identical in the Chapman--Enskog theory \citep{Chapman1917,Enskog1917}
and in the one developed by  \citet{Grad1958}. 
The main purpose of  these ad-hoc approximations was to obtain
a description of the gas as close as possible to the
 transport equations used in classical hydrodynamics, 
exclusively designed to model transport in 
collision-dominated flows. 
Another more pragmatic reason to use such truncated representation is 
dictated by the complexity of the hydrodynamic transport 
equations and of their numerical solutions.

%\textcolor{green}{\comment{JFL: to stay in line with the hystorical
%    perspective of the previous sections it would be useful to add
%    here a subsection recalling synthetically the historical outline 
%   of the development of the kinetic theory of gases}}

%\textcolor{green}{\comment{MME: in my opinion a good historical
%    approach, for both hydrodynamic and kinetic approach is given in
%    section 2 of this revised version of the manuscript. }}

%\comment{MME:moved next paragraphs into the new section inserted 
%at the end of the section}

\section{Kinetic exospheric modeling of the solar wind}
\label{sec:kineticSW}

Two years after the publication of Parker's first solar wind
model, \citet{Chamberlain60} proposed an alternative kinetic theory for
the coronal expansion.
Thus began a standing controversy between the  proponents of 
hydrodynamic solar wind models and those arguing in favor of
 kinetic models. 
%Then a kinetic exospheric ``solar breeze'' model was proposed by 
%as an alternative to.  

\subsection{On the existence of a collisionless region}

Chamberlain argued that beyond a heliospheric distance of 
$2.5~R_S$ in the solar corona, the Coulomb collision mean free path of 
thermal protons becomes larger than $H$, the atmospheric 
density scale height of the coronal plasma; $H$ is the  
characteristic range of altitudes over which the density decreases
by a factor $e=2.71$.   
Chamberlain's argument infers that $K_n$, the Knudsen number of
 plasma particles:
\begin{equation}
K_n=\frac{\lambda_c}{H}
\label{Knudsen}
\end{equation}
the ratio between $\lambda_c$, the mean free path 
of the particles and $H$, becomes larger than unity for $r  >  2.5 R_S$. 
The  density scale height in an atmosphere is defined by:
\begin{equation}
H = \frac{k T}{ \left<m\right>g}
\end{equation}
%\end{equation}
where $g$ is the gravitational acceleration, and $\left<m\right>$ is the 
mean atomic mass of the gas.
In a fully ionized Hydrogen plasma, like the solar corona, 
$\left<m\right> = m_p/2$.
%\comment{ols: This definition of the Knudsen number is only applicable
%to supersonic particles.  Since electrons are subsonic everywhere, it
%cannot be used for them.  Indeed, because most of them bounce back and
%forth in the electrostatic potential before they finally escape, they
%undergo many collisions despite that the Knudsen number is much larger
%than unity.  If the electron temperature is sufficiently low,
%electrons may even remain collision-dominated all the way to 1~AU\@.  So
%I think we need to emphasize here that we talk about protons only.}

When $K_n>1$ the VDF  deviates from a Maxwellian VDF, 
the Chapman-Enskog and Grad expansions of the velocity distribution function
(see Appendix \ref{Boltzmann_solutions}) can then become
inadequate. It is often considered that when $K_n >1 $ 
higher-order fluid models are needed to capture
the non-Maxwellian character of the velocity
distribution function. However, on top of the their complexity
reported in  Section \ref{sec:bimaxw}, it is not a priori obvious how the higher-order
models should be closed.
 In neutral classical gas hydrodynamics, it is considered that a fluid
description  is inadequate 
when ${K_n > 1}$; indeed, there is no unique way based on sound physical
arguments to close the system of moment equations \footnote{The Chapman-Enskog
and Grad expansions of the actual VDF may not be valid since
$f^{(0)}(n(\vec r, t), T(\vec r, t), \vec U(\vec r,t), \vec v)$, the zero-order approximation
of $f(\vec v, \vec r, t)$ may be a poor approximation
(see Appendix \ref{Boltzmann_solutions}). A kinetic theory is then in 
order to describe the ensemble of nearly collisionless particles.  
The actual limitation
of hydrodynamic approximations in solar wind models has been examined
and assessed from the authoritative stand point of classical
kinetic theory by \citet{Shizgal77}; 
in Appendix \ref{Boltzmann_solutions} the reader may find
a brief review of Chapman-Enskog and Grad approaches.}

Based on SW models and observations, \citet{Hundhausen68} estimates
 that $K_n > 1$  above $7~R_S$
%confirming Chamberlain's argument that above an exobase 
%altitude kinetic approaches should be used instead of hydrodynamic
%ones.  
and pointed out that, in order to explain the significant temperature 
anisotropy observed at 1~AU, the solar wind protons and heavier 
ions should be collisionless beyond 15~$R_S$. 
\citet{BrasseurLemaire77} checked that, in Parker's isothermal SW  hydrodynamic model, 
the Knudsen number becomes already larger than unity at $r = 4~R_S$;
this radial distance is lower than the sonic critical point which 
is located at $r_c = 6~R_S$ in the hydrodynamic model of the SW.

\citet{Chamberlain60} developed an exospheric model for the coronal ion-exosphere. 
His model which is known as the ``solar breeze model'' is based 
on \citet{Jeans23} theory for planetary exospheres. 
%Indeed Jeans developed the kinetic theory for the evaporation of atoms 
%from planetary atmospheres. 
Depending on their velocity and of their angular momentum 
four classes of particles can be identified above the exobase: 
\begin{itemize}
\item the \textit{escaping particles}, which have sufficiently 
large velocities to escape from the Sun's gravitational potential
well, 
\item the \textit{ballistic or captive particles} that 
fall back into the corona, 
\item the \textit{satellite or trapped particles} which are continuously bouncing up and down 
between a magnetic mirror point and a gravitational turning 
point, 
\item the \textit{incoming particles} arriving from the 
interplanetary regions and penetrating in the 
collision-dominated region below the exobase.
\end{itemize}

%In the case of charged particles spiraling along interplanetary magnetic 
%field lines, it is their magnetic moment, $\mu_s=(m_{v\bot}^2)/(e_sB)$,
% that is assumed the  second constant of motion, instead of the angular momentum.
Taking into account the conservation of the total energy and
of the magnetic moment  $\mu_s=(m_{v\bot}^2)/(e_sB)$, and using the Liouville
theorem the VDF of collisionless electrons and protons can be
determined everywhere in the exosphere provided this function is
specified at the exobase. When the VDF at the exobase is an analytical function
of $\vec v$, the VDF at any point in the exosphere is then also an
analytical function of $\vec v$. Furthermore the expressions of all
the moments of the VDF are analytical expressions of the
constant of motions (and implicitely of the electromagnetic and
gravitational potentials) and can be calculated numerically everywhere in the
exosphere  \citep{LemaireScherer71}. 
%The mass flux and the electric  current density of each
%species $s$ is an analytical function of the gravitational
%and electromagnetic potential, thus implicitly as a function of the
%radial distance from the Sun \citep{LemaireScherer71}. 

\citet{OpikSinger59}, as well as \citet{BrandtChamberlain60}, applied 
Jeans' collisionless (or zero-order) kinetic theory to calculate 
the  density distributions of neutral gases in the terrestrial exosphere.  
\citet{Chamberlain60} applied a similar exospheric model 
for the ion-exosphere of the solar corona where the protons
move without collisions along radial magnetic field lines
in the gravitational field of the 
Sun and a polarization electric field.
The latter is needed to maintain quasi-neutrality in planetary and stellar 
ionospheres in hydrostatic equilibrium.
 
\subsection{The Pannekoek-Rosseland electrostatic field and potential
  distributions}

The Pannekoek-Rosseland (PR) electrostatic field  is induced in plasmas
by the tendency of the gravitational force to polarize the plasma
by separating the ions and the less massive electrons. 
The presence of this charge separation E-field in a plasma was first 
introduced by \citet{Pannekoek1922}  and \citet{Rosseland1924} in their 
studies of ionized stellar atmospheres. 
The gravitational force acting on the massive ions is much larger 
than that acting on electrons.
% this tends to separate the positively 
% charged particles from the negatively charged one. 
Thus, like in a dielectric material, the minute charge separation 
induced by the gravitational forces polarizes the plasma and 
generates an $\vec E$-field which is oppositely directed  to the 
gravitational acceleration, $\vec g$. 
This polarization electric field prevents
the diffusion in the gravitational field of positive charges 
with respect to the negative ones,
{due to their mass difference.}

When the electrons and protons are in hydrostatic equilibrium 
in the gravitational field, and when their temperatures are equal 
and independent of altitudes (isothermal), 
the PR electrostatic field is determined by:
\begin{equation}
\vec{E}_{PR}= - \nabla \Phi_{PR}=
-\frac{\nabla p_{e}}{e n_e} =
 - \frac{(m_p - m_e) \vec g} {2e} =  \nabla \left[\frac{(m_p- m_e)
   \Phi_g}{2e}\right]
\label{E_PR}
\end{equation}
where $e$ is the electron charge; $p_e = n_e k T_e$ is the 
electron kinetic pressure; $m_p$ and $m_e$ are 
the proton and electron masses respectively; 
$\Phi_g$ and $\Phi_{PR}$ are respectively the gravitational 
and the electrostatic potential. 
These potentials are functions of the altitude in the corona.
Equation (\ref{E_PR}) determines the PR electric field 
and the electric potential in a pure hydrogen plasma. 
In the solar corona $\Delta \Phi_{PR}$, the electrostatic
potential difference between Chamberlain's exobase ($r=2.5~R_S$) 
and $r=\infty$, is equal to 150~V\@.

In the case of multi-ionic plasmas, similar analytical relationships
can be derived for the electrostatic potential, $\Phi_{PR}$, 
even when the ion temperatures ($T_i$) and the electron
temperature ($T_e$) are not equal [cf.\ Sect. 5.2.2 in
 \citeauthor{LemaireGringauz},  \citeyear{LemaireGringauz}].
The most general expressions for this polarization electric field, 
including the effect of field-aligned flows of ions and electrons 
can be found in the paper of \citet{Ganguli87}.
An expression for the parallel electric field sustained 
by two-dimensional cross-B sheared plasma flows has been derived by \citet{Echim2005}.

The density of the polarization charges responsible for 
the PR electrostatic field is extremely small: $(n_p -
  n_e)/n_e = 4\times10^{-37}$ \citep{LemaireScherer70}. 
This indicates that the quasi-neutrality equation ($n_p = n_e$) 
is indeed a good approximation in most space plasmas; 
therefore, instead of solving Poisson equation to calculate 
the electrostatic field distribution, it is rather satisfactory
(and much simpler from a computational
point of view) to solve the quasi-neutrality equation 
by an interative method in order to 
derive the spatial distribution of the electrostatic potential, 
$\Phi_E(r)$.  
%\comment{ols: It is the \emph{no current} condition that
%fixes the electric field, not the quasi neutrality condition.}

%It is remarkable that eq. (\ref{E_PR}) is applicable 
%when the plasma is permeated by a magnetic field $\vec B$, and even 
%when  $\vec B$ is parallel to
%$\vec E_{PR}$. 
%Thus, $\vec E_{PR}$ is a simple example 
%of electric field distributions  violating the ideal MHD condition:
%$\vec E \cdot \vec B= 0$. 
%Indeed it can be verified that, when  $\vec B$ 
%is parallel or anti-parallel to  $\vec g$, 
%the PR electric field (\ref{E_PR}) has a non-zero component 
%parallel to the magnetic field lines: i.e. $\vec E_{PR} \cdot \vec B \ne 0$.
%\footnote{The Pannekoek-Rosseland electric field given by eq. (\ref{E_PR}) 
%could be added to the gallery of $\vec E$-field distributions violating ideal 
%MHD: $ \vec E \cdot \vec B= 0$. 
%This gallery was opened in EOS by \citet{FalthammarMozer}
%\textcolor{red}{\textbf{E. Parker comment: Not so, e$\vec
%    E_{\parallel}$} is canceled by the opposing gravitational acceleration.}}

%This might explain why the derivation of the Pannekoek-Rosseland 
%polarization E-field is never given in textbooks concentrated on
%ideal MHD applications in plasma physics.

When (\ref{E_PR}) is applicable, the sum of the gravitational
potential energy and electric potential energy, $R(r)$, 
is the same for the protons and for the electrons: 
\begin{equation}
R_p(r) = m_p \Phi_g +  e \Phi_{PR}  = m_e \Phi_g 
-  e \Phi_{PR}  =  R_e(r) =  (m_p + m_e) \Phi_g(r)
\label{sum_potential}
\end{equation}
As a consequence of this equality,
$H_p$, the density scale height of the protons, is 
precisely equal to $H_e$, the density scale height 
of the electrons [see Sect.\ 5.2.2 in 
\citeauthor{LemaireGringauz}, \citeyear{LemaireGringauz}]. 
This has to be so, since the 
quasi-neutrality condition has to be satisfied at all altitudes.
% and therefore $H_p$ must be equal to $H_e$, to make sure that 
% $\nabla n_p = \nabla n_e$. 
% \comment{ols: It is not ``remarkable'',
%   it is by design!  We have assumed quasi neutrality, and then the
%   scale heights for protons and electrons must necessarily be equal.}

Another important consequence of the presence of the PR electric field 
is that the minimum energy for an electron to escape out of the 
potential well (i.e. the critical escape energy, 
$1/2m_e  {v_e}^2$ at given altitude), is precisely equal to that of a proton 
($0.5m_p  {v_p}^2$) at the same given altitude: 
\begin{equation}
\frac{1}{2} m_p {v_p}^2   = - R_p(r_0) =  \frac{1}{2} (m_p +
  m_e) g_0 r_0 = \frac{1}{2} m_e {v_e}^2   =  - R_e(r_0)
\label{min_energy}
\end{equation}
%These properties are implicitly used in Chamberlain's exospheric model of 
%the solar corona, since he postulated  that the electric field in 
%the coronal ion-exosphere is  the Pannekoek-Rosseland electric field, 
%inferring that the eqs.\ (\ref{E_PR})--(\ref{min_energy}) are
%applicable. 
Using Chamberlain's notation, 
$U_p = (2 k T / m_p)^{1/2}$ corresponds to the proton 
thermal velocity, and  $\lambda_{p0} = m_p {v_{p\infty}}^2 /  m_p {U_p}^2$  
is the dimensionless ratio between the critical escape energy and 
the thermal energy of a proton at the exobase altitude. 
A similar dimensionless parameter can be defined for the electrons:
 $\lambda_{e0} = m_e {v_{e\infty}}^2 /  m_e {U_e}^2$.

\subsection{The first generation kinetic/exospheric models for the coronal evaporation}

Assuming that the proton VDF at the exobase of the collisionless region can 
be approximated by a truncated Maxwellian, the Jeans evaporative flux 
of particles with velocities exceeding the critical escape 
velocity at the exobase is given by:
\begin{equation}
F_p(r_0) = n_p(r_0) U_p \left[ 1 +  \lambda_{p0}\right]
 \frac{ e^{-\lambda_{p0}}} { 2 \sqrt{\pi}}
\label{Jeans_flux}
\end{equation}
\citet{Chamberlain60} derived also analytical formulas 
for $n_p(r)$, the density of all particles 
populating the region above the exobase (i.e. the sum of the 
density of the escaping particles, of the ballistic or 
captive ones, as well as trapped or satellite particles).  
From these analytical expressions (which will not be given here), 
he was able to calculate the radial distribution of $n_p(r)$ 
and of $u_r(r) = F_p/n_p$, the average proton velocity: 
\begin{equation}
u_r(r) = (U_p/ 2 \pi^{1/2} ) \left[ 1 +  \lambda_{p0} \right] 
e^{ -\lambda_p} \left(\lambda_p/\lambda_{p0}\right)^2
\label{u_r}
\end{equation}
where
\begin{equation}
  \lambda_p(r) =\lambda_{p0} \frac{r_0}{r} = 2 \frac{R_p(r)}{m_p {U_p}^2}
\label{lambda_p}
\end{equation}
is the ratio of the total potential energy and average thermal energy
of the protons at the radial distance $r$. 
A similar function of $r$ can be defined for the electrons : 
\begin{equation}
\lambda_e(r) =\lambda_{e0} \frac{r_0}{r} = 2 \frac{R_e(r)}{m_e {U_e}^2}
\label{lambda_e}
\end{equation}
As a consequence of 
(\ref{min_energy}) and the definition of $U_i$, 
the functions $\lambda_e(r)$ and $\lambda_p(r)$  
have the same value everywhere in the exosphere
when the exobase temperature of the electrons is equal to that 
of the protons ($T_e = T_p = T$).

From (\ref{u_r}) it results that the value of the expansion
velocity of the protons is subsonic ($u_r< U_p$) everywhere above the exobase. 
This is why Chamberlain called his model ``solar breeze,''
in contrast to the supersonic bulk velocity inferred by Parker in 
his solar wind model. In the exospheric solar breeze model 
$u_r(r)$ decreases monotonically with $r$, from a maximum
value, ($U_p/ 2 \pi^{1/2} \left[1 +  \lambda_{p0}\right]
  e^{-\lambda_{0p}}$, at the exobase, to zero at $r = \infty$.

For an exobase proton temperature of $T_p = 2 \times 10^6$~K 
 and an exobase density of 
$n_p(r_0) = 10^6~\mathrm{cm}^{-3}$, one has
 $\lambda_{p0} = 2.3$, and  the average expansion velocity at 1~AU 
is 20~km/s. Chamberlain's exospheric calculations predict also 
a proton density of $370~\mathrm{cm}^{-3}$ at 1~AU, which 
he considered then to be in fair agreement with values 
derived by \citet{Behr53} from zodiacal light measurements.

It can be seen that the value of plasma bulk velocity (20~km/s)
predicted by the solar breeze model is more than an order of magnitude 
smaller than that predicted by Parker's hydrodynamic solar wind model 
at the orbit of the Earth, i.e., 500~km/s for a comparable coronal 
temperature (see models $\{7\}$ and $\{10\}$ in
Table~\ref{tab:table1}).  
The bulk velocity predicted by the exospheric solar breeze 
solution is also much too small compared to those which were measured 
later on by the first interplanetary missions (300--700~km/s) at 1~AU\@.
Furthermore, the proton density and temperature predicted at 
1~AU by Chamberlain's evaporative model occurred to be
quite different from the measurements made in 
interplanetary space in 1962 and afterwards.  
This led the space physics community to disregard the 
solar breeze model, and by the same token, all other kinetic 
or exospheric models of the solar corona like those discussed later on by 
\citet{Jenssen}, \citet{BrandtCassinelli66}. 
In these alternative exospheric models the
ballistic/captive and trapped/satellite protons
are missing; only the escaping protons contribute
to the exospheric density in \citet{Jenssen} or
\citet{BrandtCassinelli66} evaporative models.  This rather questionable
assumption lead to smaller exospheric proton densities than in the
solar breeze model  and therefore to higher values for
$u_r$ (respectively 290~km/s and 266~km/s at 1~AU).  Nevertheless,
these new exospheric models for the evaporation of the solar corona
were also based on the postulate that the polarization electric field
is given by the Pannekoek-Rosseland distribution. As a matter of
consequence they were unable to predict bulk velocities comparable
to those observed in the solar wind at 1~AU.
%Therefore, in spite of the much larger values for $\mathrm{u_r}$  their 
%contributions did not rehabilitate exospheric models, nor kinetic approaches in general. 

This failure restricted the popularity of kinetic exospheric SW
models and fed the persistent and overwhelming belief that 
only hydrodynamic solutions could 
offer satisfactory  descriptions for the supersonic coronal expansion.
It was generally  claimed that
the the exospheric solutions
do not satisfy the plasma transport equations. The latter
belief is clearly incorrect since all the moments of the VDFs obtained
by the kinetic exospheric models necessarily satisfy the entire set
of moment equations, including eqs. (\ref{eq:cont})-(\ref{eq:emom}).
Despite the undeniable evidence that beyond $4~R_S$ or so, the 
Knudsen number of the proton and electron gases become larger than 
unity, kinetic approaches were generally dismissed and merely
considered as ``academic exercises'' of no or little relevance in modeling 
the solar wind.

At about the same epoch \citet{Lemaire68} had studied 
the formal asymptotic behavior (for $r \to \infty$) 
of the radial distributions of the density, bulk velocity, and 
temperature in Chamberlain's evaporative solar breeze model, 
as well as in \citet{Jenssen} alternative ion-exosphere model.  
He compared these two different types of exospheric distributions 
with Parker's hydrodynamic solar wind models.  
This comparative study did not explain, however, why  the  first generation
kinetic models of the solar wind predict much lower expansion velocities than  
the hydrodynamic models existent at that time.  
The origin for this  basic disagreement was 
discovered later on by \citet{LemaireScherer69, LemaireScherer71}. 
An exospheric model giving a supersonic terminal velocity was found
independently by \citet{Jockers70}.
This is recalled in detail in the following section.

 \subsection{The second generation of kinetic/exospheric models}
\label{second_gen_exo}
\subsubsection{Modification of the Pannekoek-Rosseland electric field distribution by
the zero flux condition}

One common feature of the first generation exospheric models discussed in previous 
paragraphs was the Pannekoek-Rosseland electric field given by (\ref{E_PR}).
It should be reminded that this electrostatic potential distribution 
had originally been introduced  to model stellar ionospheres 
in hydrostatic equilibrium, like Chapman's conductive model of the solar corona.
 
But we presently know that the coronal plasma is not in 
hydrostatic equilibrium, and that protons and electrons 
continuously evaporate from the solar corona.
%Clear evidence for this coronal expansion is  Chamberlain's solar breeze, or
%Parker's hydrodynamic expansion. 
Since the basic hypothesis used to derive (\ref{E_PR}) 
fails to be satisfied, the polarization electric field keeping the coronal plasma 
quasi-neutral ought to be different from the 
Pannekoek-Rosseland (PR) electric field. 

\citet{LemaireScherer69} showed that the PR electric field (\ref{E_PR})
is inadequate because, according to eq. (\ref{Jeans_flux}),
the evaporative flux of the protons would be 43 
times smaller than the escape flux
of the electrons; this results from the much larger electron thermal speed.  
Indeed, when $T_e = T_p$ and $n_e(r_0) = n_p(r_0)$, 
equations (\ref{lambda_p}) and (\ref{lambda_e}) give
$\lambda_{e0} =\lambda_{p0}$ and
$\lambda_{e}(r) =\lambda_{p}(r)$, and according to (\ref{Jeans_flux}),
\begin{equation}
  F_e(r) / F_p(r) = U_p/U_e = (m_p / m_e)^{1/2}=43.
\label{e2p_flux}
\end{equation}
This  situation is  not sustainable since it would imply a huge radial electric current 
flowing out the solar corona, thus an unphysical
electric charging of the Sun.  
{Despite that} \citet{Pikelner50} 
 {discovered the same inconsistency this was not 
generally known to the community until 1969 when rediscovered
independently by Lemaire and Scherer.}
\citet{Jockers70} {developed independently
 a kinetic exospheric solar wind model quoting an earlier finding by}
 \citet{Pikelner50}  {who realized already
 two-decades earlier that the PR electric field implies 
 a larger evaporation flux of electrons than
 for the ions.}
% The model of} \citet{Jockers70}  {shows that
%proton potential is non monotonic with a maximum arround 5 $R_S$.}

In order to balance the net electric current,
%evaporative flow of the electrons and protons, 
a larger potential difference,
$\Delta\Phi_ E$, is required between the exobase and infinity 
than that corresponding to the PR electrostatic potential
distribution, $\Delta \Phi_{PR}$.
Considering that the exobase is located at a radial distance of 
$6.6~R_S$, that the exobase temperature of electrons 
is $1.4 \times 10^6$~K and $0.9 \times 10^6$~K for the protons
(these values were inferred from Pottasch's observations, \citeyear{Pottasch60}), 
\citet{LemaireScherer71} calculate that   $\Delta \Phi_{E}$  
had to be 410~V, in order to balance the net escape
 of the electrons and protons, and therefore to keep the radial
component of the electric current equal to zero. 
It can be seen that in this second generation of
exospheric models the total field-aligned electric potential 
drop is more than two times larger than that of the PR field
 implicitly inferred in Chamberlain's \citeyear{Chamberlain60} 
first generation exospheric solar breeze model \citep{LemaireScherer71}.
%This is illustrated in Fig. 1 of 
%Lemaire and Scherer (1971b), where the LSa curve corresponds 
%to their exospheric model, while the PR curve corresponds 
%to the Pannekoek-Rosseland electric potential distribution.

The much larger value of $\Delta \Phi_E$ 
compared to $\Delta \Phi_{PR}$
%for the total electric potential difference necessary
%to maintain equal escape fluxes for electrons and protons
accelerates the coronal protons to significantly larger velocities
at 1~AU than in the first generation exospheric models,
which were based on $\Delta \Phi_{PR}$ electric potential.  
For the exobase conditions listed above,
\citet{LemaireScherer71} obtained at $r = 215~R_S$ the following
values for the bulk velocity of the protons and electrons: $u_r =
320$~km/s and $n_p = n_e = 7.2~\mathrm{cm}^{-3}$ for their densities
(see model $\{3\}$ in Table~\ref{tab:table1}).  Both set of values were in
better agreement with the average quiet time solar wind observations
reported by \citet{Hundhausen70} (column $\{1\}$ in
Table~\ref{tab:table1}) than the values predicted by earlier exospheric and
hydrodynamic models listed in Table~\ref{tab:table1} (see also Table 2
in \citeauthor{LemaireScherer71}, \citeyear{LemaireScherer71}).

For completeness, it should be added that the actual values of the 
electric potential distribution, $\Phi_E(r)$, at all
altitudes in the exosphere, are uniquely determined by 
solving the quasi-neutrality equation:  
%Indeed, the quasi-neutrality equation:
\begin{equation}
n_e\left[r, \Phi_g(r), \Phi_E(r)\right] = n_p\left[r, \Phi_g(r), \Phi_E(r)\right]
\label{quasineutrality}
\end{equation}
%Indeed, (\ref{quasineutrality}) remains a good approximation 
%of Poisson's equation, since in the exospheric 
%region, the relative charge separation, $(n_p - n_e)/n_e$, is of the order of
%$10^{-37}$ \citep{LemaireScherer71}.  
%\comment{ols: In sect. 2.2 you
%  write that the value is 4e-37.  So which one is correct?}
The exospheric densities of the protons and of the electrons 
in eq. (\ref{quasineutrality}) are analytical functions of
$\Phi_E$, which are determined  by the exospheric theory. 
Therefore, the radial distribution of $\Phi_E(r_x)$ can be calculated 
for any radial distance, $r$, in the ion-exosphere.
An iterative method has first been used  to determined 
$\Phi_E (r)$ by  \citet{LemaireScherer69,LemaireScherer71};
a direct integration method was proposed later on by \citet{Lemaire91}.
%Instead of the iterative procedure initially employed by 
%\citet{LemaireScherer69, LemaireScherer71b} to solve this non-linear algebraic equation, a less cumbersome mathematical method was discovered later on by Lemaire et al. (1991). It consists in integrating numerically the gradient of eq. (11) or, more generally, the third order differential equation obtained by taking the gradient of Poisson's equation.

\begin{figure}
\center
\includegraphics{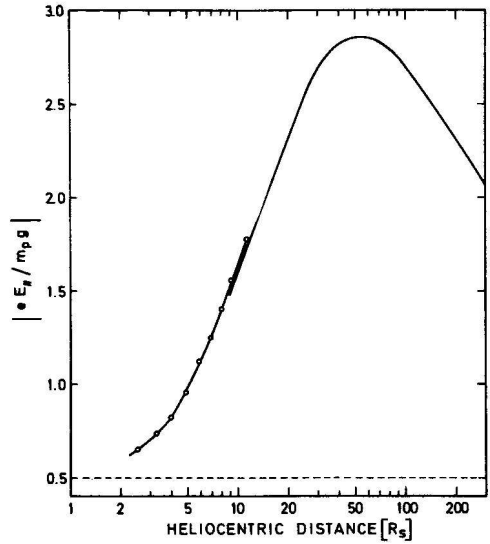}
\caption{Variation with radial distance of the ratio between
the electric and gravitational force, obtained
by the kinetic model of \citet{LemaireScherer73} \label{Fig2LS73}}
\end{figure}
Figure~\ref{Fig2LS73} shows the ratio of the electric force ($eE$) and
gravitational force ($m_pg$) acting on protons in the exospheric
model (LSd) of \citet{LemaireScherer73}.  The dots are values for
$|eE/m_pg|$ deduced from an empirical electron density distribution
derived by \citet{Pottasch60} from eclipse observations .  This
comparison shows that the values of $\left|eE / m_pg\right|$,
of the exospheric models increase in the inner corona from 0.5 (the PR
value shown by the dashed horizontal line), to a maximum value larger
than 1 at $r = 50~R_S$.  Note that beyond the exobase level in this
kinetic model, the electric force which accelerates the protons to
supersonic speed is larger than the gravitational force.

\subsubsection{Revisited second generation kinetic/exospheric models:
  comparison with previous models and data.}

\begin{figure}
\center
\includegraphics{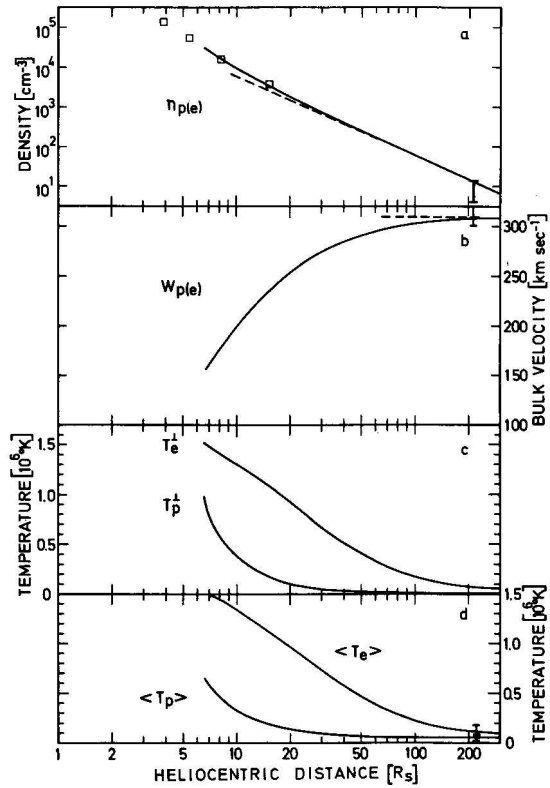}
\caption{Exospheric distribution of the density ($n_{p(e)}$), bulk
  velocity ($W_{p(e)}$), electron ($T_{e\bot}$)
and proton ($T_{p\bot}$) perpendicular temperatures and 
omnidirectional temperatures ($<T_e>$, $<T_p>$) of the LSc
model (adapted from \citeauthor{LemaireScherer72}, \citeyear{LemaireScherer72})}. 
\label{Fig3LS73}
\end{figure}

The radial distribution of the plasma density in the exospheric model
(LSc) published by \citet{LemaireScherer72} is illustrated 
by the solid line in the top panel of Fig.~\ref{Fig3LS73}. 
The dashed line indicates the $r^{-2}$ density profile 
corresponding to the asymptotic variation of the solar wind 
density at large distances where the expansion velocity is 
almost independent of $r$.  The empirical densities determined by 
\citet{Pottasch60} from eclipse observations are displayed by 
square symbols up to $r = 16~R_S$ , while the range 
of solar wind measurements at the Earth's orbit is indicated 
by a vertical bar at $r = 215~R_S$. 

The second panel in Fig.~\ref{Fig3LS73} shows the flow velocity 
in this exospheric solar wind model. The value of $u_r$ 
(labeled $W_{p(e)}$) increases from 160~km/s at the 
exobase (\mbox{$r_0 = 6.5~R_S$}), to 307~km/s at 1~AU\@. 
This flow velocity is supersonic in almost  the whole
ion-exosphere. The range of bulk speeds measured in the quiet 
solar wind regimes are indicated by a vertical bar at 
$r = 215~R_S$ \citep{Hundhausen70}. 

It can be seen that the theoretical predictions of these new kinetic
models fit well the average measurements of $n$ and $u_r$ at 1~AU\@.
These values are reported in columns $\{22\}$ and $\{1\}$ of
Table~\ref{tab:table1}.  The pitch angle averaged electron and
proton temperatures, $<T_e>$ and $<T_p>$ predicted by this kinetic
model fall within the range of values measured at 1~AU and reported
by \citet{Hundhausen70} for the quiet solar wind.  This is also
indicated in Table~\ref{tab:table1}.
% where the electron and
%proton temperatures are given at 1~AU for the LSc an d LSb
%exospheric models as well as for the solar breeze model $\{7\}$.  
The range and mean values of the corresponding parameters for the quiet solar
reported by \citet{Hundhausen70} are given in colon $\{1\}$.  
The agreement with
observations of the revisited exospheric models
 is clearly much more satisfactory than the solar breeze
model based on the Pannekoek-Rosseland model of the electric field.

Thus, the results of exospheric models LSb and LSc included in
Tables~\ref{tab:table1} and \ref{tab:table2} abrogate
Chamberlain's claim ``that Parker's supersonic solution is not
supported by some simple evaporative explanation.''  Indeed, if a
correct polarization electric field distribution would have been used
by Chamberlain, he would have obtained results which are in
surprisingly good agreement with the observations at 1~AU\@.
The data in Tables~\ref{tab:table1} and \ref{tab:table2}
contradict a persistent but false consensus that ``exospheric'' models
would only be ``academic exercises'' of little relevance for the
physics of the solar wind. The predictions of the
revisited exospheric model of \citet{LemaireScherer71,
  LemaireScherer72} were even in a better agreement with the observations
than the predictions of hydrodynamic models available at that time.

From the second order moments of the exospheric VDFs, the pitch angle averaged
temperature, \mbox{$T =(T_\parallel + 2 T_\perp)/3$}, as well as the
temperature anisotropies, $T_\parallel / T_\perp$, of the protons and
electrons can be calculated. In \citet{LemaireScherer71,
  LemaireScherer72} -- model LSb -- the averaged proton
temperature is $4.8 \times 10^4$~K at the orbit of Earth, and
the electron temperature $11.7 \times 10^4$~K\@.  As indicated above,
these theoretical values fit well the quiet solar wind observations:
$T_p = (4.4 \pm 1.8) \times 10^4$~K, $T_e = (14
  \pm 5) \times 10^4$~K \citep{Hundhausen70}.

The hybrid or semi-kinetic model of \citet{Jockers70}, denoted
$\{4\}$ in Table~\ref{tab:table1}, gives a 
less satisfactory agreement  at 
1~AU: $u_r = 288$~km/s, $n_p = n_e = 12~\mathrm{cm}^{-3}$,  $T_p = 6.7 \times 10^4$~K, 
$T_e = 46 \times 10^4$~K compared to column $\{1\}$ of Table 1.  
Similar lack of agreement exists for the hybrid/semi-kinetic models $\{5\}$
of \citet{Hollweg70}, $\{6\}$ \citet{Chen+al72}, and for the exospheric 
models $\{8\}$ and $\{9\}$ of \citet{Jenssen}, and \citet{BrandtCassinelli66}
not reported in Table~\ref{tab:table1} for lack of space.

%The asymptotic electron temperature at large distances has been determined 
%for SW exospheric models by \citet{Meyer-Vernet98}. 
%They established that the electron temperature radial profile is 
%equal to the sum of a term proportional to $r^{-4/3}$ plus a
%constant term, with both terms being of the same order of magnitude at
%about 1~AU\@.  They showed also that this remarkable asymptotic 
%behavior is independent of the velocity distribution function adopted 
%at the exobase altitude. This typical behavior corresponds remarkably 
%well to the results for $T_e(r)$ versus $r$ determined from a 
%statistical study  by \citet{Issautier98} based on Ulysses 
%observations. No hydrodynamic model of the solar wind has yet been
%able to predict this typical radial distribution for the electron 
%temperature in the solar wind. {Note also that the revisited exospheric models
%of the solar wind predict a larger  temperature for the
%electrons than for the protons at 1~AU, 
%in closer agreement with the observations.}

 {Non-Maxwellian electron distribution functions
 have been observed in  fast and slow solar wind, for a large range of
solar conditions, from quiet to active. Three main electron components
 have been identified: the ``core'' -  the thermal part of the
 electron VDF, the ``halo'' -  the suprathermal part of the electron
 VDF and the ``strahl'' - an antisunward magnetic-field-aligned
 population}
 \citep{Feldman+al1975,Pilipp+al1987}. \citet{Maksimovic+al2005} and
 \citet{Stverak2009}  {have studied the radial evolution
of the relative contribution of each of the three electron populations.  
The model of}  \citet{Stverak2009}  {based on data from Helios, Cluster
 and Ulysses suggests that  the relative density of the ``strahl'' decreases 
while the relative density of the ``halo''  increases with increasing radial
 distance, for both fast and slow solar wind regimes. It was also
 shown that in the fast SW the nonthermal tails of the electron VDF
form close to the Sun; this is a confirmation of the
assumptions made on the boundary conditions of the third generation 
exospheric models.}

Asymptotic  {total} electron temperature at large distances has been
computed  from exospheric models by \citet{Meyer-Vernet98}.
They  established that the average electron temperature radial profile
may be approximated by the  sum of a term proportional to $r^{-4/3}$ plus a
constant term, with both terms being of the same order of magnitude at
about  1~AU\@. They showed also that this remarkable asymptotic behavior is
independent of   the velocity distribution function adopted at the
exobase altitude. This typical behavior seems to be in agreement with 
the results obtained for the temperature 
of the ``core''  electrons determined from a statistical study by
\citet{Issautier98} based on Ulysses observations. 
\citet{Meyer-Vernet98}  {pointed out that their exospheric computation 
do not predict correctly the separate contributions to the temperature of arbitrary
parts of the velocity distribution function, like, for instance,
the ``halo'' population. 
%The radial profile of the latter has been
%determined experimentally by, e.g., } \cite{Stverak2009}.
\citet{Issautier+al2001} {have compared the total
  SW electron temperature from 2nd generation exospheric models and
high-latitude Ulysses data, showing a reasonable agreement.}

In summary, the density, bulk velocity, and pitch angle averaged
temperature of electrons in
the second generation of kinetic exospheric models  
are in agreement with observations. This is a 
consequence of replacing in the exospheric solution
the Pannekoek-Rosseland polarization electric field 
by the self-consistent Lemaire-Scherer (LS) electric field.
The electric potential is derived from the quasi-neutrality equation
and the additional constraint that the thermal electrons
do not evaporate out of the corona at a higher rate than the protons.

Some renewed interest for kinetic descriptions of the solar
wind appeared when \citet{Scarf1967} and \citet{Hundhausen68}
reported on the proton temperature anisotropy, with $T_{\perp}$ 
larger than $T_{\parallel}$ at
1~AU\@.  They concluded that the interplanetary
plasma should be more or less collisionless beyond a radial distance
of 10 or $20~R_S$. Indeed, none of the hydrodynamic SW models
available at that epoch did account for  the
existence of temperatures anisotropies for the proton and electron,
nor did they account for skewed and double-hump VDFs that can hardly
be fitted  by displaced Maxwellian velocity distribution
functions.

Using still another exospheric approach, \citet{GriffelDavis69} 
calculated that the temperature anisotropy resulting from
a collision-free flow of protons, along spiral interplanetary magnetic
field lines, is much larger than the corresponding values 
observed at 1~AU\@.  Thus, these authors consolidated the 
latent argument that, between 0.1~AU and 1~AU, there should be some 
relaxation mechanism producing about 2.5 collisions/~AU or pitch-angle
scattering to reduce the excessive $T_{\parallel}/T_{\perp}$ in
collisionless solar wind models. 
%This presumed scattering mechanism would decrease the too 
%large anisotropy deduced in exospheric model calculations 
%and reduce it to values comparable to those observed at 1~AU\@.  
\citet{EviatarSchulz70} and \citet{SchulzEviatar72} 
reached similar conclusions based on their own kinetic description 
of solar wind protons and electrons. 
%They also argue in favor of some physical scattering mechanisms 
%that would reduce the excessive temperature anisotropies 
%also predicted by their exospheric calculations.

%\subsection{The question of the temperature anisotropy}

\citeauthor{Hundhausen70}'s (\citeyear{Hundhausen70}) observations show that the velocity distribution
functions of electrons and  protons are not isotropic.
Similar results were obtained by  \citet{Marsch+etal1982}  and
\citet{Pilipp+etal1987c} from Helios data.
Indeed, a larger dispersion is generally observed for the velocities parallel to 
the magnetic field lines than for the perpendicular ones, i.e.
the observed electron and proton temperatures are anisotropic. 
%By convenience, in hydrodynamic models it was often assumed that 
%$T_{e\parallel}/ T_{e\perp} = 1$  and 
%$T_{p\parallel}/ T_{p\perp} = 1$;  
%the motivation behind this assumption was that collisions 
%between particles or/and wave-particle 
%interactions are frequent enough and hard enough to justify such
%hypothesizes.  
In the quiet solar wind,  measurements 
consistently indicate that 
$T_{e\parallel}/ T_{e\perp} = 1.1 - 1.2$,
$T_{p\parallel}/ T_{p\perp} = 2\pm 1$, in the average.
Similar value are obtained in fast speed streams.
Observations of polar coronal holes from the UVCS SOHO
have evidenced temperature anisotropies in protons and heavy
ions, but with $T_{p\parallel}/T_{p\perp} < 1$ at the base of the corona
\citep{Cranmer+etal1999}.

In all exospheric models, the temperature anisotropies are larger than
observed. For instance in Lemaire-Scherer's kinetic model LSa, at
1~AU: $T_{e\parallel}/ T_{e\perp} = 3.05$, and $T_{p\parallel}/ T_{p\perp}
= 164$; in \citet{Jockers70} hybrid/semi-kinetic model at 1~AU:
$T_{e\parallel}/ T_{e\perp} = 1$ and $T_{p\parallel}/ T_{p\perp} =
900$.  (see models $\{2\}$, $\{3\}$ and $\{4\}$ in
Table~\ref{tab:table1}).
\citet{Chen+al72} and \citet{Pierrard2001c} demonstrated that when a
spiral interplanetary B-field is adopted, instead of being radial, 
the anisotropy of the proton temperature at 1~AU is significantly 
reduced, but not enough, however,  to match the observed values.
%\comment{ols: Put in reference to myself here, which says that Coulomb
%collisions are insufficient.}

The extreme values predicted for the exospheric anisotropies of electron and
proton VDFs at 1~AU naturally suggest that some pitch angle scattering
mechanisms operate in the solar ion-exosphere, as inferred in the early
study of \citet{GriffelDavis69}. \citet{LemaireScherer71} suggested
for instance that Coulomb collisions, ignored in their
exospheric formulation, might indeed change the pitch angles of the
escaping particles, and will therefore attenuate the too large exospheric
anisotropies of the electron and ion VDFs.  
%Since the mean collision
%time for momentum transfer is smaller than the mean time for changing
%the energy of charged particles by Coulomb collisions, the latter may
%be good candidates to change the pitch angle distributions without
%affecting significantly the energies of the particles, i.e. without
%altering their  temperatures which fit rather well the
%observations in Lemaire-Scherer's exospheric models.
According to \citet{LemaireScherer71}, the mean pitch-angle scattering
times of solar wind protons and electrons are respectively
$6\times 10^5$~s ($7$ days), and $4\times 10^4$~s ($0.5$ days) at 1~AU\@.
These time scales are slightly smaller than the time required for
the particles to travel a distance of 0.5~AU\@ 
%and give a measure
%of the time needed for the plasma to become isotropic 
%Thus it can be
%argued that Coulomb collisions are indeed good candidates to attenuate
%the excessive temperature anisotropies predicted by exospheric models,
%without changing the values of the average temperatures, 
%$T_e$ and $T_p$.  

On the other hand, gyrotropic
fluid models of the solar wind, which did include the effect of
Coulomb collisions, predict that collisions were far from sufficient to
prevent extremely large proton temperature anisotropies in the outer
solar wind \citep{Lie-Svendsen+Hansteen+Leer+Holzer2002}.
The collision time for energy equipartition between
electrons and protons is about 100 times larger than the time required
for protons to travel a distance of 0.5~AU\@.  As a consequence, at
1~AU no significant energy can be transferred from the hotter electrons
to the solar wind protons nor to the other minor heavier ions present
in the SW.
Of course, wave-particle interactions may play a role 
in reducing the anisotropy of the VDF. 
This was already suggested by \citet{LemaireScherer71}, 
but to quantify their effect the actual frequency spectrum of 
these waves, their polarization, and radial distributions 
should be known with some confidence.

The effect of Coulomb collisions can be 
worked out by solving the Fokker-Planck equations for the 
different particle species in the coronal and solar 
wind plasma including for the minor ion species.  
Some preliminary attempts along these lines will be discussed 
later in this review. Before that, let us now examine 
some characteristic correlations observed in the solar wind
and compare them to the predictions of exospheric and hydrodynamic models.

%\subsubsection{Observed and predicted correlations of solar wind parameters}

In the LSb model developed by \citet{LemaireScherer71}, 
the exobase altitude was taken to be at $r_0 = 6.5~R_S$. 
This input parameter depends on the  coronal 
electron density  and of its scale height 
taken from eclipse observations up 
to $16~R_S$ \citep{Pottasch60}. Since the mean free path of the 
coronal electrons and protons is proportional to the square 
of the temperatures of these particles, the altitude of the 
exobase (i.e. where the pitch angle scattering mean free path 
becomes equal to the atmospheric density scale height) in Lemaire-Scherer's 
models is a decreasing function of the assumed 
exobase temperatures. Therefore, by decreasing the temperature of 
the electrons and protons at the exobase 
the corresponding exobase altitude and density
can be varied in  a self-consistent manner as described by
\citet{LemaireScherer71}.  
As a result,  a whole family of 
exospheric models have been generated by \citet{LemaireScherer71}
showing that an increase of the exobase 
temperatures in association with a decrease of the exobase altitude, 
results in an increased proton temperature, $T_p$, 
and an enhanced bulk velocity, $u_r$, at 1~AU\@.  
However,  a change of the exobase temperature did not significantly 
affect $T_e$, at the 
orbit of Earth (cf.\ Fig.\ 5 in \citeauthor{LemaireScherer71}, 
\citeyear{LemaireScherer71}).

\begin{figure}
\center
\includegraphics[width=0.98\textwidth,height=0.4\textheight]{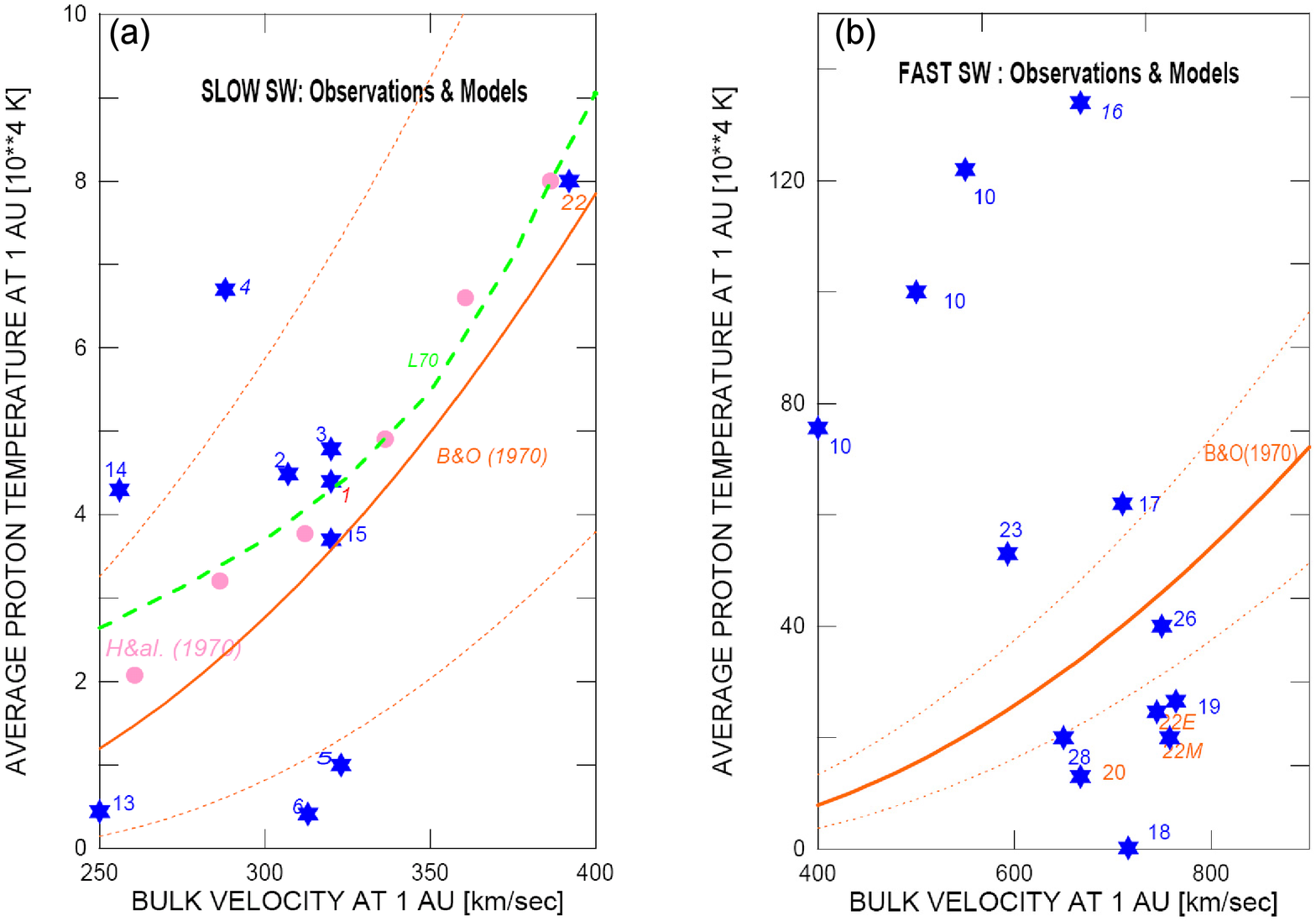}
 \caption{Correlation between average proton temperature ($T_p$) and solar wind
bulk velocity ($u_r$) at 1~AU for (a) slow solar wind, (b) fast solar
wind.   The green dashed line shows
the positive correlation between the calculated values of $u_r$ and
$T_p$ at 1~AU determined in an unpublished parametric study by Lemaire
(1971), reported in Fig. 12 of \citet{LemaireScherer73}.  The pink
dots correspond to Vela 3 measurements reported by
\citet{Hundhausen70}. The orange solid line and the two orange dotted
lines correspond respectively to the average relationship, the minimum
and maximum standard deviations, as reported by \citet{BurlagaOgilvie}
based on their statistical study of Explorer 34 measurements.  
The star symbol labeled ``1'' corresponds to the average values of the
quiet solar wind reported by \citet{Hundhausen72}; it corresponds also
to the mean values given in the column {1} of Table~\ref{tab:table1}.
The other numbered star symbols correspond to the ID numbers of
various kinetic (E), hydrodynamic (H) and semi-kinetic/hybrid (Hb)
models listed and referenced in Tables~\ref{tab:table1} and
\ref{tab:table2}. \label{fig:fig4}}
\end{figure}

The positive correlation between the SW expansion velocity and the
proton temperature found with Lemaire--Scherer's exospheric models
is illustrated in Fig.~\ref{fig:fig4}.
It can be seen   that over the range of velocities from 250~km/s to 400~km/s,
corresponding to quiet solar wind velocities
\footnote{Note that in the recent statistical study of Ulysses measurements 
by \citet{Ebert+al} the upper limit of low-speed SW 
is set to be $500$ $km/s$. ``Low-speed'' SW is also called ``quiet'' SW 
here as well as in earlier publications. The minimum bulk 
velocity of high-speed streams is 650~km/s in 
recent statistical study. In earlier studies and here, 
``high-speed'' SW was also called ``fast'' SW.}, the positive 
correlation predicted by Lemaire-Scherer's exospheric models 
(green dashed  line) fit rather well the observations of Vela 3 and 
Explorer 34.  Note that this agreement is 
well within the range of the standard deviations of observed SW 
parameters. 
% and quite
%satisfactory in  comparison to other models
%indicated by other star symbols in Fig.~\ref{fig:fig4}.
A positive correlation between $u_r$ and $T_p$ was also obtained 
by \citet{HartleBarnes70} with their set of two-fluid 
hydrodynamic models.  To obtain a good agreement with SW 
observations at 1~AU  extra heating rates of the protons had however to 
be added by  \citet{HartleBarnes70} in a region extending up 
to $25~R_S$.  Indeed, in the hydrodynamic models, the increased 
proton temperatures and wind velocity at 1~AU 
are direct consequences of the enhanced heat deposition
into the solar corona, while in Lemaire--Scherer's exospheric 
models, it is a simple consequence of a parametric increase
of the exobase temperatures.

\citet{Lopez+Freeman} had interpreted the positive correlation 
between $<T_p>$ and $u_r$ by a variation with velocity of the polytropic index 
 in their hydrodynamic SW models.  
This tended to indicate that a larger energy input is plausibly 
the cause of both higher proton temperatures and bulk velocities at 1AU.  
A similar theoretical interpretation for the observed  
(almost quadratic)  relationship between $<T_p>$ and $u_r$
has been proposed also by \citet{Matthaeus2006}. 
They suggest that MHD turbulence should provide a heating 
depending only on the age of solar wind plasma elements 
(since their release from the coronal base). 
However, after a comprehensive quantitative  analysis  
\citet{Demoulin2009} showed that  MHD turbulence gives a 
much weaker dependence on $u_r$  than  the actual observations 
at 1 AU. 
%the important acceleration of the solar 
%wind velocity in the region close to the Sun.

\subsubsection{Limitations of second generation exospheric models}

Like Euler's hydrodynamic models, exospheric models should be viewed
as zero-order approximations of more general kinetic models 
of the solar wind. 
%for the density, bulk velocity and temperatures of the electrons and protons.
However, so far the agreement shown  in Figs.~\ref{Fig3LS73} and
\ref{fig:fig4} is restricted to quiet/low-speed solar wind conditions; 
it does not apply to fast/high-speed
streams where the bulk velocity exceeds 500--650~km/s.  Such high
expansion velocity can possibly be attained for exobase electron
temperature higher than $3\times 10^6$~K; unfortunately, this is much
higher than usual coronal temperatures inferred from spectroscopic
or white light solar measurements during eclipses.

%\comment{ols: Again added that it is the \emph{electron} temperature
%  that is low in coronal holes!}  
Note that a similar restriction is applicable to
hydrodynamic models of the solar wind; indeed non-realistic high
coronal temperatures (larger than $3 \times 10^6$~K) would be required
in hydrodynamic models to attain solar wind velocities larger than
500~km/s at 1~AU\@.  Otherwise, conjectured heat deposition or/and
ad-hoc momentum boosting of the coronal plasma are needed up to
$25~R_S$ or beyond, in order to reach supersonic velocities of
700~km/s or more.  Unfortunately, this contradicts the observations
collected in fast speed streams. Indeed, it was discovered (see, e.g.
\citeauthor{Schwenn1978} \citeyear{Schwenn1978}) that
fast speed streams are consistently originating from coronal holes,
i.e., from regions of the solar corona where the electron temperature is
smaller than in the equatorial regions -- the origin of the quiet solar wind.
  
Therefore, nor the standard collision-dominated hydrodynamic models,
nor the collisionless exospheric models of the solar wind developed
so far can represent adequately the phenomenon of fast speed
solar wind streams; their applicability must be limited to the quiet
or slow solar wind flow originating from the hotter and denser
equatorial regions of the corona.  

\subsection{The third generation of kinetic/exospheric models}
%\subsubsection{More tentative improvements to model the fast solar wind}

%Since the Coulomb collision cross-section is inversely proportional to
%the fourth power of the relative velocity between charged particles
%(see Sect.~\ref{sec:prop-solar-wind}), the exobase altitude decreases
%when the energy of the colliding particles increases. Conversely,
%the exobase altitude is higher up in the corona for a subthermal proton 
%than for a particle which has a supra-thermal energy.
%

In kinetic exospheric models of the SW the 
altitude of the exobase is energy dependent.  This has
 been already pointed out by \citet{Pikelner50}.  In the exospheric
models discussed in the previous paragraphs, the mean energy of particles was considered to
define a sharp cut exobase level : a step like transition region
between the collision dominating plasma and the collisionless
exosphere. This procedure is a heritage of Jeans' exospheric
theory; to the best of our knowledge there are
no successful attempt to develop exospheric
SW models with energy dependent exobase levels.

The early kinetic exospheric models were based on the assumptions 
that: (i) the Knudsen
number changes from a small value to a value much larger than unity
over a relatively thin layer, (ii)  the velocity distribution
functions at the exobase could be approximated by truncated
Maxwellians, and (iii) the exobase level is at altitudes high
enough in the corona, so that the ratio between the electric force and
gravitational force acting on protons $|e E / m_p g|$ is larger than
unity in the whole solar ion-exosphere, i.e., that the repulsive electric
force accelerating the protons in the exosphere is everywhere larger
than the attractive gravitational force slowing them down (cf.\
Fig.~\ref{Fig2LS73}).

The kinetic modeling of the solar wind received new credit
in the second half of the 90's.
Some of the early, simplifying assumptions used in exospheric models
were questioned and less restrictive constraints have been adopted 
to reach results in better agreement with 
the challenging fast speed streams observed in 
the solar wind.

A first improvement was proposed by \citet{PierrardLemaire96}, who 
developed an exospheric theory based on a Lorentzian VDF 
(or ``kappa'' function), instead of the usual Maxwellian VDF\@.  
Isotropic Lorentzian VDFs, $f_\kappa(v)$ , 
are characterized by three independent parameters: 
 a ``kappa index,'' $\kappa$, a density, $n_0$, 
and by a most probable  thermal speed, $v_{th}$,  defined by :
\begin{equation}
  v_{th}^2 =  \frac{\left(2 \kappa -3\right) k T_0} {\kappa m}
\label{Vth_kappa}
\end{equation}
where, $T_0$, is the exobase temperature.
The Lorentzian velocity distribution function is written:
\begin{equation}
f_\kappa(v, r, t) =  n_0 
\frac{\Gamma(\kappa+1)}{\Gamma (\kappa - 0.5 ) (\pi \kappa  {v_{th}}^2)^{3/2}} 
\left( \frac{1+ v^2}{\kappa {v_{th}}^2} \right)^{-(\kappa+1)}.
\label{kappa_VDF}
\end{equation}
where $\Gamma$ is the Gamma function.
Similar expressions with different values for $\kappa_e$
and  $\kappa_p$  can be employed for the electron and proton 
VDFs at the exobase.

%%%%%%%%%%

Lorentzian (kappa) VDFs decrease as a power law of $v$, instead of varying 
exponentially like a Maxwellian;  a 
relatively  large fraction of particles populate the tail of the VDF 
(i.e. with velocities larger than the critical escape velocity).  
In the limit $\kappa \to \infty$ the 
Lorentzian function approaches a Maxwellian distribution function
with the same values of $n_0$ and $T_0$. A review of
kappa distribution functions and their applications in
space plasma physics has been recently published by
\citet{PierrardLazar2010}.

The existence of Lorentzian VDF characterized by given values of 
$\kappa$, is treated in third generation exospheric models
without discussing fundamental physical mechanism that would be able to 
account the non-Maxwellian power law tails. That fundamental yet unresolved 
question has been addressed in a few recent theoretical papers.
A comprehensive study reporting earlier attempts can be found in
\citet{Shizgal2007}.
% see also the review by \citet{PierrardLazar2010}.
It is shown that in a uniform plasma, the relaxation toward equilibrium 
does not necessarily tend to kappa distributions, even when a certain type of wave-particle interactions
is taken into account. But this does not exclude that such suprathermal tails 
could not be produced by other processes, including the presence of a gravitational force in
association with vertical density gradients. Another recent study by \citet{Liva2009} is addressing this
fundamental question: what could energize the particles inside the solar corona, and produce 
a population of suprathermal electrons and protons at the exobase, or even closer to the Sun? 
The latter indicates how such non-Maxwellian distributions may arise 
naturally from Tsallis statistical mechanics.
 { Another explanation for
the formation of suprathermal electrons, in particular the
magnetic-field-aligned component (or ``strahl'', see below) 
was proposed by}  \citet{Vocks+al2005} 
 {in terms of resonant interaction with 
anti-sunward propagating whistler waves.}

\citet{Scudder1992a,Scudder1992b} has been the first to introduce lorentzian VDFs instead of Maxwellians 
in modelling the distribution of plasma in the solar corona.
Assuming that the velocity distribution functions of corona ions have enhanced  
suprathermal tails, like the lorentzian VDFs, \citet{Scudder1992a,Scudder1992b} 
showed that the ion temperature is an increasing function of altitude in the inner corona. 
This coronal "heating" mechanism is caused by the increasing of the ratio of suprathermal 
over thermal particles as a function of altitude, without deposition of wave energy 
or magnetic field energy into the gas (Dorelli and Scudder, 1999). 
It was then given the name of "velocity fitration effect".
These results were subsequently confirmed and extended by \citet{PierrardLamy2003}.

As a consequence of a postulated enhanced population of suprathermal electrons
with energies above the critical energy, the Jeans evaporation flux of 
electrons  will be larger for a Lorentzian VDF than for 
a Maxwellian VDF given by eq. (\ref{Jeans_flux}). 
Since the evaporation flux of coronal electrons emitted by 
the corona is then larger, a larger electrostatic potential 
difference, $\Delta \Phi_E$, is needed to cancel the net
electric current out of the corona.  
As a consequence of the larger electric potential difference 
between the exobase and infinity, the protons are accelerated to 
higher supersonic velocities.

%%%%%%%%%%

%Lorentzian VDFs decrease as a power law of $v$, instead of varying 
%exponentially like a Maxwellian;  a 
%relatively  large fraction of particles populate the tail of the VDF 
%(i.e. with velocities larger than the critical escape velocity).  
%In the limit $\kappa \to \infty$ the 
%Lorentzian function approaches a Maxwellian distribution function
%with the same values of $n_0$ and $T_0$.

%As a consequence of the increased population of suprathermal electrons
%with energies above the critical energy, the Jeans evaporation flux of 
%electrons  will be larger for a Lorentzian VDF than for 
%a Maxwellian VDF given by eq. (\ref{Jeans_flux}). 
%Since the evaporation flux of coronal electrons emitted by 
%the corona is then larger, a larger electrostatic potential 
%difference, $\Delta \Phi_E$, is needed to cancel the net
%electric current out of the corona.  
%As a consequence of the larger electric potential difference 
%between the exobase and infinity, the protons are accelerated to 
%higher supersonic velocities.

\begin{figure}
\center
\includegraphics[width=0.8\textwidth,height=0.45\textheight]{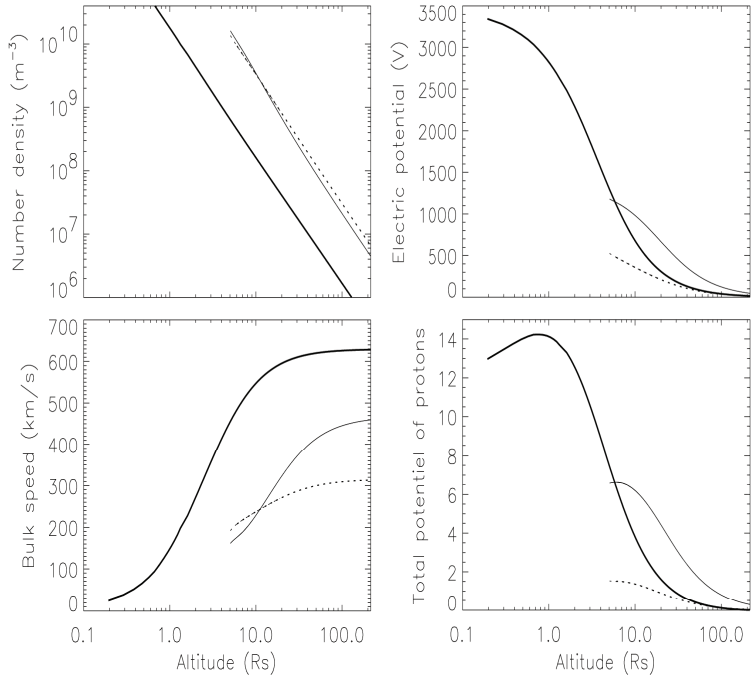}
\caption{Exospheric distribution of the density (upper left panel), 
electric potential (upper right panel),
bulk velocity (lower left panel), and total potential (lower right) for solar protons
from the model by  \citet{Lamy2003a} \label{FigLamy}.  
The dotted curves refer to a second generation exospheric model based
on a Maxwellian VDF at the exobase  ($r_0 = 6~R_S$); 
the thin solid curves correspond to a third generation
exospheric model for which the exobase 
is at the same altitude but for which the electron and proton VDF are
Lorentzian functions whose  index $\kappa =
3$. The thick solid line corresponds to an exospheric model
whose VDF are also Lorentzian functions ( $\kappa =3$)
but for which the exobase altitude is at $0.2$ $R_S$, i.e. below the
altitude where the total potential energy of the protons has its
maximum value. In all the three models the coronal temperature
of electrons and protons is the 
same, $T_0=1.5\times 10^6$ K.}
\end{figure}

This is illustrated in the four panels of Fig.~\ref{FigLamy} taken 
from \citet{Lamy2003b}, where the curves  
 show respectively the exospheric distributions of the density, $n(r)$, 
electric potential, $\Phi_E(r)$, plasma bulk velocity,  $u_r(r)$, and 
total potential  (the sum the 
gravitational and electric potential) of protons, $R_p(r)$,  in three 
different exospheric models. 
It can be seen that the bulk velocity in the exospheric model 
based on kappa VDF provides larger supersonic velocities
($>450$~km/s)  than in the case 
of a Maxwellian VDF for which $\kappa \to \infty$.  
Note that changing the value of $\kappa$ does not change 
significantly the plasma density in the ion-exosphere: the latter remains 
in excellent agreement with the average density observed both in 
the quiet or slow solar wind and in fast/high-speed solar wind regimes 
\citep{Hundhausen70, Ebert+al}.

An interesting parametric study has been published by 
\citet{Zouganellis2004} where non-thermal VDFs have been used 
at the exobase of the solar corona, instead of truncated Maxwellians.  
Their study illustrates the key role played by the population 
of supra-thermal electrons in the process of acceleration of 
the solar wind ions to large terminal speeds. 
\citet{Zouganellis2004} showed that the high terminal SW speeds 
obtained in their revisited exospheric models using Lorentzian (kappa) 
VDF at the exobase, is not just an artifact of the type of kappa 
function, but is a general feature when the tail of the electron VDF 
is parametrically enhanced. Indeed, similar high terminal speeds are 
obtained by using a sum of two Maxwellian VDFs : i.e. a first one 
with normal  coronal temperature, representing the core electrons, 
and a second one with a higher temperature corresponding to the 
halo electrons observed in the SW.  They found that 
 when the kappa index is reduced, or 
when the relative abundance of the halo electrons is enhanced 
with respect to the core electrons, the SW bulk speed at 1AU is 
systematically enhanced.
% in their exospheric models at least those 
%for which the exobase altitude is located beyond where the 
%outward directed electrostatic force becomes larger than the 
%gravitational force on a solar wind proton :  $|eE/M_pg| > 1$.

By reducing the index $\kappa$ from large values to less than 3, or by
increasing $T_0$, the exobase temperature, above $2\times 10^6$~K, it
is possible to increase the solar wind velocity at 1~AU to even higher
values.  However, it seems unrealistic to push these two parameters to
such extreme values in order to obtain bulk velocities as high as
700--800~km/s reported in fast speed solar wind streams.
Since fast speed streams originate from coronal holes where the
electron temperature is less than $10^6$~K it appears that some
additional modifications is required to boost exospheric bulk
velocities up to 1000~km/s.  This is what is tentatively proposed
in the studies reported in the next section.

%\subsubsection{Lowering the exobase altitude}

%In the previous paragraphs it has been emphasized that 
The mean free
path of the particles is larger in coronal holes than in the
equatorial corona at the same altitude and for the same temperature. 
As a matter of consequence the
exobase is located at a lower altitude in coronal holes than in the
equatorial region of the corona.  This implies that the gravitational
force there is larger than the electric force acting on protons.
This is illustrated in Fig. 2.  As a consequence of the lowering of
the exobase altitude the total potential energy of protons, $R_p(r)$,
remains dominated by the gravitational potential energy at the base of
the solar ion-exosphere.  Therefore, just above the exobase, $R_p(r)$ is
increasing with $r$.  It reaches maximum value $R_{p,max}$ at $r =
r_{max}$ where $|e \vec E / m_p \vec g| = 1$; beyond this heliocentric
distance, $R_p(r)$ decreases with $r$, just as in the earlier
exospheric model where the exobase was located higher in the
equatorial corona. If the exobase altitude is
significantly lowered, $R_p(r)$ is no more a monotonically decreasing
function of $r$, as assumed in the second generation exospheric models of
Lemaire-Scherer.
% $R_p(r)$ has now a maximum at some radial distance 
%($r_{max}$) above the exobase. 
The non-monotonicity of the total
potential energy of the protons,   {anticipated by the
  early kinetic exospheric model of } \citet{Jockers70},
is illustrated by the thick solid line
in the lower right hand side panel of Fig.~\ref{FigLamy}. For this
coronal hole exospheric model, $r_0 = 1.2~R_S$ and $r_{max} =
1.9~R_S$.  Below $r_{max}$, protons are attracted toward the Sun by
the dominating gravitational force; ballistic or captive protons are
then populating this region in addition to the escaping ones. The
former class of particles don't have a kinetic energy large enough to
reach over the maximum potential barrier, $R_{p,max}$.

In the third generation of kinetic SW models,
new analytical expressions were found for the densities, escape fluxes, expansion velocity, 
pressure tensors, temperatures, energy and heat fluxes by
\citet{PierrardLemaire96, Lamy2003a}.  {These authors 
performed an exact exospheric treatment of a
non-monotonic total (electric and gravitational) potential for the
protons, also treated by the kinetic exospheric model of} \citet{Jockers70}.
% in such  more 
%sophisticated exospheric models, new analytical expressions had to 
%be developed to determine the radial distributions for the moments of 
%the exospheric proton and electron VDFs.  
The analytical expressions derived by \citet{PierrardLemaire96, Lamy2003a}  
depend, of course, on the value of $r_{max}$  and of $R_{p,max}$. 
These parameters must first be calculated by an iterative procedure which has 
been developed and described by \citet{Lamy2003a}.  
This is how the exospheric distributions illustrated by 
the thick solid curves in Fig.~\ref{FigLamy} were obtained
\footnote{This new kind of exospheric SW model is freely available 
on the European Space Weather Portal (www.spaceweather.eu).}.

The thick solid line in the third panel of Fig.~\ref{FigLamy} shows
the radial distribution of $u_r$ obtained in this new exospheric model
for which the exobase is here at $r_0 = 1.1~R_S$, the exobase
temperature is $1.5 \times 10^6$~K and $\kappa=3$ for both the VDFs of
the electrons and the protons.  It can be seen that the supersonic
expansion velocity reaches now more than 600~km/s, which is larger
than the values obtained in earlier exospheric models with exobase
altitudes much higher up in the equatorial region of the corona.
Nevertheless, values as large as 1000~km/s can hardly be obtained even
with this ultimate improvement of the exospheric models for the solar
wind expansion. The plasma density in this new kind of exospheric model for
coronal holes is shown in the upper left panel of Fig.~\ref{FigLamy}.
As a consequence, of the lower exobase density adopted to match
coronal holes densities, the solar wind density at 1~AU is also lower
in this fast speed stream exospheric model, which is consistent with
the observations.   {One limitation of the  third
  generation exospheric models is a relatively high maximum temperature
reached by SW electrons at some distance from the Sun,
for highly nonthermal electron VDFs at the exobase.
%producing large terminal SW velocities at 1 AU.
 This problem could be due to the approximations used 
by the model to compute the self-consistent electric potential
or a possible superradial expansion of the wind}
\citep{Zouganellis2004}. 
 {Nevertheless, the third generation of exospheric models  by
\citet{Lamy2003a,Zouganellis2004} give a description of the transonic region, 
a major advance that enabled also to obtain large SW 
terminal velocities at 1 AU. Further improvement will provide a
more accurate representation of the solar wind, including of the
electron  temperature.}
%\comment{ols: The density at the lower boundary of
%  the model in Fig.~\ref{FigLamy} is extremely low and much lower than
%  coronal hole observations, which indicate densities at this altitude of order
%  $10^{13}-10^{14}~\mbox{m}^{-3}$.  Should we comment on this?}

As a consequence of the wide variety of adjustable 
parameters, the modern exospheric models behold a  great 
flexibility and allow to explain not only qualitatively, but
quantitatively as well, a broad range of SW features observed at 1AU; 
despite the rather artificial truncation of the exospheric VDFs,
exospheric models offer the substantial advantage of giving clues to
 understand the physics of the acceleration of the solar wind
  flow to supersonic terminal speeds.  Such an advantage is
  limited, however, in single-fluid hydrodynamic SW models, where the
  electrostatic force accelerating the protons to such speeds is
 actually hidden behind the gradient of the kinetic electron pressure
\citep{Parker2010}.
% in single-fluid hydrodynamic equations of the solar wind plasma. 
% This is why it is still often claimed within the MHD community that 
% the physical mechanism accelerating the solar wind is not well 
% understood. But, if properly reconsidered, the exospheric 
% theory of the SW, does help to dissipate this popular impression. 

Note that in any exospheric models of the solar wind, no
adjusted/ad-hoc heat or momentum deposition rate has been postulated, so
far.  But, based on the results reviewed above, additional heating 
mechanisms and acceleration of coronal protons seem indeed to be 
in order to drive fast speed streams up to bulk velocities of 1000
km/s.  Let this then be a provisional conclusion of the present 
review after half a century efforts to model the solar wind flow in 
the framework of exospheric theory which is inherited from 
J.H. Jeans.

\subsection{The fourth generation of SW kinetic models: Fokker-Planck solutions}
\label{kinetic_FK}

%An indisputable advantage of exospheric models is that they are 
%based on algebraic and analytical expressions instead of 
%solutions of non-linear  ordinary differential equations:
% the hydrodynamic or moment equations. 
%But most importantly, they have 
%served to reveal the key role of the polarization electric 
%field in the acceleration of solar wind ions to supersonic bulk
%velocities.  Inappropriate and inconsistent  distribution of this
%$\vec E$-field distribution 
%was indeed the major reason why the solar breeze model of the corona 
%did not provide results compatible with Parker's hydrodynamic solar 
%wind, nor with in-situ measurements at 1~AU.

Like any physics-based model, the exospheric 
approach has its limitations: (1)  it is restricted to  stationary models;
%Hydrodynamic transport equations can deal with 
%time-dependent aspects more readily, but the price to pay is 
%then to solve partial differential equations using finite difference numerical schemes. 
(2) the subtle effects of
rare collisions and of
wave-particle interactions above the exobase are ignored and (3) it cannot account
for processes near and below the exobase.

Fortunately, the effect of infrequent Coulomb collisions can 
be dealt with by solving the Boltzmann equation with
the Fokker-Planck (FP) collision term 
(see section \ref{Fokker-Planck}), for electrons or the protons, 
or for both species together. The FP collision term is, indeed, more appropriate 
than the Boltzmann collision integral (\ref{eq:boltzmann_J}).
In plasmas the cumulative 
effects of long range Coulomb collisions is more efficient than 
close binary collisions to produce momentum transfer, pitch 
angle deflection, and thermalization of the different species 
of charged particles. 

\citet{Lie-Svendsen+Hansteen+Leer1997} solved the 
Boltzmann equation with the Fokker-Planck collision term 
(\ref{eq:Fokker_Planck_Rosenbluth})
to develop the first collisional model of solar wind. 
They determined the VDF of the test electrons in the 
exobase transition region. They used a standard finite difference
numerical method to calculate the VDF at preset altitudes.  
A similar Fokker-Planck equation for the solar wind electrons was
solved by \citet{Pierrard+Maksimovic+Lemaire}. They developed a spectral method to
determine the polynomial expansion of the electron VDF across the 
exobase transition and adapted a  numerical code \citep{PierrardPhD97}
developed earlier to solve the FP equation for the polar wind.
 {The model by} \citet{Pierrard+Maksimovic+Lemaire} 
 {used correct boundary
conditions at the exobase, satisfying the regularity conditions
determined by  \citet{PierrardPhD97}.}
In a more recent application \citet{Pierrard2001b} added collisions
with protons. \citet{Pierrard+Maksimovic+Lemaire} integrated 
the FP equation for the solar wind electrons from 1~AU to $4~R_S$ in 
the solar corona. A typical VDF of core-halo electrons, 
observed at 1~AU in high speed solar wind stream by the WIND spacecraft, 
was used as  boundary conditions 
at the top of the integration domain in their FP model calculation.
Fig.~\ref{FigPierrard2001} illustrates the VDF of electrons 
given by this model at two
different radial distances in the solar wind ($r = 215~R_S$ and
and $4~R_S$).    The asymmetry of the
electron VDF observed at 1~AU, which can be interpreted as forming the
strahl population of high speed SW electrons, maps down into the solar
corona as a slight asymmetry of the coronal VDF.
%\begin{figure}
%\center
%\includegraphics[width=0.7\textwidth,height=0.45\textheight]{core.pdf}
%\caption{ The contour plots show the iso-density contours of
%$f_e(y_\perp, y_\parallel)$, at three different radial distances:
%$r=4~R_S$ (lower left panel), $r=180~R_S$ (right panels),
%$r=215~R_S$ (upper left panel); $y_\perp$, $y_\parallel$ denote the
%perpendicular respectively the parallel component of the velocity.
%The dashed circles correspond to the thermal speed of the electrons. The
%line-plots show the cross sections of $f_e(v_\perp, v_\parallel)$ in
%the radial direction, which is assumed here to be the direction of the
%interplanetary magnetic field; the dashed lines in these three panels
%show the corresponding cross-sections of the electron VDF in the
%direction perpendicular to the magnetic field;
%figure from Pierrard et al. (2001) \label{FigPierrard2001}}
%\end{figure}

%\begin{figure}[ht]
\begin{figure}
%\begin{minipage}[b]{0.5\linewidth}
\begin{minipage}{0.5\linewidth}
\centering
\includegraphics[scale=0.18]{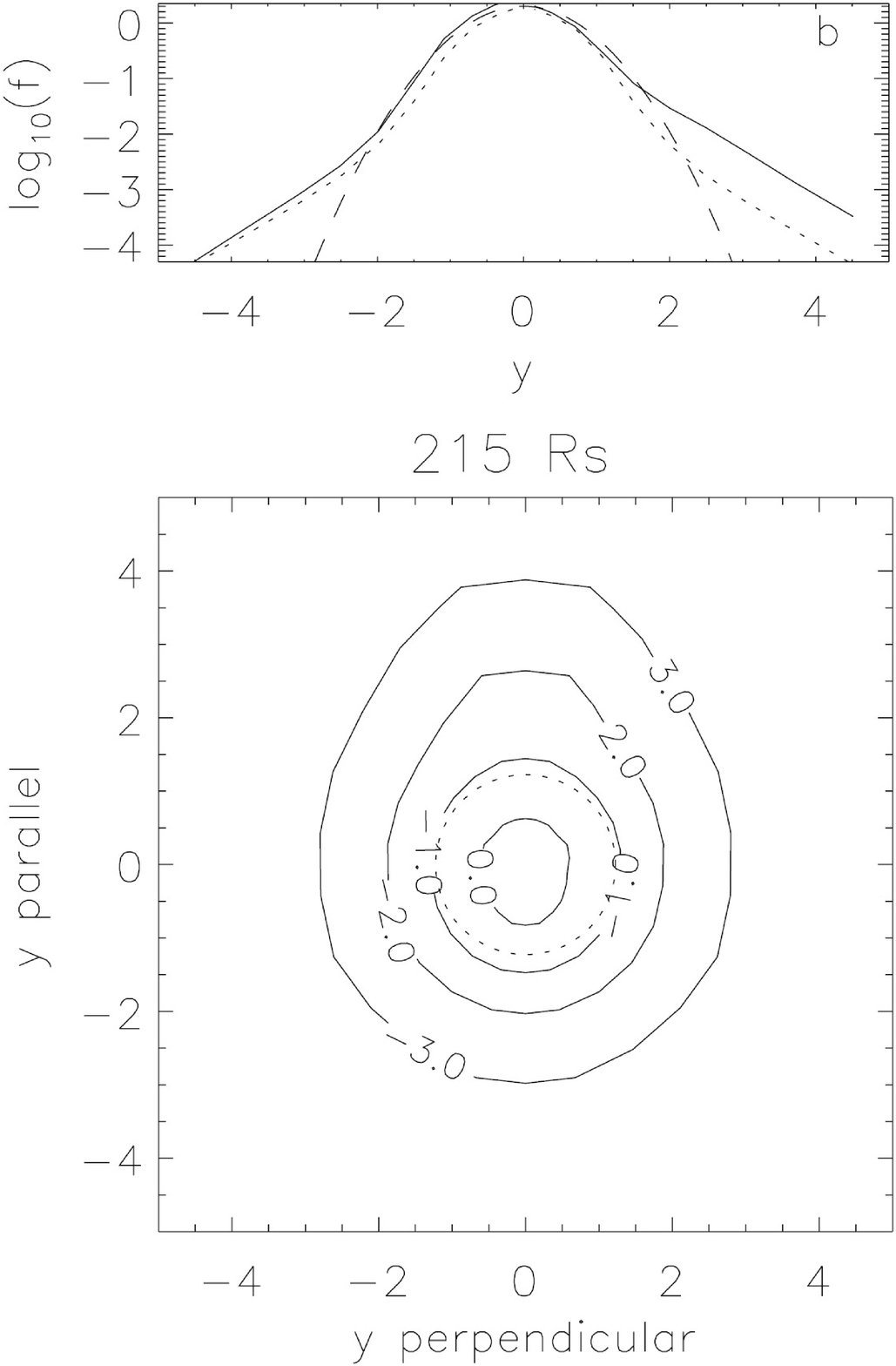}
%\label{fig:figureViv1}
\end{minipage}
\hspace{0.2cm}
\begin{minipage}{0.5\linewidth}
\centering
\includegraphics[scale=0.18]{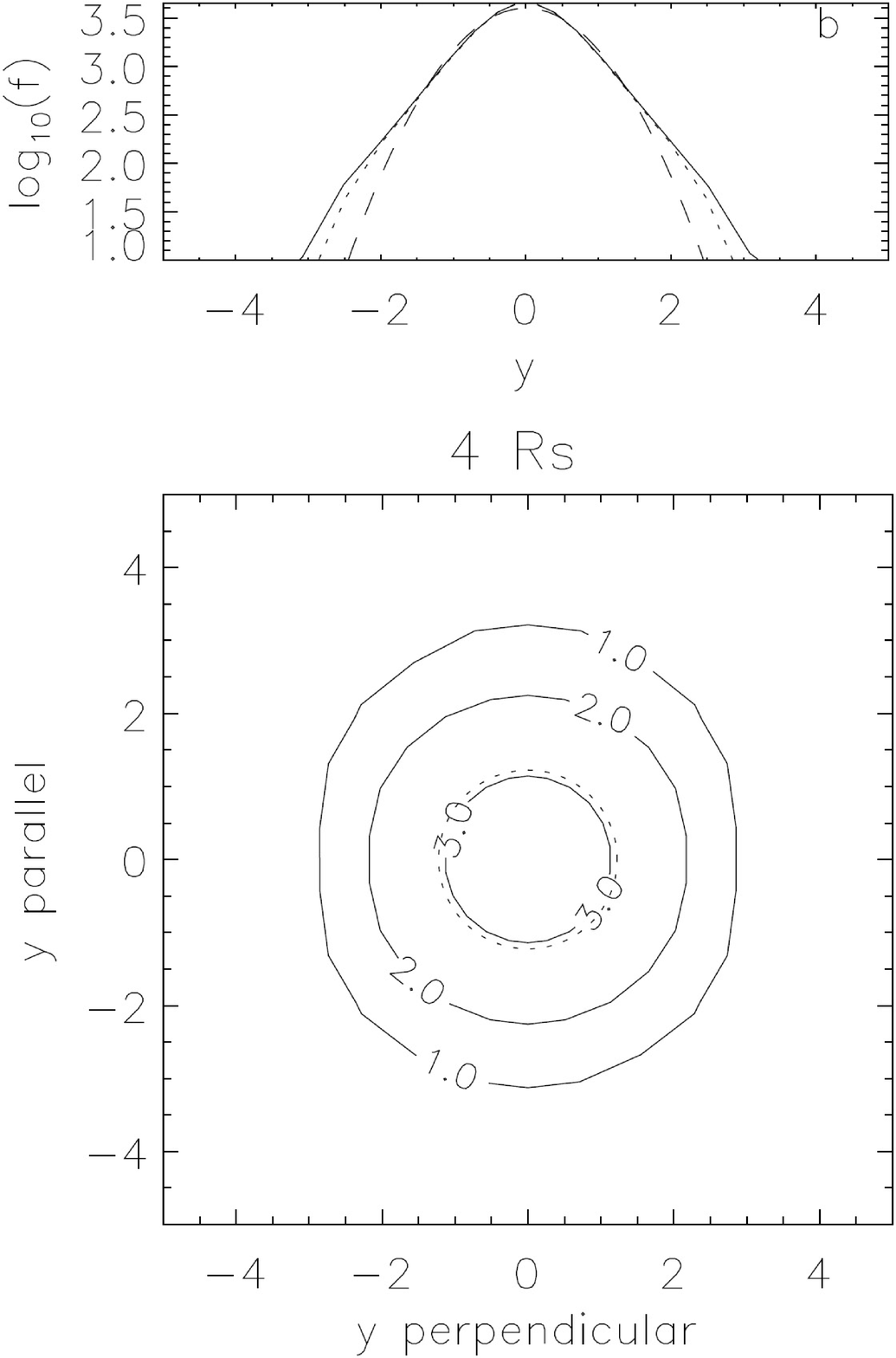}
%\label{fig:figureViv2}
\end{minipage}
\caption{The contour plots show the iso-density contours of
$f_e(y_\perp, y_\parallel)$, at two different radial distances:
$r=215~R_S$ (left panels), $r=4~R_S$ (right panels);
$y_{perpendicular}$, $y_{parallel}$ denote the
perpendicular respectively the parallel component of the velocity.
The dashed circles correspond to the thermal speed of the electrons. The
line-plots show the cross sections of $f_e(v_\perp, v_\parallel)$ in
the radial direction, which is assumed here to be the direction of the
interplanetary magnetic field; the dashed lines in these two panels
show the corresponding cross-sections of the electron VDF in the
direction perpendicular to the magnetic field;
figure from \citet{Pierrard+Maksimovic+Lemaire} \label{FigPierrard2001}}
\end{figure}

From their study based on the Fokker-Planck equation
\citet{Pierrard+Maksimovic+Lemaire} infer that, in order to match the halo and
strahl electron VDFs observed at 1~AU, a suprathermal tail should be
present in the VDF of the electrons deep into the solar corona.  In
other words, the features characterizing the suprathermal
electrons observed at 1~AU cannot not be generated within the
interplanetary medium solely by Coulomb collisions on the way between
the corona and 1~AU; according to their results an origin deep in the
corona has to be postulated. Alternatively, kinetic instabilities or
wave-particle interactions may be responsible for producing specific
features  in high speed SW streams.

Models of solar wind VDFs based on coupled Fokker-Planck equations for
electrons and protons remain to be developed in the future.  
Using coupled and time dependent Fokker-Planck
equations for solar wind electrons and protons, is a widely open field
of investigation: an unexploited avenue for scientists
interested in advancing basic plasma physics for applications to space
weather issues.
%Motivation for such basic kinetic theory and kinetic modeling
%studies needs revival; it is of basic importance in kinetic space
%plasma physics and current space weather predictions.  Let us hope
%that a few skilled scientists and engineers will be concerned by these
%challenging problems which have been neglected during the last years.
%The computing power to address this problem and solve the coupled
%kinetic Fokker-Planck equations is now available, but what is mostly
%missing is the appropriate stimulation to undertake such in-depth
%studies.

\section{Fluid models of the Solar Wind}
\label{sec:fluid-models}

Instead of considering direct solutions to the fundamental Boltzmann equation
(with varying degrees of simplification), to calculate the velocity
distribution $f(\vec r, \vec v,t)$ and its moments, one can consider models that
``only'' provide averages of the VDF, such as particle density,
mean flow velocity and temperature (or mean kinetic energy).  This is
obviously a much more ``coarse grained'' picture than provided by the
full VDF\@.  One may therefore be misled to think that fluid models
are inferior to kinetic models.  This is in general not the case,
however.  Although the Boltzmann equation (\ref{eq:boltzmann})
%\begin{equation}
%  \label{eq:boltzmann}
%  \pderiv{f}{t} + \vec v \cdot \nabla_r f + \frac{\vec F}{m} \cdot
%  \nabla_v f = \frac{\delta f}{\delta t}_\mathrm{coll},
%\end{equation}
looks deceptively simple, obtaining actual numerical solutions for a
realistic physical system can be a formidable task.
Even in the
simplest, spherically symmetric solar wind outflow, at least two
velocity components must be included, and obtaining a steady state
solution entails solving a partial differential equation in three
dimensions (one space, two velocity space).  With the full collision
term, (\ref{eq:boltzmann_J}) or (\ref{eq:Fokker_Planck_Rosenbluth}), 
included, it even becomes a nonlinear integro-differential
equation.  As we have seen in the previous section, it can only be
solved numerically (or analytically) in very simple situations, such
as steady state, one-dimensional, collisionless flow.

\subsection{Advantages and limitations of fluid models}

Fluid models, by providing a more coarse grained picture of the
plasma, can afford to do much more:
\begin{itemize}
\item Collisions can be included (although the terms can become very
  complicated and approximations have to be made, particularly for
  higher-order fluid models).
\item Processes such as ionization and recombination may easily be
  included.
\item Many particles species, and their interactions, may be
  included simultaneously.
\item Interaction of matter with radiation may be accounted for.
\item The fluid equations may be coupled with Maxwell's equations,
  creating e.g.\ magnetohydrodynamic (MHD) equations, and thus
  accounting for the interaction between the plasma and the electric
  and magnetic fields.
\item A very large range of spatial scales may be included
  simultaneously, ranging from the solar upper atmosphere and to the
  heliopause, if need be.
\item One is not restricted to stationary one-dimensional flow, and
  with today's computational resources obtaining time-dependent
  solutions for three-dimensional outflow, e.g., to model coronal mass
  ejection, is feasible.
\end{itemize}
Because fluid models allow us to include more processes and spatial
scales, they are well suited as ``global'' models of the solar wind,
while kinetic models are more suited to specific processes that cannot
easily or reliably be described by fluid models.  For instance, the
actual evolution of the VDF in a collisionless plasma requires solving
the Boltzmann (or more correctly, Vlasov) equation.  As we shall see,
fluid models have their limitations and uncertainties, which can only
be overcome in a kinetic description.

The list above also illustrates that the particle transport
equation(s) (either Boltzmann or fluid equations) is but one ingredient
in a model describing the solar wind system.  Including these other
processes requires that the transport equations themselves must be
kept as simple as possible.  That also applies to the fluid models.
Depending on the assumptions made when deriving them, they may quickly
attain a complexity rivalling that of the Boltzmann equation,
particularly so-called higher-order fluid models.  What is potentially
gained by their more detailed description of the plasma may quickly be
lost in the difficulties obtaining numerical solutions to them, or
because we no longer can afford to include some of the other processes
listed above.
Fluid equations have an additional advantage over kinetic models: We
have many decades of experience developing algorithms for solving them
numerically.  In fact, the vast field of computational fluid dynamics
(CFD) is devoted to this.

Fluid equations describe the evolution of the  \textit{moments} of the VDF,
such as the density, (mean) flow velocity, and temperature.  As
discussed in Sect. \ref{Transport_eqs}, these
equations may be formally derived by multiplying (\ref{eq:boltzmann})
by some velocity moment $v_i v_j\ldots$ and integrate over $\vec v$.
However, because of the $\vec v \cdot \nabla_r f$ term in
(\ref{eq:boltzmann}), the equation for a moment of order $N$ (in
velocities) will contain a moment of order $N+1$.  Hence this
procedure leads to an infinite series of coupled partial differential
equations, which has to be \textit{closed} by expressing a higher-order
term (moment) as some analytic function of lower-order terms.  In a
collision-dominated gas the closure assumption need not be
problematic; for instance, one may safely assume that the heat
flux (a higher-order moment) is proportional to the temperature
gradient (a lower-order moment).  However, in collisionless gases,
where the departure from a Maxwellian VDF may be large, it is not
obvious how the set of equations should be closed.

A second problem with fluid equations is how to evaluate the collision
term.  Because $\delta f/\delta t_\mathrm{coll}$ is an integral over the
product of the VDFs of two colliding species, it will necessarily
involve higher order moments.  Moreover, except for the case of
Maxwell molecule interactions (in which case the collisional cross
section is inversely proportional to the relative velocity of the
colliding particles, $\sigma \propto 1/|\vec v - \vec v'|$), the
collision term cannot be evaluated without making explicit assumptions
about the shape of the VDFs \citep{Burgers1969}.  Since Maxwell
molecule interactions are not relevant for a fully ionized plasma (but
rather for collisions between ions and neutrals), this means that
solar wind fluid equations cannot be developed without assuming an
explicit, analytic form for the VDF\@.  This analytic form will also
provide the closure for the moment equations.  Except for the simplest
possible approximation, the so-called 5-moment approximation, which is
not adequate for the corona and solar wind as we shall see, the
collision terms can become unwieldy and further approximations have to
be made in order to obtain analytical expressions.  Typically one has
to assume that the flow speed difference between species is much
smaller than thermal speeds, $|\vec u_s .- \vec u_t| \ll
\sqrt{kT_{s(t)}/m_{s(t)}}$, which limits the applicability to the
transition region and corona where collisions are sufficiently strong
to prevent large flow speed differences.  Finally, particularly for
Coulomb collisions the collision terms will be very sensitive to the
shape of the VDF\@.  Without a careful choice for the VDF, large
errors may result, e.g. in the forces acting between particles and in
the heat flux coefficient, even in the lower corona where particles
are subsonic.

Although fluid models have their advantages over kinetic models, and
for computational reasons we often have no choice but using a fluid
description, the above discussion shows that fluid models also have
their shortcomings.  They depend strongly on the 
assumptions made on the VDF and the closure relationship.
One must carefully select the ``right'' model for
the problem one wants to investigate.

The presentation below must necessarily be short, and will barely do
justice to the topic.  For the reader searching for a thorough
presentation of the topic, the classic textbook of \citet{Chapman1970}
provides a detailed account of diffusion, thermal conduction, and
viscosity in non-uniform gases.  The monograph by \citet{Burgers1969}
provides a thorough description of collisional effects, and presents
the techniques necessary to evaluate the collision integrals
analytically.  More recently, \citet{Gombosi1994} provides an
introduction to the kinetic theory of gases, including so-called
generalized transport equations.  In addition to the extensive
treatment of \citet{Grad1958}, discussed in the Appendix,
\citet{Schunk1977} offers a more ``practical'' and concise
presentation of transport equations for use in aeronomy, at various
levels of complexity, while \citet{Barakat+Schunk1982} review
transport equations for anisotropic space plasmas.

\subsection{Properties of the solar wind and upper solar atmosphere
  {relevant for fluid modeling}}
\label{sec:prop-solar-wind}

%Before we can proceed, we need to know something about the system we
%wish to describe.  
%The solar wind has it source, of course, in the
%upper solar atmosphere.  
Many of the basic properties of the solar
wind are set in the subsonic region of the flow.  Models suggest that
the elemental abundances of the solar wind may even be set in the
upper chromosphere, where the gas becomes ionized
\citep[e.g.][]{Henoux1998}.  Observations indicate that the wind is
rapidly accelerated within a few solar radii from the Sun
\citep{Habbal+etal1995,Grall1996,Antonucci+Dodero+Giordano2000},
implying that most of the energy flux carried by the solar wind has
been deposited in the corona.

Moreover, the solar wind mass flux is not only set in the subsonic
region of the flow, but will largely be determined by the energy
balance in the solar transition region (the interface between the
chromosphere and corona). The mass flux depends on the coronal
pressure, which is approximately equal to the transition region
pressure.  In the absence of flow the transition region pressure is in
turn directly proportional to the downward heat flux density from the
corona \citep{Landini+Fossi1975}.  Hence the coronal density, and thus
the solar wind mass flux, depends sensitively on the energy balance of
the transition region \citep{Lie-Svendsen+Leer+Hansteen2001}.

This implies that if obtaining, e.g., the elemental composition or the
mass flux of the solar wind is an important goal of the model, it must extend all the
way down to the upper chromosphere.  In such SW models it is of course
essential that Coulomb collisions are described accurately, as e.g.,
heat is transported in a plasma dominated by collisions.  Since heat
conduction is essential in the transition region and corona, one must
also choose a fluid description that allows for heat
conduction (but as we shall see later, the simplest possible fluid description
does not).

The Coulomb interaction 
%is special because of its long
%range and because it is dominated by small-angle collisions (this is
%the reason for which the Fokker-Planck collision term 
%(\ref{eq:Fokker_Planck_Rosenbluth}) is used).  In addition, it
differs from the other collision types (neutral-neutral and
neutral-ion collisions) by being strongly energy dependent.  In fact,
the cross section for Coulomb collisions is proportional to the
velocity difference of the colliding particles to the inverse fourth
power, Eq. (\ref{Coulomb_crosssec})
% {
%\begin{equation}
%  \label{eq:coulombcross}
%  \sigma \propto \frac{1}{|\vec g|^4} = \frac{1}{|\vec v - \vec v_1|^4}.
%\end{equation}
%(see also eq. \ref{eq:boltzmann_J}). }
As a consequence, the collision frequency, assuming Maxwellian VDFs,
is strongly temperature dependent, $\nu \propto T^{-3/2}$. 
The temperature increases,
particularly for protons and heavier ions, 
from the transition region into the corona, while the density
decreases with altitude.  The combined effect is a rapidly decreasing collision
frequency, resulting in a steep transition from
collision-dominated to collisionless flow for protons and ions,
somewhere in the corona.

Another effect of the velocity dependence of the Coulomb cross section
is the strong temperature dependence of heat conduction in a fully
ionized, collision-dominated gas, where the electron heat flux is
\citep{Spitzer+Harm1953,Braginskii1965}
\begin{equation}
  \label{eq:qe}
  \vec q_e = - \frac{\kappa_e'}{\ln\Lambda} T^{5/2} \nabla T,
\end{equation}
with $\kappa_e' = 1.84 \times
10^{-10}~\mbox{W~m}^{-1}~\mbox{K}^{-7/2}$ 
%from                                %Killie et al.
and $\Lambda$ the plasma parameter specified in section \ref{Fokker-Planck} 
accounting for the Debye shielding of
the electric field.  The $T^{5/2}$-dependence of $\vec q_e$ means that
heat conduction is virtually absent in a cold plasma, while it becomes
``suddenly'' very efficient at coronal temperatures of the order of
$10^6$~K\@.  Heat conduction thus acts like a thermostat for the
electrons: little heating is required to attain temperatures of order
$10^6$~K, while it is difficult to attain temperatures much higher
than this (except, perhaps, under extreme conditions such as in a
solar flare).  This is also in agreement with, e.g., observations of
coronal holes, the source regions of the fast solar wind, indicating
electron temperatures slightly below $1 \times 10^6$~K
\citep{Wilhelm98}.  The efficient heat conduction at these
temperatures also implies that electrons will not become collisionless
in the corona, and as the temperature decreases outwards in the solar
wind they may even remain collisional far into the solar wind, despite
the decreasing density.  Note also from (\ref{eq:qe}) that heat
conduction is almost independent of the coronal density (with only a
very weak density dependence through the $\ln\Lambda$ term).

Collisions quickly cease to be important for protons and ions at
larger distance from the Sun, where the solar wind is instead influenced
by external forces (gravity and electric and magnetic forces),
pressure forces within the plasma, and wave-particle interactions.
Here a fluid model  needs to ensure that the effects of collisions are sufficiently
small to be unimportant or small first order corrections.  
However, the lack of collisions means that
the VDF may, and will, become highly anisotropic 
(and possibly non-Maxwellian).  In fact,
observations of coronal holes from the SOHO satellite show that
protons and oxygen ions are highly anisotropic even in the corona,
with the temperature along the line of sight ($T_\perp$) higher
than the temperature perpendicular to the line of sight
($T_\parallel$) \citep{Cranmer+etal1999}.  

In situ observations by the Helios spacecraft show that protons 
are highly non-Maxwellian and anisotropic with
$T_\perp > T_\parallel$ also at 0.3~AU in the fast solar wind
\citep{Marsch+etal1982}. $T_\perp$ and $T_\parallel$ refer to
directions relative to the SW magnetic field.
Since adiabatic expansion of the wind should have the opposite effect
at large distances,
leading to $T_\perp \ll T_\parallel$ at $r$, because of magnetic moment
conservation, this observation shows that some form of wave-particle
interaction could have affected the proton VDF\@.  Electrons on the
other hand remain close to Maxwellian even at such large distances
from the Sun, with the core of the VDF being quite isotropic but with
an antisunward high-velocity ``strahl'' in high-speed solar wind
streams \citep{Pilipp+etal1987c}.  To capture such 
proton anisotropies in a fluid
description, one must consider so-called gyrotropic equations, in
which the VDF is expanded about a bi-Maxwellian with different
temperatures parallel and perpendicular to the magnetic field.

Below we shall separate the fluid descriptions into models in which
the VDF is expanded about a Maxwellian, and models in which it is not
(including expansion about a bi-Maxwellian).  Although there is
definitely overlap between the two classes of models, one may regard
the former as more suited to describe the transition region, corona,
and inner solar wind, while the latter group is mostly
suited to the collisionless, supersonic solar wind where VDFs
deviate strongly from isotropic Maxwellians.
We shall only include elastic Coulomb
collisions, although models extending into the low corona and
transition region should include ionisation, recombination, and
charge exchange reactions, as well as neutral-neutral and
neutral-ion collisions. This is certainly the case when minor elements are
taken into account in the model. 
Since the VDF should be close to a Maxwellian where and when
these processes are important, the corresponding collision terms can
be derived fairly straightforwardly if required.

\subsection{Models for the corona and inner solar wind}
\label{sec:models-corona-inner}

When collisions, which restore a Maxwellian VDF in a uniform gas, are important,
it is generally considered as reasonable to assume a VDF for species $s$ of the form
\begin{equation}
  \label{eq:fiso}
  f_s(\vec r, \vec v, t) = f_{s}^{(0)}(\vec r, v) \left( 1 + \phi_s(\vec r, \vec
    v, t) \right)
\end{equation}
where $f_{s}^{(0)}$ is a drifting Maxwellian,
\begin{equation}
  \label{eq:fmaxw}
  f_{s}^{(0)} \equiv n_s \left( \frac{m_s}{2\pi k T_s} \right)^{3/2}
  \exp\left( - \frac{m_s c_s^2}{2kT_s} \right).
\end{equation}
Here $m_s$ is the mass of a particle and $k$ is the Boltzmann's constant,
while $\vec c_s \equiv \vec v - \vec u_s(\vec r, t)$ is the ``peculiar''
velocity as measured in the reference frame moving at the species
drift velocity $\vec u_s$.

Irrespective of the form of $\phi_s$, we shall always insist that
$n_s$, $\vec u_s$ and $T_s$ are the density, flow velocity and
temperature of species $s$ (moment definitions are summarized in
Appendix A):
\begin{eqnarray}
  n_s &=& \int \int \int f_s\label{eq:nmom}\,  d\vec v \\
  \vec u_s &=& \frac{1}{n_s} \int \int \int \, \vec v f_s \,  d\vec v \, \label{eq:umom}\\
  T_s &=& \frac{m_s}{3kn_s} \int \int \int \, (\vec v - \vec u_s)^2
  f_s\,  d\vec v\,. \label{eq:tmom}
\end{eqnarray}

The correction term $\phi_s$ accounts for the departure from a
Maxwellian, and will generally be some polynomial of order $N$ in
$\vec v$ (or more correctly, in $\vec c \equiv \vec v - \vec u_s(\vec
r, t)$).  However, because it
is postulated that $\vec u_s$ should be the mean flow velocity of the gas, higher
order terms will have to be included to ensure this.  Moreover,
to be able to evaluate the collision terms analytically one generally
has to assume that $|\phi_s| \ll 1$, so that terms proportional to
$\phi_s \phi_t$ may be neglected in the collision integrals.  This
requires that flow velocity differences
are small in comparison with thermal velocities.

\subsubsection{The 5-moment approximation, $\phi_s=0$}
\label{sec:5-moment-appr}
%\comment{ols: Eqs.~(\ref{eq:cont})--~(\ref{eq:emom}) are not the
%  5-moment equations, but rather general equations valid for any
%  closure assumption.  So I have changed back to the original,
%  5-moment, equations here, as these are not identical to those in
%  Sect.\ 1.}

The simplest choice for the VDF that one assumes first is $\phi_s=0$.  
Multiplying~(\ref{eq:boltzmann}) by 1, $\vec v$, and
$v^2$, and using eqs.~(\ref{eq:nmom})--(\ref{eq:tmom}), one obtains
the moment equations (\ref{eq:cont}) and (using the notation of
\citet{Schunk1977}),
\begin{eqnarray}
 m_s n_s \frac{D_s \vec u_s}{Dt} + \nabla p_s - m_s n_s \vec g - e_s
 n_s ( \vec E + \vec u_s \times \vec B) &=& \frac{\delta \vec
   M_s}{\delta t}\label{eq:mom5} \\
 \frac{D_s}{Dt} \left( \frac{3}{2} p_s \right) + \frac{5}{2} p_s
 \nabla \cdot \vec u_s &=& \frac{\delta E_s}{\delta t}.\label{eq:e5}
\end{eqnarray}
where $p_s=n_skT_s$ is the thermal pressure. When summed over the
number of component species, equations
(\ref{eq:mom5})-(\ref{eq:e5}) give the Euler equations 
(\ref{Euler:momentum})-(\ref{Euler:energy}), see
definitions of moments in Appendix \ref{definitions}.
% where we have introduced the convective derivative,
% \begin{equation}
%  \label{eq:convderiv}
%  \frac{D_s}{Dt} \equiv \pderiv{}{t} + \vec u_s \cdot \nabla ,
% \end{equation}
% $P_s = n_s k T_s$ is the pressure, $\vec g$ is the gravitational
% acceleration, $e_s$ is the electric charge, and $\vec E$ and $\vec B$ are
% electric and magnetic fields.  
When $\phi_s=0$ the collision terms can
be obtained analytically without making further approximations, and
the terms on the right hand side of (\ref{eq:mom5})--(\ref{eq:e5})
may be written. These expressions, given below,
are valid for arbitrary temperatures and flow velocities:
\begin{eqnarray}
  \label{eq:dmdt}
  \frac{\delta\vec M_s}{\delta t} &=&  m_s n_s \sum_t \nu_{st} (\vec
  u_t - \vec u_s) \Phi_{st} \\
  \frac{\delta E_s}{\delta t} &=& \sum_t \frac{m_s}{m_s+m_t} n_s \nu_{st}
  \left[ 3 k (T_t-T_s) \Psi_{st} + m_t (\vec u_s - \vec u_t)^2
    \Phi_{st} \right],\label{eq:dedt}
\end{eqnarray}
where the sums extend over the species with which the $s$-specie
collides; $\nu_{st}$ is the collision frequency 
between species $s$ and $t$, and $\Phi_{st}$ and $\Psi_{st}$
are correction factors.

For Coulomb collisions the collision frequency reads (in SI units):
\begin{equation}
  \label{eq:nu}
  \nu_{st} = \frac{1}{3} n_t \frac{m_t}{m_s+m_t} \left( \frac{2\pi k
      T_{st}}{\mu_{st}} \right)^{-3/2} \frac{e_s^2
    e_t^2}{\epsilon_0^2\mu_{st}^2} \ln\Lambda
\end{equation}
The reduced mass and temperature are defined by:
\begin{eqnarray}
  \label{eq:mu}
  \mu_{st} &\equiv& \frac{m_s m_t}{m_s+m_t} \\
  T_{st}   &\equiv& \frac{m_tT_s+m_sT_t}{m_s+m_t}
\end{eqnarray}
%and $\epsilon_0$ is the permitivity of vacuum.  
For Coulomb collisions the velocity dependent correction factors read
\begin{eqnarray}
  \label{eq:Phi}
  \Phi_{st} &=& \frac{3}{2\epsilon_{st}^2} \left( \frac{\sqrt{\pi}}{2}
    \frac{\mbox{erf}(\epsilon_{st})}{\epsilon_{st}} -
    \exp(-\epsilon_{st}^2) \right)\\
  \Psi_{st} &=& \exp(-\epsilon_{st}^2),\label{eq:Psi}
\end{eqnarray}
where $\mbox{erf}(x)$ is the error function and
\begin{equation}
  \label{eq:epsst}
  \epsilon_{st} = \frac{|\vec u_s - \vec u_t|}{\sqrt{2kT_{st}/\mu_{st}}}.
\end{equation}

To a good approximation the solar wind is a fully ionized
electron-proton plasma.  Assuming quasi neutrality and no current, and
neglecting all terms proportional to the electron mass $m_e$, the
electron and proton momentum equations (\ref{eq:mom5}) may be added to
read simply
\begin{equation}
  \label{eq:momep}
  m_p n \frac{D\vec u}{Dt} + \nabla(nk(T_e+T_p)) - m_p n \vec g = 0,
\end{equation}
where now $n_e=n_p=n$, $\vec u_e=\vec u_p=\vec u$ and the right-hand
side is zero from Newton's 3rd law (there may still be collisions).

The most serious shortcomings of the 5-moment approximation are (i) 
that the temperatures have to be prescribed according to a
polytropic law and (ii) the lack of heat
conduction.  As emphasized in Sect.~\ref{sec:prop-solar-wind}, heat
conduction, particularly for the electrons, is essential in the corona,
as the high temperature there makes it extremely efficient.  Mainly
for that reason, this description is not acceptable for the corona and
solar wind.  It has other shortcomings as well; for instance, the
frictional force between charged particles is too large (for a given
$\vec u_s-\vec u_t$), implying a too large electric resistivity.  For
an electron-proton solar wind plasma these terms vanish in any case, as
assumed in (\ref{eq:momep}), in cases where electric currents or
other particles (such as helium) are included, the errors will be
large.

%\comment{ols: I really mean $\alpha$-particles below, ionized helium,
%  not ``$s$''-particles!}

The 5-moment approximation is  attractive because it is
simple, and the collision terms are exact
and valid for all flow conditions.  When one allows for $\phi_s \neq 0$, to
account for heat conduction the collision terms will only be valid
for flow speeds much smaller than the thermal speed, $\epsilon_{st} \ll
1$.  When the solar wind becomes supersonic this condition may not be
met.  

In solar wind models with $\alpha$-particles (or minor ions)
included, spurious coupling between
$\alpha$-particles and protons may occur in the supersonic solar wind.
Because the temperature tends to decrease as the wind is accelerated (even with
heat conduction), and $\nu_{st} \propto T^{-3/2}$ from (\ref{eq:nu}),
the frictional coupling between protons and $\alpha$-particles may
become large if the flow speed difference is large.  However,  the
Coulomb cross section (\ref{Coulomb_crosssec}) shows  that the
coupling should be small if the velocity difference is large, which is
handled by the $\Phi_{st}$ term in the 5-moment description.  To
overcome this problem,  the collision terms are ``repaired'' by
introducing ad hoc velocity corrections.  An obvious solution is to
apply the 5-moment velocity corrections (\ref{eq:Phi}) and
(\ref{eq:Psi}), also to higher-order fluid models where they are not
formally correct.  Similarly, the second term on the right-hand side
of (\ref{eq:dedt}) represents frictional heating, which can also be
important when, e.g., $\alpha$-particles stream faster than protons in
the outer corona.  Again, this effect will be absent in the
following higher-order approximations with $\phi_s \neq 0$ because
the computation of the collision terms is difficult when flow speed differences
become large.  This illustrates that a procedure that is formally
(mathematically) correct, is not necessarily physically correct, and
that the seemingly more accurate higher-order approximations that we
shall discuss next may be, in reality, no more accurate than even the
``crude'' 5-moment approximation.

\subsubsection{The 8-moment approximation}
\label{sec:8-moment}
To account for heat conduction, one must have a skewness in the
VDF, so that $\phi_s \neq 0$.  The simplest choice is a function
linear in $\vec c_s$, $\phi_s \propto \vec a \cdot \vec
c_s$. {Recall that $\vec c_s$ is the  velocity
of particles $s$ measured in the frame moving with the average
velocity, $\vec c_s=\vec v - \vec u_s$.} Since
$\phi_s$ must be a scalar function, one takes the scalar product
with some vector $\vec a$.  This may be regarded as the first order
Taylor expansion, in powers of $c_{si}$, of the ``true'' $\phi_s$.
However, in order to enforce that the average thermal velocity
is equal to zero, i.e.   $\int\, \vec c_s
f_{s0}(1+\phi_s)  d\vec c_s\, = \vec 0$, so that $\vec u_s$ is the mean flow
velocity, one must add a third order term as well.
The 8-moment assumption for $\phi_s$ may then be written
\begin{equation}
  \label{eq:8momvdf}
  \phi_s = - \frac{m_s}{kT_sp_s} \left( 1 - \frac{m_s c_s^2}{5kT_s}
  \right) \vec q_s \cdot \vec c_s,
\end{equation}
where the coefficient in front is chosen such that $\vec q_s$ is the
heat flux density,
\begin{equation}
  \label{eq:qdef}
  \vec q_s \equiv \frac{1}{2} m_s n_s \int\, c_s^2 \vec c_s \,  d\vec c_s.
\end{equation}

With this choice, the left-hand side of the momentum
equation~(\ref{eq:mom5}) remains unchanged, while the heat flux
divergence is added as a source or sink of thermal energy to
(\ref{eq:e5}):
\begin{equation}
  \frac{D_s}{Dt} \left( \frac{3}{2} p_s \right) + \frac{5}{2} p_s
  \nabla \cdot \vec u_s + \nabla \cdot \vec q_s = \frac{\delta
    E_s}{\delta t}.\label{eq:e8}
\end{equation}
Taking the appropriate moment of (\ref{eq:boltzmann}), the equation
for $\vec q_s$ becomes
\begin{equation}
  \label{eq:q}
  \frac{D_s \vec q_s}{Dt} + \frac{7}{5} (\vec q_s \cdot \nabla) \vec
  u_s + \frac{7}{5} \vec q_s (\nabla \cdot \vec u_s) + \frac{2}{5}
  (\nabla \vec u_s) \cdot \vec q_s + \frac{5}{2} \frac{kp_s}{m_s}
  \nabla T_s - \frac{e_s}{m_s} \vec q_s \times \vec B =
  \frac{\delta\vec q_s'}{\delta t},
\end{equation}
where $\nabla \vec u_s$ denotes a tensor, and the right-hand side
again denotes the contribution from the collision term.

The corresponding collision terms have only been evaluated when flow
velocity differences are small compared with species thermal speeds,
and are given by \citet{Burgers1969}.  If furthermore, it is assumed that
temperature differences are small, the collision terms 
at the right hand-side of the  equations of momentum and energy
conservation (\ref{eq:mom5}) and (\ref{eq:e8})
%transfer collision
%term for Coulomb collisions and the energy collision term 
become
\begin{eqnarray}
  \label{eq:dmdt8}
  \frac{\delta \vec M_s}{\delta t} &=& m_s n_s \sum_t \nu_{st} (\vec
  u_t - \vec u_s) + \frac{3}{5} \sum_t \nu_{st}
  \frac{\mu_{st}}{kT_{st}} \left( \vec q_s - \frac{m_s n_s}{m_t n_t}
    \vec q_t \right)\\
  \frac{\delta E_s}{\delta t} &=& m_s n_s \sum_t
  \frac{\nu_{st}}{m_s+m_t} 3 k (T_t-T_s)\label{eq:dedt8}
\end{eqnarray}
With the same approximation, $\delta \vec q_s'/\delta t$ is given in a
concise form by \citet{Schunk1977}.  As mentioned at the
end of Sect. \ref{sec:5-moment-appr} the frictional heating is
absent in (\ref{eq:dedt8}), because of the approximations that had to
be made when evaluating the collision terms.

By allowing for heat conduction in the formulation, not only
the description of the energy balance in the corona and solar
wind is improved, but also the description of forces, by allowing for
thermal diffusion.  Moreover, by having a separate equation
(\ref{eq:q}) for the heat flux density, this set of equations display
a self-consistent transition to collisionless flow: Classical heat
conduction ($\vec q_s \propto -\nabla T_s$) is only retained in the
collision-dominated limit of small flow speeds when the term
proportional to $\nabla T_s$ dominates the left-hand side of
(\ref{eq:q}), and there is thus no need for some ad hoc transition
from classical to non-classical heat conduction in this equation set.

Although heat conduction and thermal diffusion are accounted for in
this description, they are not accounted for well.  In the
collision-dominated limit with a small temperature gradient, the
electron heat conduction coefficient $\kappa_e'$ (see eq.~(\ref{eq:qe}))
obtained from this approximation is less than half of the correct
value obtained, e.g., by \citet{Spitzer+Harm1953}, based on a
numerical solution of  the Boltzmann equation with Fokker-Planck collision
term.  Thermal diffusion is off by a similarly large factor.  The
reason for the discrepancy lies in the choice of a first order (in
$\vec c_s$) correction term in (\ref{eq:8momvdf}).  In a
collision-dominated region with a small temperature gradient, the actual
electron VDF for small values of $c_e$ is rather proportional to
$c_e^3$ or $c_e^4$ \citep{Spitzer+Harm1953}.  Since the Coulomb cross
section is so sensitive to the relative velocity
(eq.~(\ref{Coulomb_crosssec})), this inaccuracy leads to large errors in the
collision integrals.  To improve the description of Coulomb collisions
it is therefore necessary to include higher-order terms in $\vec c_s$
in $\phi_s$.  Note also that the term of order $c_s^3$ in
(\ref{eq:8momvdf}) is not really a third order term as it does not
have an independent coefficient; recall that it was merely introduced
to ensure that $\vec u_s$ remained the mean flow velocity.
%\comment{MME: are the last sentences, regarding the necessity to 
%develop $f_s$ in terms of higher powers of $c_s$, a 
%justification/introduction for the 13-moments approximation? 
%I do not immediately see the relevance as in (\ref{eq:13momvdf})
%$c_ic_j$ does not have an independent coefficient; or I missed the
%argument above ? In what sense the coefficients need to be
%independent ?}

The second sum in (\ref{eq:dmdt8}) accounts for thermal
diffusion, which is important for minor ions (and He ions) in the
transition region and inner corona \citep{Nakada1969}, but also in
coronal loops. In regions with steep temperature gradients, heavy,
ionized particles will collide more often with ``cold'' electrons and
protons streaming upwards than with the (relatively) hotter particles
streaming down the temperature gradient, the net effect being that the
heavy particles feel a strong upward force which is even stronger than
gravity in the transition region.  In a pure electron-proton plasma
with no currents the thermal diffusion term merely modifies the
electric field, but has no impact on the electron-proton flow.

\subsubsection{13-moment approximation}
\label{sec:13-moment}
The next level of approximation is to add a second order (in $\vec
c_s$) term to $\phi_s$ \citep{Grad1958},
\begin{equation}
  \label{eq:13momvdf}
  \phi_s = - \frac{m_s}{kT_sP_s} \left( 1 - \frac{m_s c_s^2}{5kT_s}
  \right) \vec q_s \cdot \vec c_s + \frac{\tau_{sij}}{2kT_sP_s} c_i c_j,
\end{equation}
where $\tau_{sij}$ are components of the stress tensor
$\tensor{\tau_s}$, given as
\begin{equation}
  \label{eq:tau}
  \tau_{sij} \equiv m_s \int\left( c_{si} c_{sj} -
    \frac{1}{3} c_s^2 \delta_{ij}\right) f_s \,  d\vec c_s ,
\end{equation}
and a sum over indices $i$ and $j$ is implied in (\ref{eq:13momvdf}).
Because $\tau_{sij}=\tau_{sji}$ and $\tau_{sii}=0$ it has 5
independent components, hence this is a 13-moment approximation.
In addition to the pressure gradient of the
momentum equation (\ref{eq:mom5}), we now must add the divergence of
$\tensor{\tau_s}$,
\begin{equation}
    m_s n_s \frac{D_s \vec u_s}{Dt} + \nabla p_s + \nabla \cdot
    \tensor{\tau_s} - m_s n_s \vec g - e_s n_s ( \vec E + \vec u_s
    \times \vec B) = \frac{\delta \vec M_s}{\delta t},\label{eq:mom13}
\end{equation}
where $\nabla \cdot \tensor{\tau_s} \equiv \vec e_k \cdot
\partial(\tau_{sij} \vec e_i \vec e_j)/\partial x_k$.  The equation
for the thermal energy reads
\begin{equation}
  \label{eq:e13}
  \frac{D_s}{Dt} \left( \frac{3}{2} p_s \right) + \frac{5}{2} p_s
  \nabla \cdot \vec u_s + \nabla \cdot \vec q_s + \tau_{sij}
  \pderiv{u_{sj}}{x_i} = \frac{\delta E_s}{\delta t}.
\end{equation}
Again collision terms are hard to evaluate without making severe
approximations (except for the non-relevant case of Maxwell molecule
interactions).  The momentum and energy collision terms in the limit
of small flow speed differences and small relative temperature
differences are identical to (\ref{eq:dmdt8}) and (\ref{eq:dedt8}) of the
8-moment description.  \citet{Zamlutti1998} has proposed approximate
collision terms for the 13-moment approximation also to be used for
large flow speed differences.  However, these collision terms have
only been obtained from the Boltzmann collision integral, and are not
directly applicable to charged particle interactions which are
dominated by small-angle collisions (described by the Fokker-Planck
collision term).

The equations for $\tensor{\tau_s}$ and $\vec q_s$ are given, e.g., by
\citet{Schunk1977}.
By including the stress tensor, this formalism includes viscous
effects in addition to heat conduction, and in a collision-dominated
gas the Navier-Stokes equations may be retrieved
\citep[see][]{Gombosi1994}.

By including a higher-order component (the stress tensor), this
formalism should better describe a weakly collisional solar wind
plasma, in which departure from a Maxwellian  will be larger.
However, since the collision terms have only been evaluated in a limit
which requires a \emph{strongly} collisional plasma, it is far from
obvious that these equations will provide a better description of a
weakly collisional plasma.  And in the fully collisionless solar wind
departures from a Maxwellian may become large, so that there is no
guarantee that even a second (or higher) order approximation to
$\phi_s$ is any better than the first order approximation.  In that
case, it may be advantageous to switch to gyrotropic equations in
which the VDF is expanded about a bi-Maxwellian, which we shall
discuss in the next section.

Note also that the assumption (\ref{eq:13momvdf}) still does not
accurately describe the electron VDF in a collision-dominated plasma,
where, as we have discussed, $\phi_s$ should rather be of third or
fourth order in $\vec c_s$.  Hence it is expected that the errors in
the collision terms are still large; since $\delta \vec
M_s/\delta t$ remains unchanged from the 8-moment approximation,
thermal diffusion will still not be accurately described.
%\textcolor{green}{\comment{JFL: ? this is jeopardizing the classical
%    point of view !}}
%
%\comment{MME: maybe we should stress here that long range, Coulomb
%  collisions are envisaged.}

To improve the description of collisions, one should at least include
third order terms in $\phi_s$.  This results in the so-called
20-moment approximation \citep[see][]{Gombosi1994}, in which 20
coupled partial differential equations have to be solved.  Needless to
say, these are so complex that they seem to be of little practical
use.

The higher-order moments, such as the stress tensor, are usually of
little physical importance in themselves for the solar wind system.
Mostly we are concerned with the basic properties, such as the mass
flux, flow speed, and energy flux.  The higher-order moments are only
of interest as they affect the low-order moments, either through the
closure assumption or because they modify the collision terms.  One
may therefore ask whether solving higher-order equations is
worthwhile, given the tremendous effort required to obtain a numerical
solution to them, keeping in mind that we still have to use highly
simplified collision terms.

\subsubsection{8-moment approximation with improved Coulomb collision terms}
\label{sec:8-moment-improved}

If describing collisions reasonably correctly is important, instead of
trying to solve the almost intractable 20-moment set of equations one
may instead make a simple assumption for $\phi_S$ which still captures
the essence of Coulomb collisions.  \citet{Killie+etal2004} suggested
that the 8-moment approximation (\ref{eq:8momvdf}) be replaced by an
8-moment approximation which is \emph{third} order in $\vec c_s$,
\begin{equation}
  \label{eq:8improvedvdf}
  \phi_s = - \frac{m_s c_s^2}{5k^2T_s^2P_s} \left( 1 - \frac{m_s c_s^2}{7kT_s}
  \right) \vec q_s \cdot \vec c_s,
\end{equation}
where the fifth order term and the coefficients have been chosen such
that $\vec u_s$ and $\vec q_s$ remain the flow speed and heat flux
density, respectively.  
%\textcolor{green}{$[$ again this is
%ad-hoc, i.e. arbitrary, but not justified by physical principles
%nor experimental evidence$]$} 
With this, still 8-moment, approximation, it
turns out that the left-hand sides of the transport equations are
identical to those of the original 8-moment approximation, given by
eqs.~(\ref{eq:cont}), (\ref{eq:mom5}), (\ref{eq:e8}), and
(\ref{eq:q}).  However, the collision terms $\delta\vec M_s/\delta t$
and $\delta \vec q_s'/\delta t$ will change.  Collision terms still
have to be evaluated in the ``semi-linear'' approximation of small flow
speed differences.  In the linear limit of small relative temperature
differences as well, one obtains
\begin{eqnarray}
  \label{eq:dmdtimproved}
  \frac{\delta\vec M_s}{\delta t} &=&  n_s m_s \sum_{t} \nu_{st}
  (\vec{u}_t - \vec{u}_s) + \sum_{t}\nu_{st} \frac{3}{5}
  \frac{\mu_{st}}{kT_{st}}
  \left[\vec{q}_s\left( 1 - \frac{5}{7} \frac{m_t}{m_s+m_t} \right)
  \right. \nonumber\\
  && \left. - \vec{q}_t \frac{m_s n_s}{m_t n_t} \left( 1 - \frac{5}{7}
  \frac{m_s}{m_s+m_t} \right) \right].
\end{eqnarray}
The modified terms in the second sum (compared with (\ref{eq:dmdt8}))
means that thermal diffusion is now reduced (when the heat flux
remains unchanged).  The heat flux collision term, $\delta \vec
q_s'/\delta t$, also changes substantially.

In the collision-dominated limit with small temperature gradients, the
new collision terms lead to a heat conduction coefficient $\kappa_e'$
that does not deviate by more than 20\% from the correct value
(see the expression (\ref{eq:qe}) of the Spitzer-H\"{a}rm
heat flux); note also that the original 8-moment assumption described 
in section \ref{sec:8-moment} lead to a more than factor 2 deviation.
The coefficients of thermal diffusion agree within approximately 30\%.  
%\textcolor{green}{\comment{JFL: what is the correct value 
%for $\kappa_e'$ ?}}
Hence this equation set should be well suited to the part of
the solar wind strongly influenced by collisions, while the equations
also contain a self-consistent transition to collisionless flow.  
In the collisionless regime the new equations are identical to the
original 8-moment equations.

%\textcolor{green}{\comment{JFL: is this correct? This is saying that
%    the classical transport equations are not valid in the 
%    collision-dominated region!}}

%\textcolor{green}{\comment{MME: I think it is saying that the 8-moment
%approach with collision terms evaluated \'{a} la Burgers provides
%less good results than the 8-moments solutions with improved collision
%terms.}}

This shows how it is possible to improve the model without greatly
increasing the complexity, by using a carefully chosen $\phi_s$.  As a
formal derivation it seems strange to start the Taylor expansion at
third order --- what happened to the first and second order terms?
A formal expansion to third order implies solving the
formidable 20-moment set of equations, as well as deriving the
appropriate collision terms.  Because the complexity increases
drastically when the order is increased, it is imperative to keep the
order as low as possible.  Hence, in this case it is better to skip
the first two terms in the Taylor expansion altogether, knowing that
they must be negligible in a collision-dominated plasma.

Note that this approximation is tailored to a fully ionized,
collision-dominated plasma.  In a weakly ionized or neutral gas, where
Coulomb collisions do not dominate, this equation set should offer no
improvement over the original 8-moment set.

\subsection{Models for the supersonic solar wind}
\label{sec:bimaxw}
Since collisions quickly cease to be important as the solar wind
expands and becomes supersonic (except for the electrons), describing
collisions is no longer essential.  Rather, the fluid description
should describe the (nearly) collisionless evolution reasonably well,
and be able to accommodate effects of wave-particle interactions.
Since collisions are nearly absent, the departure from a Maxwellian
($\phi_s$ in (\ref{eq:fiso})) may become large.  One may use
higher-order fluid models, described in the previous section, but it
can be argued that the cost of doing that may be prohibitive.  Moreover,
if the departure from a Maxwellian is large, it may no longer make
sense to expand about a Maxwellian and it may be advantageous to
consider expansions about other zero-order assumptions for the VDF\@.

The magnetic field in the corona and inner solar wind is strong (it is
a low-$\beta$ plasma) and the rapid gyration of charged particles in
the magnetic field implies that the VDF should be nearly
gyrotropic (i.e., isotropic in the plane perpendicular to $\vec
B$), at least in the rest frame of the plasma (the plasma as a whole
may undergo $\vec E \times \vec B$ drift).  If
collisions are few, the motion along and perpendicular to $\vec B$ may
become partly decoupled.  A more accurate description of such a
magnetized plasma will be provided if  different
``temperatures'' are considered along and perpendicular to the magnetic field.
Moreover,  as also discussed in Sect. \ref{second_gen_exo},
 observations of ions in the corona by the UVCS instrument on
the SOHO satellite show that even in this region, where collisions
should be frequent, the temperature along the line of sight (assumed
to be nearly perpendicular to $\vec B$ in the corona) is much higher
than the parallel temperature \citep{Kohl+etal1997, Cranmer+etal1999,
  Antonucci+Dodero+Giordano2000, Zangrilli+etal2002}.  To capture such
effects, without going to prohibitively high order in $\phi_s$,
the moment expansion about a zeroth order velocity
distribution has to allow for temperature anisotropies and to
be sufficiently simple that collision terms can still be developed 
(since they should not be negligible in the corona).

In the gyrotropic formulation  the VDF is expanded about a
\emph{bi-Maxwellian},
\begin{equation}
  \label{eq:bimaxw}
  f_s^\mathrm{bM} \equiv n_s \frac{m_s}{2\pi kT_{s\perp}}
  \sqrt{\frac{m_s}{2\pi k T_{s\parallel}}} \exp\left[ - m_s \left(
      \frac{c_{s\perp}^2}{2kT_{s\perp}} +
      \frac{c_{s\parallel}^2}{2kT_{s\parallel}} \right) \right].
\end{equation}
Here $\vec c_{s\parallel}$ and $\vec c_{s\perp}$ are the components of
the peculiar velocity $\vec c_s \equiv \vec v - \vec u_s$ parallel and
perpendicular to the magnetic field, $\vec c_{s\parallel} \equiv \vec
e_3 \vec e_3 \cdot \vec c_s$ and $\vec c_{s\perp} \equiv (\tensor{I} -
\vec e_3 \vec e_3) \cdot \vec c_s$, where $\tensor{I} \equiv \vec e_1
\vec e_1 + \vec e_2 \vec e_2 + \vec e_3 \vec e_3$ is the unit tensor.
The unit vectors $\vec e_1$, $\vec e_2$, and $\vec e_3$ are
orthogonal, with $\vec e_3$ chosen to be parallel to $\vec B$.  The
expansion then proceeds as in Sect.~\ref{sec:models-corona-inner},
assuming the full VDF to be
\begin{equation}
  \label{eq:fgyro}
  f_s = f_s^\mathrm{bM} (1 + \chi_s)
\end{equation}
where $\chi_s$ is assumed to be small (at least when evaluating
collision terms).

This choice has the advantage that the base function $f_s^\mathrm{bM}$
is quite simple, and that it reduces to the isotropic Maxwellian
(\ref{eq:fmaxw}) when $T_{s\parallel} = T_{s\perp} = T_s$.  Hence it
has the potential that a reasonably correct VDF may be retrieved in
the collision-dominated limit, and thus that it can reproduce,
reasonably well, the transport coefficients of classical transport
theory.  Hence fluid equations based on (\ref{eq:fgyro}), with a
proper choice for $\chi_s$, should give a reasonable description of
the solar wind flow all the way from the chromosphere and into
interplanetary space.

%At the end we shall also consider an expansion for the collisionless solar
%wind based on a different VDF than $f_s^\mathrm{bM}$, instead using
%VDFs observed in situ in the solar wind as the basis for the
%expansion.

\subsubsection{Gyrotropic equations with $\chi_s=0$}
\label{sec:gyrotr-chi0}

%\comment{ols: I notice that I use upper case for the pressure tensor
%  ($\tensor{P}$) while lower case $\tensor{p}$ is used in Sect.\ 1.
%  So this must be made consistent.}

Again, the simplest possible choice is just to use (\ref{eq:bimaxw})
as the closure assumption for the VDF\@.  In addition to the continuity
equation~(\ref{eq:cont}), this leads to five equations for the
expansion coefficients $\vec u_s$, $T_{s\parallel}$, and $T_{s\perp}$.
In compact notation these may be written
\begin{eqnarray}
  \label{eq:mombM1}
  n_s m_s \frac{D\vec u_s}{Dt} + \nabla \cdot \tensor{p_s} - n_s m_s
  \vec g - n_s e_s (\vec E + \vec u_s \times \vec B) &=&
  \frac{\delta\vec M_s}{\delta t} \\
  \frac{D \tensor{p_s}}{Dt} + \tensor{p_s} (\nabla \cdot \vec u_s) +
  2( \tensor{p_s} \cdot \nabla) \vec u_s &=& \frac{\delta
    \tensor{E_s}}{\delta t},
\end{eqnarray}
where $\tensor{p_s} = p_{s\perp} (\tensor{I}-\vec e_3 \vec e_3) +
p_{s\parallel} \vec e_3 \vec e_3$ is the partial pressure tensor
of species $s$, with
$p_{s\parallel(\perp)} = n_s k T_{s\parallel(\perp)}$.  Equivalently,
expressed in terms of $p_{s\parallel}$ and $p_{s\perp}$ they may be
written, %Barakat+Schunk1982, tau=q=0
\begin{eqnarray}
  \label{eq:mombM2}
  n_s m_s \frac{D\vec u_s}{Dt} + \nabla_\perp p_{s\perp} +
  \nabla_\parallel p_{s\parallel} + (p_{s\parallel} - p_{s\perp})
  \nabla \cdot (\vec e_3 \vec e_3) && \nonumber\\ - n_s m_s
  \vec g - n_s e_s (\vec E + \vec u_s \times \vec B) &=&
  \frac{\delta\vec M_s}{\delta t} \\
  \frac{D p_{s\parallel}}{Dt} + p_{s\parallel} (\nabla \cdot \vec u_s
  + 2 \nabla_\parallel \cdot \vec u_s) &=& \frac{\delta
    E_{s\parallel}}{\delta t} \\
  \frac{D p_{s\perp}}{Dt} + p_{s\perp} (\nabla \cdot \vec u_s
  + \nabla_\perp \cdot \vec u_s) &=& \frac{\delta
    E_{s\perp}}{\delta t}  \label{eq:momE2}
\end{eqnarray}
where $\nabla_\parallel \equiv \vec e_3 \vec e_3 \cdot \nabla$ and
$\nabla_\perp \equiv (\tensor{I} - \vec e_3 \vec e_3) \cdot \nabla$.

This set of equations was first derived by \citet{Chew}, although they
did not make the assumption (\ref{eq:bimaxw}) explicitly, and they used
the ideal MHD approximation $\vec E = - \vec u_s \times \vec B$ and
Maxwell's equations to cast the Lorentz force term in
(\ref{eq:mombM1}) in terms of $\vec u_s$ and $\vec B$.

With $\chi_s=0$ the VDF is sufficiently simple that the collision
terms have been calculated without having to make further approximations,
even for Coulomb collisions and even valid for (arbitrarily) large
flow speed differences and temperature anisotropies
(see, \citeauthor{Barakat+Schunk1981}, \citeyear{Barakat+Schunk1981}).
A closed form for the Coulomb collisions 
terms for a bi-Maxwellian plasma drifting in the
direction parallel to the magnetic field has been recently found
 by \citet{Hellinger2009}. The terms are sufficiently complicated
that they will not be reprinted here, though.

Solving the full 3D set of equations, even this fairly simple set, can
be a challenging task.  Often it is sufficient to consider the flow in
only one dimension.  For instance, the steady state fast solar wind
from polar coronal holes may to a good approximation be regarded as a 1D
problem, in which the plasma flows along field lines.  In that case
the only effect of the magnetic field is to specify the flow
geometry.  Moreover, far from the Sun the solar wind plasma pressure
is substantially larger than the magnetic pressure, in which case the
magnetic field plays no role in the wind expansion, which may be
regarded as approximately radial.  In such cases we may imagine that
the flow occurs along a flow tube which has a cross section of,
say, $A_0=1~\mbox{m}^2$ at the solar ``surface'' and which is centered on
a radial magnetic field line.  At a distance $r$ from the center of
the Sun the flux tube cross section is $A(r)$.  Equations
(\ref{eq:cont})  and (\ref{eq:mombM2})--(\ref{eq:momE2}) then
become
\begin{eqnarray}
  \label{eq:contrad}
  \pderiv{n_s}{t} + \frac{1}{A} \pderiv{(n_su_s A)}{r} &=& 0\\
  \pderiv{u_s}{t} + u_s \pderiv{u_s}{r} + \frac{k}{m_s n_s}
  \pderiv{(n_s T_{s\parallel})}{r} + \frac{1}{A} \deriv{A}{r}
  \frac{k}{m_s} (T_{s\parallel}-T_{s\perp}) - g - \frac{e_s}{m_s} E
  &=& \frac{1}{m_s n_s}\frac{\delta M_s}{\delta t} \label{eq:momrad}\\
  \pderiv{T_{s\parallel}}{t} + u_s \pderiv{T_{s\parallel}}{r} + 2
  T_{s\parallel} \pderiv{u_s}{r} &=& \frac{1}{n_s k} \frac{\delta
    E_{s\parallel}}{\delta t} \label{eq:eparrad}\\
  \pderiv{T_{s\perp}}{t} + u_s \pderiv{T_{s\perp}}{r} + \frac{1}{A}
  \deriv{A}{r} u T_{s\perp} &=& \frac{1}{n_s k} \frac{\delta
    E_{s\perp}}{\delta t}.\label{eq:eperprad}
\end{eqnarray}
A purely radially expanding outflow (i.e., spherically symmetric
outflow) corresponds to $A(r) = A_0 (r/R_S)^2$.

Note from~(\ref{eq:momrad}) that $T_{s\perp} > T_{s\parallel}$ leads
to acceleration of the wind.  This is a consequence of the Lorentz
force: When the magnetic field changes slowly with radial distance (as
we assume), magnetic moment is conserved, leading to conversion of
perpendicular motion into parallel motion and hence acceleration.
With an isotropic VDF as in Sect.~\ref{sec:5-moment-appr}, we are not
able to include this effect of the Lorentz force.  Note also that in a
steady state with no collisions, (\ref{eq:eparrad}) and
(\ref{eq:eperprad}) simplify to
\begin{eqnarray}
  \label{eq:paradiab}
  \deriv{(u_s^2 T_{s\parallel})}{r} &=& 0\\
  \deriv{(A T_{s\perp})}{r} &=& 0,\label{eq:perpadiab}
\end{eqnarray}
stating that as the wind is accelerated, $T_{s\parallel}$ will
decrease with distance, while $T_{s\perp}$ decreases as the flow tube
expands.  If the wind is accelerated fast near the Sun, and fast
compared to the flow tube expansion, we expect that $T_{s\parallel} <
T_{s\perp}$.  On the other hand, in the outer solar wind, where
$du_s/dr \approx 0$, eventually $T_{s\parallel} \gg T_{s\perp}$.  This
illustrates that, in the absence of collisions or wave particle
interactions, anisotropies \emph{must} arise in the solar wind,
showing why it is necessary to go beyond the isotropic formulation of
Sect.~\ref{sec:models-corona-inner} describing the supersonic solar
wind far from the Sun.

This equation set is simple, collision terms valid for any flow
conditions are available, and they capture important effects of the
Lorentz force.  Their main limitation is that they do not allow for
heat fluxes.  Although heat conduction is not as critical from the
corona and outwards as it is in the transition region, it nevertheless
plays a role, particularly for the electrons near the Sun.

Another shortcoming is that the temperature evolution they predict
cannot be entirely correct.  Imagine that in the corona $T_{s\perp}
\gg T_{s\parallel}$, implying that some process (e.g., cyclotron
waves) has heated the particles perpendicularly to $\vec B$, and also
supported by observations.  Neglecting collisions and interaction with
waves, magnetic moment is conserved, and the large perpendicular
motion in the corona must be translated into parallel motion in the
solar wind.  However, all of the perpendicular motion cannot be
translated into the flow speed $u_s$ (that is, wind acceleration): a
high $T_{s\perp}$ implies that the VDF is very {broad} in the
perpendicular direction, with some particles having a small
perpendicular velocity and others a large velocity.  Since the
magnetic field changes only   the direction of the velocity vector, 
the magnetic field expansion should lead to a VDF that
is similarly broad in the parallel direction.  If one solves the
collisionless Boltzmann equation, mapping the VDF from the corona and
outwards, one would thus find that the true VDF in the solar wind is
narrow in the perpendicular direction, but very broad in the parallel
direction, which means that $T_{s\parallel}$ in the solar wind must
become large, as shown by the collisionless kinetic models
(Sect.~\ref{second_gen_exo}).  This effect is not captured at all with
$\chi_s=0$.  In fact, (\ref{eq:paradiab}) predicts the opposite
effect: $T_{s\parallel}$ should {decrease} monotonically outwards
as the wind is accelerated, and there is no coupling that converts a
high $T_{s\perp}$ in the corona into a high $T_{s\parallel}$ in the
solar wind.  The remedy for this flaw is also to allow for heat flux
by having a nonzero $\chi_s$.

\subsubsection{The 16-moment approximation}
\label{sec:16-moment}

Hydrodynamic equations based on (\ref{eq:fgyro}) that allow for heat
fluxes as well as stresses, were first developed by
\citet{Oraevskii1968} for a collisionless plasma.  Their assumption
for $\chi_s$ is, when written in the form used by \citet{Barakat+Schunk1982},
\begin{eqnarray}
  \label{eq:chi16}
  \chi_s &=& \frac{\beta_{s\perp}}{2 m_s n_s}\left[ \beta_{s\perp}
    (c_{s1}^2-c_{s2}^2) \tensor{\tau_{s}} : \vec e_{1} \vec e_{1} + 2\beta_{s\perp}
    \tensor{\tau_{s}} : \vec e_{1} \vec e_{2} c_{s1} c_{s2} + 2\beta_{s\parallel}
    \tensor{\tau_{s}} : \vec c_{s\perp} \vec c_{s\parallel} \right] \nonumber\\
  && - \frac{\beta_{s\perp}^2}{m_s n_s} \left( 1 - \frac{\beta_{s\perp}
      c_{s\perp}^2}{4} \right) \vec q_{s}^\perp \cdot \vec c_{s\perp}
  - \frac{\beta_{s\perp} \beta_{s\parallel}}{m_s n_s} \left( 1 -
    \frac{\beta_{s\perp} c_{s\perp}^2}{2} \right) \vec q_s^\perp \cdot
  \vec c_{s\parallel} \nonumber\\
  && - \frac{\beta_{s\parallel}^2}{2m_s n_s} \left( 1 -
    \frac{\beta_{s\parallel} c_{s\parallel}^2}{3} \right) \vec
  q_s^\parallel \cdot \vec c_{s\parallel} - \frac{\beta_{s\perp}
    \beta_{s\parallel}}{2 m_s n_s} ( 1 - \beta_{s\parallel}
  c_{s\parallel}^2) \vec q_s^\parallel \cdot \vec c_{s\perp}
\end{eqnarray}
where $\beta_{s\parallel} \equiv m_s/(kT_{s\parallel})$ and
$\beta_{s\perp} \equiv m_s/(kT_{s\perp})$ and we have introduced the
tensor contraction $\tensor{A} : \tensor{B} \equiv A_{ij} B_{ji}$.  In
this approximation there are 16 moments of the VDF to be solved for.
Note that there are now two heat flux densities: $\vec
q_{s}^\parallel$ denote the transport of parallel thermal energy while
$\vec q_{s}^\perp$ denote the transport of perpendicular thermal
energy.  Since $T_{s\parallel}$ and $T_{s\perp}$ may be very
different, the conduction of the two forms of thermal energy
may also be different.
% {thermal conduction along the parallel and perpendicular
%direction  may also be very different.}

With this assumption, the left-hand side of the momentum equation acquires
an additional stress tensor term (compared with the $\chi_s=0$ case
(\ref{eq:mombM2})) while the equations for
$p_{s\parallel}$ and $p_{s\perp}$ get additional terms from the heat
flux vectors and the stress tensor:
\begin{eqnarray}
  n_s m_s \frac{D\vec u_s}{Dt} + \nabla_\perp p_{s\perp} +
  \nabla_\parallel p_{s\parallel} + (p_{s\parallel} - p_{s\perp})
  \nabla \cdot (\vec e_3 \vec e_3) + \nabla \cdot \tensor{\tau_s} &&
  \nonumber\\ - n_s m_s
  \vec g - n_s e_s (\vec E + \vec u_s \times \vec B) &=&
  \frac{\delta\vec M_s}{\delta t} \label{eq:mom16}\\
   \frac{D p_{s\parallel}}{Dt} + p_{s\parallel} (\nabla \cdot \vec u_s
  + 2 \nabla_\parallel \cdot \vec u_s) + 2 \vec e_3 \vec e_3 : (
  \tensor{\tau_s} \cdot \nabla \vec u_s ) + \nabla \cdot \vec
  q_s^\parallel && \nonumber\\
  - \tensor{\tau_s} : \frac{D(\vec e_3 \vec e_3)}{Dt}
  -\tensorc{Q_s} \threedot \nabla(\vec e_3 \vec e_3) &=& \frac{\delta
    E_{s\parallel}}{\delta t}  \label{eq:ppar16} \\
  \frac{D p_{s\perp}}{Dt} + p_{s\perp} (\nabla \cdot \vec u_s
  + \nabla_\perp \cdot \vec u_s) + (\tensor{I} - \vec e_3 \vec e_3 ) :
  (\tensor{\tau_s} \cdot \nabla \vec u_s) + \nabla \cdot \vec
  q_s^\perp && \nonumber\\
  + \frac{1}{2} \tensor{\tau_s} : \frac{D(\vec e_3 \vec
    e_3)}{Dt} + \frac{1}{2} \tensorc{Q_s} \threedot \nabla(\vec e_3 \vec
  e_3) &=& \frac{\delta E_{s\perp}}{\delta t} \label{eq:pperp16}
 \end{eqnarray}
where $\tensorc{Q_s}$ is a rank 3 tensor given as
\begin{equation}
  \label{eq:q3}
  \tensorc{Q_s} = \vec q_s^\parallel \vec e_3 \vec e_3 + e_3 \vec
  q_s^\parallel \vec e_3 + \vec e_3 \vec e_3 \vec q_s^\parallel - 2
  \vec q_{s\parallel}^\parallel \vec e_3 \vec e_3
\end{equation}
and $\tensorc{C} \threedot \tensorc{D} \equiv C_{ijk} D_{kji}$.  The
equations for the stress tensor $\tensor{\tau_s}$ and the heat flow
vectors are given by \citet{Barakat+Schunk1982} and will not be listed
here.

The corresponding Coulomb collisions terms have been derived by
\citet{Chodura+Pohl1971}, again in the limit of small flow
speed differences, and are too complex to be listed here.  Collision
terms for other types of interactions were later derived by
\citet{Demars+Schunk1979}.

If we limit ourselves to one spatial dimension, with flow along a
radial magnetic field with a specified flow tube area $A(r)$, we
retrieve the $\chi_S=0$
equations~(\ref{eq:contrad})--(\ref{eq:eperprad}), but with additional
source terms for $T_{s\parallel}$ and $T_{s\perp}$, and additional
equations for $q_s^\parallel$ and $q_s^\perp$ (the radial flow of
parallel and perpendicular thermal energy, respectively):
\begin{eqnarray}
  \label{eq:tpar16rad}
    \pderiv{T_{s\parallel}}{t} + u_s \pderiv{T_{s\parallel}}{r} + 2
  T_{s\parallel} \pderiv{u_s}{r} + \frac{1}{n_s k}
  \pderiv{q_s^\parallel}{r} + \frac{1}{A} \deriv{A}{r}
  \frac{q_{s}^\parallel}{n_s k} - \frac{2}{A} \deriv{A}{r}
  \frac{q_s^\perp}{n_s k} &=& \frac{1}{n_s k} \frac{\delta
    E_{s\parallel}}{\delta t}\\
  \pderiv{T_{s\perp}}{t} + u_s \pderiv{T_{s\perp}}{r} + \frac{1}{A}
  \deriv{A}{r} u T_{s\perp} + \frac{1}{n_s k} \pderiv{q_s^\perp}{r} +
  \frac{2}{A} \deriv{A}{r} \frac{q_s^\perp}{n_s k} &=& \frac{1}{n_s k}
  \frac{\delta E_{s\perp}}{\delta t}\label{eq:tperp16rad}\\
  \pderiv{q_{s}^\parallel}{t} + u_s \pderiv{q_{s}^\parallel}{r} + 4
  q_{s}^\parallel \pderiv{u_s}{r} + u_s q_{s}^\parallel \frac{1}{A}
  \deriv{A}{r} + 3 \frac{k^2n_sT_{s\parallel}}{m_s}
  \pderiv{T_{s\parallel}}{r} &=& \frac{\delta q_{s}^\parallel}{\delta
    t}'  \label{eq:qpar16rad} \\
  \pderiv{q_{s}^\perp}{t} +u_s \pderiv{q_{s}^\perp}{r} + 2
  q_{s}^\perp \pderiv{u_s}{r} + 2 u_s q_{s}^\perp \frac{1}{A}
  \deriv{A}{r} + \frac{k^2n_sT_{s\parallel}}{m_s}
  \pderiv{T_{s\perp}}{r} && \nonumber\\
  + \frac{1}{A} \deriv{A}{r}
  \frac{k^2n_sT_{s\perp}}{m_s} (T_{s\parallel}-T_{s\perp}) &=&
  \frac{\delta q_{s}^\perp}{\delta t}'.  \label{eq:qperp16rad}
\end{eqnarray}

Comparing (\ref{eq:eparrad}) and (\ref{eq:tpar16rad}), note that
$q_s^\perp$ has now become a ``source'' for $T_{s\parallel}$.  Because
of this term, if $T_{s\perp} \gg T_{s\parallel}$ in the corona,
magnetic moment conservation leads to an increasing $T_{s\parallel}$
with increasing radial distance.  This effect is also demonstrated in
numerical solar wind solutions using this equation set
\citep{Olsen+Leer1999}, and it is in agreement with solutions
to the collisionless Boltzmann equation, and the opposite of the
behavior predicted by (\ref{eq:paradiab}).  Hence by including heat
fluxes, also the description of the collisionless flow is improved.

In situ observations of the proton VDF between 0.3 and 1~AU by the
Helios spacecraft \citep{Marsch+etal1982} show that in most cases
$T_{p\parallel}/T_{p\perp}>1$, and that the ratio tends to increase
with radial distance from the Sun, consistent with the effect of
magnetic moment conservation described above.  However, there are
certainly cases where $T_{p\parallel}/T_{p\perp} \le 1$, particularly
for the fast solar wind at small heliocentric distances.  This shows
that the proton magnetic moment is not strictly conserved.
Since Coulomb collisions should be quite unimportant for protons so
far from the Sun, it indicates that wave-particle interactions must
take place in the solar wind.  We should also point out that if
$T_{p\parallel} \gg T_{p\perp}$ the firehose plasma instability may be
triggered, which would tend to lower the temperature anisotropy.  Such
effects are not included in the models considered here, but can
certainly be included by appropriate choices for the
collision terms on the right-hand side of the equations (which would
then account for both Coulomb collisions and wave-particle
interactions).  To determine the magnitude of these plasma processes,
required to reproduce the observed VDF, it is important that the fluid
models describe reasonably well the effect of the expanding magnetic
field and the wind acceleration on the temperature anisotropy.

With the actual temperature anisotropies and radial variation of
temperatures measured by Helios, \citet{Holzer1986} found that the
anisotropy of the pressure tensor had little influence on the solar
wind acceleration and heating of the plasma, compared with the
pressure gradient force and the pressure work term, respectively.

Although the description of the collisionless solar wind is improved
by the 16-moment approach, compared to the $\chi_s=0$ case, 
the description of the heat conduction in a collision-dominated
plasma is not accurate.  The heat conduction terms in (\ref{eq:chi16})
are still first order in $\vec c_s$, as it is in the isotropic
13-moment approximation (\ref{eq:13momvdf}).  In a 
{multi-fluid model of a} collision-dominated
plasma the electron heat conduction will thus not be described much better
than in the simple 8-moment approximation of Sect.~\ref{sec:8-moment}.

\subsubsection{Gyrotropic equations with improved collision terms}
\label{sec:gyrotr-improved}

To improve the description of Coulomb collisions it is possible to
repeat the approach of Sect.~\ref{sec:8-moment-improved}, namely by
letting $\chi_s$ be of third order in $\vec c_s$ instead of first
order.  \citet{Janse+etal2005} developed gyrotropic transport
equations assuming
\begin{equation}
  \label{eq:chiimproved}
  \chi_s = \alpha_{s \perp} \vec q_s \cdot \vec c_{s
    \perp} c^2_s \left( 1 + \gamma_{s \perp} c^2_s \right)
  + \alpha_{s \parallel} \vec q_s \cdot \vec c_{s
   \parallel} c^2_s \left( 1 + \gamma_{s \parallel}
  c^2_s\right),
\end{equation}
where the coefficients are chosen such that $\vec u_s$ and $\vec q_s$
are the flow velocity and the heat flux density, leading to
\begin{eqnarray}
  \gamma_{s \perp} &=& -\frac{m_s}{kT_{s \perp}}\frac{4 T_{s
      \perp}^2 + T_{s \perp} T_{s \parallel}}{24 T_{s \perp}^2 + 8 T_{s
      \perp} T_{s \parallel} + 3T_{s \parallel}^2} \\
  \gamma_{s \parallel} &=& -\frac{m_s}{kT_{s \perp}}\frac{2T_{s
      \perp}^2+3T_{s \perp}T_{s \parallel}}{8T_{s \perp}^2+12T_{s
      \perp}T_{s \parallel}+15T_{s \parallel}^2}\\
  \alpha_{s \perp} &=& - \frac{m_s^2}{n_s k^3} \frac{1}{T_{s
      \perp}}\frac{24T_{s \perp}^2+8T_{s \perp}T_{s \parallel}+3T_{s
      \parallel}^2}{96T_{s \perp}^4+48T_{s \perp}^3T_{s
      \parallel}+4T_{s \perp}^2T_{s \parallel}^2+24T_{s \perp}T_{s
      \parallel}^3+3T_{s \parallel}^4}\\
  \alpha_{s \parallel} &=& - \frac{m^2_s}{n_s k^3} \frac{1}{T_{s
      \parallel}}\frac{8 T_{s \perp}^2+12T_{s \perp}T_{s \parallel} +
    15 T_{s \parallel}^2}{16 T_{s \perp}^4+48 T_{s \perp}^3 T_{s
      \parallel} + 6 T_{s \perp}^2 T_{s \parallel}^2 + 60 T_{s \perp}
    T_{s \parallel}^3 + 45 T_{s \parallel}^4}.
\end{eqnarray}
Note that, compared with (\ref{eq:chi16}) the stress tensor has been
omitted and there is only one heat flux vector.

For the case of one-dimensional, radial flow, the resulting momentum
and energy equations
become formally identical to (\ref{eq:momrad}), (\ref{eq:tpar16rad}),
and (\ref{eq:tperp16rad}), while the equation for the heat flow
becomes
\begin{eqnarray}
  \label{eq:qradimproved}
  \pderiv{q_s}{t} + u_s \pderiv{q_s}{r} + 2 q_{s}^\parallel
  \pderiv{u_s}{r} + \frac{1}{2} q_{s}^\parallel u_s \frac{1}{A}
  \deriv{A}{r} + 2 q_{s}^\perp \pderiv{u_s}{r} + 2 q_{s}^\perp u_s
  \frac{1}{A} \deriv{A}{r} &&\nonumber\\
  + \frac{k^2 n_s T_{s\parallel}}{m_s} (
  T_{s\parallel} - T_{s\perp} ) &=& \frac{\delta q_s'}{\delta t},
\end{eqnarray}
where
\begin{eqnarray}
  q_{s}^\parallel &=& 30 q_s 
  \frac{T_{s \parallel}^3(4 T_{s \perp}+3 T_{s \parallel})}{16 T_{s
    \perp}^4+48T_{s \perp}^3T_{s \parallel}+6T_{s \perp}^2T_{s
    \parallel}^2+60T_{s \perp}T_{s \parallel}^3+45T_{s \parallel}^4} \\
q_{s}^\perp &=& 2 q_s 
\frac{T_{s \perp}^2 (8 T_{s \perp}^2+24T_{s \perp} T_{s \parallel} + 3
    T_{s \parallel}^2)}{16T_{s \perp}^4  + 48 T_{s \perp}^3 T_{s
    \parallel} + 6 T_{s \perp}^2 T_{s \parallel}^2 + 60 T_{s \perp}
    T_{s \parallel}^3 + 45 T_{s \parallel}^4}
\end{eqnarray}
still represent the flow of parallel and perpendicular thermal energy
in the radial direction, satisfying $q_s = q_s^\parallel/2 +
q_s^\perp$.

In the isotropic limit $T_{s\parallel} = T_{s\perp}$ these equations
simplify to the equations of Sect.~\ref{sec:8-moment-improved}.  
%\comment{MME: maybe we can skip the last two sentences and cite
%Killie et al. in the last sentence, that basically restates what
%is said in the deleted ones.}
%\sout{For this reason, the collision terms developed by Killie et al (2004)
%may to a good approximation also be used here, although they are
%formally not consistent with the assumption (\ref{eq:chiimproved}).
%The collision terms $\delta E_{s\parallel}/\delta t$ and $\delta
%E_{s\perp}/\delta t$ do not contain terms proportional to $\chi_S$,
%and will therefore be identical to those derived by Chodura and Pohl (1971).}
%\citet{Chodura+Pohl1971}}.
Collision terms for the momentum and heat flow
equations consistent with the approximation (\ref{eq:chiimproved})
have not been derived.  The corresponding isotropic collision terms  
\citep{Killie+etal2004}
will not be entirely correct in the outer corona where temperature
anisotropies become large.  However, the collision terms derived by
\citet{Chodura+Pohl1971} are also questionable in this region as they
assume that flow speed differences are small.

In the collisionless regime these equations give essentially the same
behavior as the 16-moment equations of Sect.~\ref{sec:16-moment}
(without the stress tensor).  In particular, the conversion of
$T_{s\perp}$ in the corona into $T_{s\parallel}$ and wind
acceleration is well described also with this equation set.

These equations are thus intended to be used for models of a fully
ionized solar wind, extending all the way from the collision-dominated
transition region and into the magnetized flow of interplanetary
space.

%\comment{ols: Added next two paragraphs.  These could be part of the
%  conclusion of the fluid section, or part of the conclusion of the
%  whole paper (we need one)?}

\subsubsection{Higher order gyrotropic equations tailored to spherically
  symmetric plasma outflow}
\label{sec:cuperman}
\citet{Cuperman1980,Cuperman1981} developed transport equations that account for
temperature anisotropies, heat conduction, and a higher-order term that
leads to non-thermal tails in the assumed VDF\@.  The
equations are specifically tailored to one-dimensional (i.e., purely
radial), spherically symmetric outflow, where, e.g., heat conduction
is strictly in the radial direction.  Although temperature
anisotropies are included, the expansion is actually about a pure
Maxwellian, not a bi-Maxwellian, and is hence based on the same
starting point as the equations of
Sect.~\ref{sec:models-corona-inner}.

In spherically symmetric outflow, the VDF can only depend on three
independent variables: the radial distance $r$, the speed $v=|\vec
v|$, and $\cos\theta$ where $\theta$ is the angle between the velocity
and radius vectors.  Hence Cuperman et al.\ assume that the VDF for species
$s$ has the form
\begin{equation}
  \label{eq:cupvdf}
  f_s(r,c_s,\cos\theta) = f_s^{(0)}(r,c_s) + \sum_{n=0}^2 b_n(r,c_s)
  f_s^{(0)}(r,c_s) P_n(\cos\theta)
\end{equation}
where $c_s \equiv |\vec v - \vec u_s(r,t)|$ is still the peculiar
speed, $f_s^{(0)}$ is the Maxwellian defined by (\ref{eq:fmaxw}) and
$P_n$ are Legendre polynomials.

In addition to the familiar moments $n_s$, $u_s$, $T_{s\parallel}$,
$T_{s\perp}$, $T=(T_\parallel + 2T_\perp)/3$, and $q_s$, this
expansion introduces the fourth order moment $\xi_s(r,t)$ defined as
\begin{equation}
  \label{eq:xi}
  \xi_s \equiv \frac{1}{n_s} \int\int\int (c_s\cos\theta)^4 f_s d\vec
  v - 3 \left( \frac{kT_{s\parallel}}{m_s} \right)^2.
\end{equation}
$\xi_s$ is thus a moment that can account for tails in $f_s$.  In
terms of these, the $b_n$ polynomials in (\ref{eq:cupvdf}) are
\begin{eqnarray}
  b_0(c_s) &=& c_0 \left( 1 - \frac{2m_sc_s^2}{3kT_s} + \frac{4 m_s^2
      c_s^4}{15 (kT_s)^2} \right) \\
  b_1(c_s) &=& c_1 \sqrt{\frac{m_s}{kT_s}} c_s \left( -1 + \frac{m_s
      c_s^2}{5kT_s} \right)\\
  b_2(c_s) &=& c_2 \frac{m_s c_s^2}{kT_s},
\end{eqnarray}
and the coefficients contain the moments of the VDF,
\begin{eqnarray}
  c_0 &=&\frac{5}{8} \left( \frac{m_s}{kT_s} \right)^2 \xi_s +
  \frac{4}{3} \left( \frac{T_{s\parallel} - T_{s\perp}}{T_s}
  \right)^2\\
  c_1 &=& \left( \frac{m_s}{kT_s} \right)^{3/2} \frac{q_s}{m_s n_s} \\
  c_2 &=& \frac{T_{s\parallel} - T_{s\perp}}{3T_s}.
\end{eqnarray}
Notice that the $b_1$ term, which accounts for heat conduction, is
identical to the 8-moment assumption (\ref{eq:8momvdf}).

The resulting momentum equation is identical to the gyrotropic
momentum equation (\ref{eq:momrad}) when $A(r) \propto r^{-2}$, while
the other equations are (omitting subscript $s$ to simplify notation):
\begin{eqnarray}
  \label{eq:cuptpar}
  \pderiv{T_\parallel}{t} + u \pderiv{T_\parallel}{r} + 2 T_\parallel
  \pderiv{u}{r} + \frac{6}{5} \frac{1}{nk}\pderiv{q}{r} + \frac{4}{5}
  \frac{1}{nk} \frac{q}{r} &=& \left( \frac{\delta T_\parallel}{\delta
      t} \right)_c \\
  \pderiv{T_\perp}{t} + u\pderiv{T_\perp}{r} + 2 T_\perp
  \frac{u}{r} + \frac{2}{5} \frac{1}{nk}\pderiv{q}{r} + \frac{8}{5}
  \frac{1}{nk} \frac{q}{r} &=& \left( \frac{\delta T_\perp}{\delta
      t} \right)_c \label{eq:cuptperp}\\
  \pderiv{q}{t} + u \pderiv{q}{r} + \frac{2}{r} u q + 4q \pderiv{u}{r}
  + \frac{5}{2} \frac{k^2nT_\parallel}{m} \pderiv{T_\parallel}{r}
  &&\nonumber\\
  \mbox{} - \frac{10}{3} \frac{k^2 n}{m} (T_\parallel - T_\perp)^2 + \frac{5}{6}
  m \pderiv{(n\xi)}{r} &=& \left( \frac{\delta q}{\delta t}
  \right)_c \label{eq:cupq}\\
  \pderiv{\xi}{t} + u \pderiv{\xi}{r} + 4\xi \pderiv{u}{r} +
  \frac{36}{5} \frac{k}{n m^2} q \pderiv{T}{r} - \frac{16}{5}
  \frac{k}{nm^2} q \pderiv{(T_\parallel - T_\perp)}{r} &&\nonumber\\
  \mbox{} + \frac{24}{5} \frac{kT}{m} \pderiv{}{r} \left( \frac{q}{nm} \right)
  - \frac{24}{5} \frac{k}{m} (T_\parallel - T_\perp) \pderiv{}{r} \left( \frac{q}{nm}
  \right) &&\nonumber\\
  \mbox{} - 8 \frac{k}{(nm)^2} q (T_\parallel - T_\perp) \pderiv{n}{r}
  - \frac{64}{5} \frac{k}{nm^2} \frac{q}{r} (T_\parallel - T_\perp)
  &=& \left( \frac{\delta \xi}{\delta t} \right)_c, \label{eq:cupxi}
\end{eqnarray}
where the right-hand side contain collision terms.  Comparing with the
16-moment equations~(\ref{eq:tpar16rad})--(\ref{eq:tperp16rad}) (when
$A(r) \propto r^{-2}$), we notice the same temperature dependence in
the two sets, while the heat conduction terms differ.

The collision terms for Coulomb collisions, using the Fokker-Planck
collision approximation, have been derived in explicit form in the
(usual) limit when relative flow differences are much smaller than
thermal speeds ($(u_s-u_t)^2 \ll kT_{s(t)}/m_{s(t)}$), and by
neglecting the contribution from the higher order term $b_2$
\citep{Cuperman1981}.  The expressions are sufficiently complex that
they will not be reproduced here, but they are still in a form that
should be straightforward to include in a numerical model.

This equation set accounts for heat conduction and thermal forces in
the collision-dominated regime, and allows for a transition to
collisionless flow.  It may thus be an alternative to the equation set
of Sect.~\ref{sec:gyrotr-improved} for a solar wind model that extends
all the way from the solar transition region to interplanetary space.
A disadvantage is that the first order correction $b_1$ is still first
order in $c_s$, and it is therefore not clear how well this set
describes heat conduction and thermal forces in the transition region.
On the other hand it has the advantage (compared to the set in
Sect.~\ref{sec:gyrotr-improved}) that a higher-order correction
($\xi_s$) is included, which can accommodate a high-energy tail in the
VDF\@.  A comparison of the two sets would also be of interest; since
the closure assumptions are not the same, it would be useful to know,
e.g., how the proton heat flux differ in the two sets in the
transition to collisionless flow in the outer
corona.

To our knowledge, the full equation set (\ref{eq:momrad}),
(\ref{eq:cuptpar})--(\ref{eq:cupxi}) has so far not been implemented
and solved in any numerical model.

\subsubsection{Gyrotropic models based on observed proton VDFs}
\label{sec:leblanc}

So far all models have been expansions about Maxwellians or
bi-Maxwellians, with the underlying assumption that collisions are
sufficiently frequent that the departure from a Maxwellian will not be
very large.  However, far from the Sun collisions are so few that the
VDF may depart significantly from a Maxwellian; indeed, the in situ
solar wind proton observations confirm this \citep{Marsch+etal1982}.
Expanding about a base function that is far from the observed VDF
requires that we may have to include a large number of terms in the
correction term $\phi_s$ or $\chi_s$ to be able to reconstruct the
observed VDF\@.  We have already argued that increasing the order of
the correction term will drastically increase the complexity of the
equations.  Moreover, with the forms for $\phi_s$ and $\chi_s$
discussed above, the VDF may even become negative in parts of velocity
space when, e.g., the heat flux is large.

\citet{Leblanc+Hubert1997, Leblanc+Hubert1998} and \citet{Leblanc+Hubert2000}
have therefore proposed to use a base function that can reproduce the
basic properties of some of the proton VDFs observed by the Helios
spacecraft \citep{Marsch+etal1982}, and particularly the high-energy
tail of the VDF\@.  The full VDF is assumed to have the shape
\begin{equation}
  \label{eq:LH}
  f_s = f_s^{G} (1 + \psi_s)
\end{equation}
where
\begin{equation}
  \label{eq:LHbase}
  f_s^G = n_s \frac{m_s}{4\pi k T_{s\perp} D_s^*} \exp\left( -
    \frac{m_s}{2kT_{s\perp}}  c_{s\perp}^2 -
    \frac{c_{s\parallel}+D_s^*}{D_s^*} + \frac{1}{E_s^*} \right) \,
  \mbox{erfc} \left[ \sqrt{E_s^*} \left( \frac{1}{E_s^*} -
      \frac{c_{s\parallel} + D_s^*}{2D_s^*} \right) \right],
\end{equation}
and $\mbox{erfc}$ is the complementary error function, with
\begin{eqnarray}
  D_s^* &=& \sqrt[3]{\frac{q_s^\parallel}{2 m_s n_s}}\\
  E_s^* &=& \frac{2 m_s D_s^{*2}}{kT_{s\parallel} - m_s D_s^{*2}}.
\end{eqnarray}
In the limit $D_s^* \rightarrow 0$, $f_s^G$ becomes identical to the
bi-Maxwellian $f_s^{bM}$ (\ref{eq:bimaxw}).

The expansion $\psi_s$ is then written in a form very similar to
(\ref{eq:chi16}), except that the coefficients are somewhat different
because the base function is different.  For one-dimensional flow
along the magnetic field --- neglecting the stress tensor and assuming
that all heat flow is along the magnetic field --- $\psi_s$ simplifies
to
\begin{equation}
  \label{eq:psi1d}
  \psi_s = - \beta_{s\parallel} \beta_{s\perp} \left( 1 -
    \frac{\beta_{s\perp}}{2} c_{s\perp}^2 \right)
  \frac{q_{s\parallel}^\perp}{m_s n_s} c_{s\parallel},
\end{equation}
where $\beta_{s\parallel}$ and $\beta_{s\perp}$ are defined in
Sect.~\ref{sec:16-moment}.

This expansion leads to an equation set very similar to the 16-moment
set of Sect.~\ref{sec:16-moment}, with the main difference in the
collision terms.  The main drawback seems to be that collision terms
have not been presented in a closed form suitable for inclusion in a
numerical model.

To our knowledge, solar wind solutions with this equation set have not
been obtained, neither with nor without the collision terms.  Without
actual numerical solutions, it is difficult to quantify the
improvement offered by this approximation.

\section{Summary and conclusions}
%\comment{ols: Ok, something has to be written here\ldots}

 {In this paper we summarized the main 
arguments and limitations of  the
kinetic-exospheric and multi-fluid models of the solar wind. 
Our aim was: on  one hand to outline a brief history of more than 
50 years of solar wind modeling,  as well as to 
provide a comparative review of 
kinetic and multi-fluid approaches emphasizing
 their advantages and limitations as well as
their common theoretical roots: the fundamental equations of
plasma state.}

{%By recalling the history of the solar wind models
 % we reviewed the main paradigms adopted to explain the
 % supersonic expansion of the solar corona. 
  On the one hand 
  we discussed the  purely   mechanical, or hydrodynamic, 
  approach  that explains the main properties of the solar wind  by its
  macroscopic evolution in terms of density, bulk velocity and
  temperature. This description  was the
  first to predict   successfully  the main bulk properties of the solar
  wind \citep{Parker58}. 
  On the other hand we emphasized the arguments  of the holders of a more detailed description, the
  kinetic theory,  that provides a detailed view on how the energy is
  distributed among plasma particles, pitch angles, from the collisional
  region out to the collisionless supersonic stream. This approach 
  is built on the fundamental concept of the velocity   distribution 
  function, of each of the species streaming in the
  solar wind. It would be false, however, to state that the kinetic
  theory disregards the macroscopic plasma transport. 
  Indeed, the entire hierarchy of the transport equations is
  satisfied by the kinetic solutions. Although the 
  first generation kinetic exospheric models were less succesfull to obtain the
  supersonic solar wind expansion, the corrections found for the
  Pannekoek-Rossland electric field \citep{LemaireScherer69} 
  eventually provided the second generation kinetic
  exospheric models in agreement with observations and with the results
  of the hydrodynamic approach
  \cite{LemaireScherer71}. At the same epoch \citet{Jockers70} developed independently a
  kinetic exospheric model for the supersonic solar wind.
  The third generation kinetic exospheric models \citet{Lamy2003a,
  Zouganellis2004} treats the effects
  of non-Maxwellian electron velocity distribution function in the corona and give clues
  about the kinetic processes in the transonic region.
  The fourth generation kinetic exospheric models \citet{Pierrard+Maksimovic+Lemaire,
  Pierrard2001a} include Coulomb collisions and their effects on the
  velocity distribution functions and macroscopic parameters of the
  solar wind. The kinetic treatment of wave-particle interactions
  \citep{Marsch2006,Aschwanden2009}, is not discussed in this review,
  but gives insight on the anisotropies observed in the
  solar wind.
  After several decades of fluid and kinetic solar wind modeling, 
  often leading to controversies, scientists
  now realize that both approaches are complementary and not opposed 
  \citep{LemaireEchimEOS2008, Parker2010}.} 

{Fluid and kinetic models stem from the
same theoretical root: the Boltzmann equation.
%and perform well in both spatial and time resolution, being suitable
%for applications like, for instance, space weather modeling. 
%Although fluid approaches are based on a truncated physical representation
%of the velocity distribution function, implying incomplete
%representations of plasma processes like, for instance, the heat transfer,
% they perform rather well in both spatial and time resolution,
%and are  suitable for applications, e.g. the space weather modeling.}
 The Chapman-Enskog and
  the Grad theories (see Appendix \ref{Boltzmann_solutions}) are
  examples of solutions of the Boltzmann equation bridging the way between microscopic and 
  macroscopic descrition.
  The  Chapman-Enskog and   the Grad solutions
  are based on the assumption that the zero-order
  approximation of the velocity distribution function  of the protons and electrons,
  $f(v, r, t)$, is  necessarily a displaced Maxwellian, $f_M^{(0)}(v, r, t)$,
  characterized by a given number density $n(r, t)$, a given average velocity
  $\vec u(r,  t)$, and a given isotropic temperature $T(r, t)$;
  the values of these lowest order moments of the actual VDF
  do not change when higher order approximations are defined and determined for
  $f(v, r, t) = f_M(v, r, t) \left[1 + g_1(v, r, t) +
      \ldots\right]$; in other words nor in the Chapman-Enskog
  approximation nor in Grad's one the series of functions
  $g_k$ contribute to changing the value of $n$, $\vec u$ and $T$; 
  this means that the higher order terms, $f_M(v, r, t)g_1(v, r, t)$ in the
  expansion of $f$, affect only the higher order moments, e.g. the
  non-diagonal components of the pressure tensor $p_{ij}$,
  the components of the energy flux tensor, etc. 
%  In Appendix B we
%  discuss how these requirements on $f(v, r, t)$ limit seriously the
%  applications of the two approaches to collisionless region of the
%  solar wind.

Another key-aspect of kinetic and fluid modeling is the treatment 
of the heat flux. In the
kinetic treatment the heat flux is
correctly computed from the moments of the velocity distribution function.
By having a separate equation for the time evolution of the heat flux
vector, thus treating the heat flux on an equal footing to density,
flow speed, and temperature, the multi-fluid isotropic equations sets in
Sects.~\ref{sec:8-moment}--\ref{sec:8-moment-improved}, and the
gyrotropic multi-fluid equations in Sects.~\ref{sec:16-moment} and
\ref{sec:gyrotr-improved}, allow for a transition from classical heat
conduction, proportional to $-\nabla T$ to a ``collisionless'' heat
flux that will not be proportional to the temperature gradient.  This
is particularly important in multi-fluid solar wind models driven mainly by proton
heating (as the recent UVCS/SOHO observations indicate), since protons
will quickly become collisionless due to the strong heating.  Whether
the proton heat flux is classical or non-classical (collisionless) can
have a large impact on the energy budget of the corona/solar wind
system: If classical heat conduction dominates in a large region of the
corona, much of the energy received by the protons may be conducted
downwards and subsequently, through collisions, converted into
electron heat conduction which is lost as radiation in the transition
region.  On the other hand, if the protons quickly become
collisionless, more of the heating may be used to accelerate the
wind.  In solar wind models based on the 16-moment approximation
\citep{Lie-Svendsen+Hansteen+Leer+Holzer2002} it was found that the
protons would become almost collisionless quite low in the corona, so
that the proton heat flux was positive (pointing outwards from the
Sun) even below the proton temperature maximum, leading to a very
high speed wind with a very low mass flux.

But the critical question remains: How well do these higher-order
fluid models describe the proton heat flux in the transition to
collisionless flow?  Since heat flux is a higher-order moment, one
might expect that it is sensitive to the shape assumed for the VDF (the
closure assumption).  At present we have no independent way of
evaluating the accuracy of the proton heat flux description of these
equations in this transition regime.  This is one area in which
kinetic models could provide valuable insight.  By actually solving
the Boltzmann equation for protons in the
outer corona, with proton-electron and proton-proton collisions
included through the Fokker-Planck collision
term, one could evaluate how well the
higher-order fluid models do the energy bookkeeping in this region.
This is a challenging task for the kinetic models, however: since
collisions are still essential in this region, collisionless kinetic
models (exospheric models) are inadequate.  Moreover, proton
self-collisions are needed.  Since the proton VDF is expected to
deviate strongly from a Maxwellian, in particular if the heating is
caused by cyclotron waves heating mostly perpendicular to the magnetic
field, the test particle approach developed by
\citet{Lie-Svendsen+Hansteen+Leer1997} and
\citet{Pierrard+Maksimovic+Lemaire} for electron self-collisions is
most likely not directly applicable to protons.  Despite the
challenges, or because of them, this is a problem that definitively
deserves attention in order to improve our understanding of the energy
budget of the corona and solar wind.

{Our review emphasizes that the kinetic  approach is fundamental 
and provides a rigorous treatment  of the physics at all levels, 
going  beyond the limits of an academic exercise.
Indeed, kinetic models describe microscopic and macroscopic processes,
addressing self-consistently the complex aspects of the interactions
between plasma particles and fields.
Such a detailed level of description is obtained at the expense
of a lower spatial resolution and is currently limited to
stationary situations. Future development of new kinetic 
Vlasov and Fokker-Planck simulations are expected to give new momentum
to kinetic modeling of the solar wind.}
% the theoretical solutions
%as initial conditions, are rapidly developing, taking
%profit of the huge available computing power.}

\appendix

\section{Appendix}
\label{definitions}
\setcounter{equation}{0}
\renewcommand{\theequation}{A-\arabic{equation}}

The following list defines the first moments of the velocity distribution function
of each species $s$, and the main macroscopic properties of the plasma itself.
A similar list can be found in \citeauthor{Montgomery+Tidman}
(\citeyear{Montgomery+Tidman}, p.\ 197), the only difference is that
we define the peculiar, or thermal, velocity $\vec c_s=\vec v - \vec u_s$, 
following  \citet{Grad1958} and \cite{Schunk1977}, i.e. in the reference frame
moving with the average  velocity, $\vec u_s$, of each species:
\begin{eqnarray}
n_s & = & \int\int\int{f_s\, d \vec v}, \hspace{0.3cm} \mbox{partial number
  density of species $s$} \\
n & = & \sum_{s}n_s, \hspace{0.3cm} \mbox{total number
  density of the plasma} \\
\vec u_s & = &  \frac{1}{n_s}\int\int\int{f_s\, \vec v\, d \vec v},  \hspace{0.3cm}
  \mbox{average velocity of species $s$} \\
\rho_m & = & \sum_{s}{n_s}m_s,  \hspace{0.3cm}  \mbox{total mass
  density  of the plasma} \\
\vec U & = & \frac{1}{\rho_s}\sum_{s}{m_s n_s \vec
  u_s},  \hspace{0.3cm}  \mbox{center of  mass velocity or bulk
  velocity  of the plasma} \\
\vec w & = & \vec v - \vec U,  \hspace{0.3cm}  \mbox{particle velocity with
  respect to the center of  mass velocity} \\
\vec c_s & = & \vec v - \vec u_s,  \hspace{0.3cm}  \mbox{random  velocity} \\
\vec U_s & = &  \frac{1}{n_s}\int\int\int{f_s \vec w\, d \vec v},  \hspace{0.15cm}
  \mbox{average velocity of species $s$ with respect to center of
  mass velocity} \\
\vec J & = & \sum_s{e_s n_s \vec u_s},   \hspace{0.3cm}
  \mbox{total electric current density  of the plasma} \\
\rho_c & = & \sum_s{e_s n_s},   \hspace{0.3cm}
  \mbox{total electric charge density  of the plasma} \\
\tensor{p_s} & = & m_s \int\int\int{\vec c_s
  \vec c_s\, f_s\, d\vec v} \hspace{0.3cm}  \mbox{kinetic pressure tensor of
  species $s$} \\
\tensor{P} & = &  \sum_s \tensor{p}_s,  \hspace{0.3cm}
  \mbox{total pressure of the plasma} \\
\tensor{\pi} & = &  \sum_s m_s \int\int\int {\left(\vec c_s
  \vec c_s - 1/3c_s^2 \tensor{1} \right)f_s\, d\vec v},  \hspace{0.3cm}
  \mbox{stress tensor, traceless part of $\tensor{P}$} \\
\tensorc{Q_s} & = & m_s \int\int\int{\vec c_s
  \vec c_s \vec c_s\, f_s\, d\vec v} \hspace{0.3cm}  \mbox{heat tensor of
  species $s$} \\
\vec q_s & = &  \frac{m_s}{2} \int\int\int{c_s^2\,
  \vec c_s\, f_s\, d\vec v} \hspace{0.3cm}  \mbox{heat flux vector of species
  $s$}\\
\vec q & = & \sum_{s} \vec q_s  \hspace{0.3cm}  \mbox{total heat flux
  vector of the plasma} \\
T_{s\parallel} &=& \frac{m_s}{kn_s} \int\int\int{(c_s)_{\parallel}^2\,f_s\, d\vec v} \hspace{0.3cm}  \mbox{the partial
  parallel temperature of species $s$} \\
T_{s\bot} &=& \frac{m_s}{2kn_s} \int\int\int{(c_s)_{\bot}^2\,f_s\, d\vec v} \hspace{0.3cm}  \mbox{the partial
  perpendicular temperature of species $s$} \\
T_s & = & \frac{1}{3}(T_{s\parallel} + 2T_{s\bot})
  \hspace{0.3cm}  \mbox{the omnidirectional
  temperature of species $s$} \\
T_{\parallel} & = & \frac{1}{n} \sum_s n_s  T_{s\parallel}  \hspace{0.3cm}
  \mbox{the plasma parallel temperature} \\
T_{\bot} & = & \frac{1}{n} \sum_s n_s  T_{s\bot}  \hspace{0.3cm}
  \mbox{the plasma perpendicular temperature} \\
T & = & \frac{1}{3}(T_{\parallel} + 2T_{\bot})  \hspace{0.3cm}  \mbox{the 
  plasma temperature} 
\end{eqnarray}
%In the above relation the distribution function  $f_s$ of each species is
%normalized such that $n_{0s}$ is the average density. 
%\comment{ols: The definition of $\tensor{p_s}$ here is not agreement
%  with the pressure tensor used in eq.~(\ref{eq:mom}), which uses
%  $\vec c_s$ in the integrand, not $\vec w$.  This needs to be
%  resolved.  I have not done it since I think Marius should decide
%  which convention to use.  However, since these moments are mostly
%  used in Sect.~\ref{sec:fluid-models}, I think the least amount of
%  work will be to use Schunk's convention (with $\vec c_s$ not $\vec w$ as
%  the integrated variable).}

%\comment{ols: $n_{0s}$ is not defined.  I suggest that also $\vec q_s$
%(not just the total heat flux) and $T_{s\parallel}$ and $T_{s\perp}$
%be defined here for completeness (makes it easier for the reader to
%find all the definitions in one place).}

\section{Appendix}
\label{Boltzmann_solutions}
\setcounter{equation}{0}
\renewcommand{\theequation}{B-\arabic{equation}}

In this Appendix we briefly review
% discuss theoretical aspects
%relevant for solving the Boltzmann equation, which is a
%fundamental equation for solar wind modeling.
%The reader interested in a formal derivation of the Boltzmann equation
%from the Liouville theorem in the general framework of  statistical mechanics, 
%may find the demonstration in the seminal paper by \citet{Grad1958} and
%the classical references therein. 
{blue}{two methods adopted} to solve the
Boltzmann equation, relevant for kinetic and fluid modeling
of the solar wind.
% others are presented here
%to outline relevant 
%theoretical aspects which might be of interest in future
%fundamental studies for solar wind models. 
A historical review of the approaches adopted to solve the Boltzmann
equation may be found in the  introduction of 
Chapter 8 of the monograph by \citet{Jancel+Kahan}.

In order to illustrate the process 
by which one can reduce the solutions of the Boltzmann 
equation to fluid solutions, and thus switch from the
formalism based on the velocity distribution functions to the
one based on their first moments, it is useful to
introduce  a smallness parameter, $\epsilon$:
\begin{equation}
  \label{eq:boltzmann_epsilon}
\frac{\mathcal{D}f_s}{\mathcal{D}t}= \frac{1}{\epsilon}J_s
\end{equation}
where  $\mathcal{D}f_s/\mathcal{D}t$ is identical to the left-hand
side of (\ref{eq:boltzmann}) 
%\sout{the operator $\mathcal{D}/\mathcal{D}t$ is defined by the
%spatio-temporal derivatives given in the left hand-side of
%equations (\ref{eq:boltzmann}) and  (\ref{eq:boltzmann_actual_v}).}
%In equation (\ref{eq:boltzmann_epsilon}) $\vec F_s$ denotes the external force,
and $J_s$ is the collision integral defined in (\ref{eq:boltzmann_J}).
%\comment{ols: no, (\ref{eq:boltzmann_J}) is (currently) only valid for
%hard-sphere interactions}; 
$\epsilon$ is a smallness  parameter associated  to the
smallest spatial scale (can be the mean-free-path) or the smallest
time period (the mean collision time) or both. 
%The normal solutions
%are obtained for $\epsilon \ll 1$.
In this Appendix we shall also denote the left hand side of the equation
(\ref{eq:boltzmann_epsilon}) by $\dot{f}$.

A special class of the Boltzmann equation are the
so-called ``normal solutions'' that are valid only when the mean
free path, $\lambda_c$, and the collision time,
$\tau_c$ are small with respect to the spatial and temporal scale, $L$
or $\mathcal{T}$ respectively:
i.e.  $\epsilon \ll 1$. 
The normal solutions are obtained
by expanding $f_s$ as a power series of  $\epsilon = \lambda_c/L$
or $\epsilon = \tau_c/\mathcal{T}$. 
The class of normal solutions is built on the
rigorous mathematical proof given by \citet{Hilbert} for 
the existence of a solution for a Fredholm type equation
like the Boltzmann equation. The expansions proposed first by
\citet{Chapman1917} and \citet{Enskog1917} give normal
solutions of interest for physicists.

\subsection{Chapman-Enskog expansion}
\label{Chapman-Enskog}

\citet{Enskog1917} and \citet{Chapman1917} developed 
a method to solve the Boltzmann equation suited for 
practical situations that requires a 
smaller number of initial conditions, as in fluid dynamics approaches.
In addition to the expansion
\begin{equation}
   \label{eq:Hilbert_expansion}
  f_s = f_s^{(0)} + \epsilon f_s^{(1)} +  \epsilon^2 f_s^{(2)} \ldots 
\end{equation}
Enskog adds also an expansion of the temporal derivative operator:
\begin{equation}
   \label{eq:Enskog_expansion}
  \pderiv{}{t} \equiv \sum_{r=0}^\infty\epsilon^r\pderiv{^{(r)}}{t} 
\end{equation}
and also defines an expansion for the moments
 $\overline{M}$, based on (\ref{eq:Hilbert_expansion}) 
as given below:
\begin{equation}
\label{eq:moment_VDF_enskog}
 \overline{M} =  \overline{M}^{(0)} + \epsilon  \overline{M}^{(1)} +  \epsilon^2
   \overline{M}^{(2)} +\dots =  \frac{1}{n}\int \int \int M{\left[  f_s^{(0)} + \epsilon f_s^{(1)} +
  \epsilon^2 f_s^{(2)} \ldots \right]} d\vec c_s
\end{equation}
When the expansion (\ref{eq:moment_VDF_enskog}) is inserted into the
conservation equations (continuity, momentum and energy, 
see eq. \ref{eq:cont}--\ref{eq:emom}) the
corresponding results define the formal expansion of the time
derivative operator (\ref{eq:Enskog_expansion}) as a function of spatial
derivatives only:
\begin{eqnarray}
  \pderiv{^{(0)} n^{(0)}}{t} & = & -\vec \nabla_r \cdot
  \left(n^{(0)} {\vec u^{(0)}} \right),  \label{Enskog_exp_1}\\
  \pderiv{^{(0)} {\vec u^{(0)}}}{t} & = & -\left({{\vec u^{(0)}}}\cdot \vec
  \nabla_r \right) {\vec u^{(0)}} + \frac{\vec F}{m} -\frac{1}{\rho}\vec
  \nabla_rp^{(0)}, \\
 \pderiv{^{(n)}\vec u^{(n)}}{t} & = & - \frac{1}{\rho} \vec \nabla_r\cdot
  \tensor{\overline{p}^{(n)}}, \hspace{0.5cm} \mbox{$n>0$} \\
\pderiv{^{(0)}T^{(0)}}{t} &=& -\vec u^{(0)} \cdot \vec \nabla_r T^{(0)} -
  \frac{2T}{3}\vec \nabla_r \cdot \vec u^{(0)}, \\
\pderiv{^{(n)}T^{(n)}}{t} &=& -
  \frac{2}{3kn^{(0)}}\left[\tensor{\overline{p}^{(n)}}: 
\vec \nabla_r\vec u + \vec \nabla_r\cdot \overline{ \vec
  q}^{(n)}\right],  \hspace{0.5cm} \mbox{$n>0$} \label{Enskog_exp_n}
\end{eqnarray}
%where the subscript $s$ indexing the species has been 
%dropped to keep the equations simpler. 
The above equations are specific for each species but in order
to keep the notation simple the index $s$ corresponding to the
species has been dropped.   $p=\overline{p^{(0)}}$ is the isotropic
 pressure corresponding to the isotropic
Maxwellian $f^{(0)}$, while $\tensor{{p}^{(n)}}$
and  ${ \vec q}^{(n)}$ are  defined respectively by
(\ref{eq:moment_VDF_enskog}) for $M=\vec c \vec c$ and 
$M=c\vec c$ (see also Appendix \ref{definitions} for the definitions
of $\tensor{p}$ and $\vec q$, note that the index $s$ has been dropped
in equations \ref{Enskog_exp_1}-\ref{Enskog_exp_n}).
% When equations
%(\ref{Enskog_exp_1})-(\ref{Enskog_exp_n}) are replaced
%in  (\ref{eq:boltzmann_enskog}) one obtains a form of the Boltzmann
%equation that does not
%involve anymore time derivatives of $n$, $\vec U$ and  $T$.
From the so-called integrability conditions 
\citep{Hilbert}  one obtains a set of equations equivalent to :
\begin{eqnarray}
\label{Enskog_constraint}
\overline{n}^{(r)}&=&0, \mbox{$r\ge1$} \label{n_Enskog}\\
\overline{\vec v}^{(r)}&=&0, \mbox{$r\ge1$}  \label{v_Enskog}\\
\overline{T}^{(r)}&=&0, \mbox{$r\ge1$}  \label{T_Enskog}
\end{eqnarray}
The equations (\ref{n_Enskog})-(\ref{T_Enskog}) state that
\emph{at any time} the higher
order corrections of $f$ do not modify the density, average velocity and
temperature. The latter are fully determined by $f^{(0)}$: 
the zero order approximation of $f$ which is postulated to be an isotropic
Maxwellian distribution function. 
The equations (\ref{n_Enskog})--(\ref{T_Enskog}) are restricting
arbitrarily the class of normal solutions of the Boltzmann equation.
In certain situations, e.g. when the Knudsen number gases or
plasmas is of the order of unity or larger , 
the restrictive conditions may not be appropriate and
justified. This appears to be the case in the collisionless region
of the interplanetary medium where the VDF departs
significantly from displaced Maxwellians.

%Taking into account the properties of the
% Boltzmann collision integral 
%the integrability conditions can be rewritten as:
%\begin{equation}
%\label{integrability_Enskog}
%\int\int\int \psi_i \mathcal{D}^{(r)} d \vec v =0
%\end{equation}
%with $\mathcal{D}^{(r)}$ defined by (\ref{Dr}).
%In the zero order approximation ($n=0$) the condition (\ref{integrability_Enskog})
%gives the Euler equations for fluid dynamics, in the 
%first order ($n=1$) they give the Navier-Stokes approximation,
%and for  ($n=2$) the Burnett description is obtained.

The Chapman-Enskog solutions of the Boltzmann equation 
are necessarily based on the
assumption that $\epsilon$ is small and on the restrictive conditions
(\ref{n_Enskog})--(\ref{T_Enskog}). 
The mathematical procedures outlined 
above describe how  the detailed kinetic description can be 
coarse grained to  a multi-fluid one.  As put forward by \citet{Grad1958}  
an essential consequence of the assumptions
of Chapman and Enskog expansion is the existence of two
well separated time scales: (1) a (small) collision time, i.e. the time
necessary for the non-Maxwellian flow to relax and (2) a  macroscopic
decaying time where dissipation can be described in terms
of traditional thermodynamic processes/variables like diffusion, 
viscosity, and heat conductivity.  When $\epsilon$ is not small, i.e.
when the mean-free-path or the collision frequency are no
longer small compared to the dimensions of the system 
the results obtained based on
the expansion (\ref{eq:Hilbert_expansion}) 
and on the constraints (\ref{n_Enskog})--(\ref{T_Enskog})  are no longer applicable.
%Other methods have then to be employed to solve the kinetic
%equation and resulting moment equations of plasma.
%Examples are given in section \ref{Grad_section}, \ref{sec:gyrotr-improved},
%and \ref{sec:leblanc}.

In the Chapman-Enskog kinetic theory outlined above,  
displaced Maxwellian VDFs are adopted as the zero order approximation
of $f(\vec v, \vec r, t)$. 
Nevertheless, when $K_n > 1$, the VDFs of collisionless particles
are not well approximated by displaced Maxwellian functions, which
correspond to equilibrium VDFs in a uniform collision dominated gas or
plasma.  For instance in a Knudsen gas where $K_n > 1$, the kinetic pressure
tensor (i.e. the second order moments of the VDF) departs from its
classical Navier-Stokes approximation;
furthermore, the expression for the heat flux departs then from the
classical Fourier law: $\vec q = - \lambda \nabla T$, where $\lambda$
is the thermal coefficient of the gas.  
\citet{Scudder1992b} stressed that the Chapman--Enskog approach is 
self-consistent only for (i) subsonic flows with (ii) a rate of collisions high enough
to sustain a spatially homogeneous Maxwellian velocity distribution
function, and (iii) weak gradients of 
the macroscopic moments. 
All the three conditions must be satified cumulatively \citep{Scudder1992b}.
%The latter expression for the
%heat flux is a standard formula applicable in all materials where the
%continuum ansatz can be postulated.  The mathematical expression for
%$\lambda$ is a well determined function of the density,
%temperature and interactions forces between the particles only when
%the VDFs are sufficiently close to Maxwellians.

\subsection{Expansion about a local Maxwellian --- Grad's method}
\label{Grad_section}

When the smallness factor $\epsilon$ used to expand the distribution 
function, $f\left( \vec v, \vec r, t \right)$ is no
longer much smaller than 1 the normal solutions are not valid.
In Grad's method the expansion  of the
VDF $f$ is again based on a local isotropic Maxwellian in the frame
of reference moving with the bulk velocity $\vec u_s(\vec r,t)$:
\begin{equation}
\label{f0}
f^{(0)}=n\left( \vec r, t \right) \left(\frac{m}{2\pi kT(\vec r, t)}
\right)^{\frac{3}{2}} \exp\left[{-\frac{m\left( c_s(\vec r, t)
    \right)^2}{2kT(\vec r,t)}}\right]
\end{equation}
The density, $n(\vec r, t)$, temperature,  $T(\vec r,t)$, as
well as  $\vec u_s(\vec r,t)$ are
not constants but functions of time and spatial coordinates.
This may be justified when  there  is a clear separation
between spatio-temporal scales specific to  collisions and the 
hydrodynamic evolution of the fluid \citep{Grad1958}. 
This implies that the mean-free path of particles is much
smaller than the scale length of the gas, and that the
characteristic collision time is much smaller than the time
over which $\vec u$, $n$ and $T$ change significantly.
When this is no longer the case  
an expansion about an absolute Maxwellian (i.e. with $n$, 
$\vec U$ and   $T$ independent of $\vec r$) is in order
as pointed out by \citet{Grad1958}.
%In the following we
%describe the main features of the expansion about a
%local Maxwellian, of the Grad solution \citep{Grad1958}.

The method of Grad (\citeyear{Grad1958}) consists in expanding  the distribution
function using the orthogonal base of
the Hermite polynomials, with the local Maxwellian, $f^{(0)}$, as a weight function:
\begin{equation}
\label{Grad_expansion}
f=f^{(0)}\sum_{n=0}^{\infty}\frac{1}{n!}a_{i_n}^{(n)}\mathcal{H}_{i_n}^{(n)}
\end{equation}
where the term $a_{i_n}^{(n)}\mathcal{H}_{i_n}^{(n)}$ stands for the
sum 
\[
\sum_{i_1i_2\ldots i_n}a_{i_1i_2\ldots
  i_n}^{(n)}\mathcal{H}_{i_1i_2\ldots i_n}^{(n)}.
\]
$\mathcal{H}_{i_n}^{(n)}$ denotes a tensor with $n$ subscripts
$i_n=i_1i_2\ldots i_n$ whose components are polynomials of $n$-th
degree as defined below:
\begin{eqnarray}
\label{Hermite_polynomials}
\mathcal{H}^{(0)} &=& 1, \hspace{0.5cm}
\mathcal{H}_i^{(1)} = v_i, \hspace{0.5cm}
\mathcal{H}_{ij}^{(2)} = v_iv_j - \delta_{ij}, \nonumber \\
\mathcal{H}_{ijk}^{(3)} &=& v_iv_jv_kv_l   - \left(v_i\delta_{jk} +
v_j\delta_{ik} + v_k\delta_{ij} \right), \hspace{0.5cm}
\ldots  \nonumber
\end{eqnarray}
The coefficients $a_{i_n}^{(n)}$ are tensors determined 
by an inversion formula:
\begin{equation}
\label{a_i_n}
a_{i_n}^{(n)}= \frac{1}{n}\int\int\int{f\mathcal{H_i^{(n)}}} d\vec v
\end{equation}
giving a sequence of moments of increasing order:
%\begin{eqnarray}
\begin{equation}
\label{Hermite_moments}
{a}^{(0)} = 1, \hspace{0.5cm} {a}_i^{(1)} = 0,  \hspace{0.5cm} {a}_{ij}^{(2)} = \pi_{ij}/p,
 \hspace{0.5cm} {a}_{ijk}^{(2)} = Q_{ijk}/{p\sqrt{kT}} 
%\mathcal{a}_{ijk}^{(3)} &=& Q_{ijkl}/{pkT}  - \frac{1}{p}\left(p_{ij}\delta_{kl} +
%p_{ik}\delta_{jl} + p_{il}\delta_{jk} + p_{kl}\delta_{kl} \right), \nonumber \\
\ldots 
%\end{eqnarray}
\end{equation}
where the definitions of moments 
from Appendix \ref{definitions} have been used.

%\comment{ols: A bit puzzled that $a_i^{(1)}=0$. Does that mean that
%  ordinary heat flux is not included (in Schunk it appears as this
%  first order correction)?}
%\comment{MME: no, it means that the average velocity is equal to zero.}

The formulas found for ${a}^{(0)}$, ${a}_i^{(1)}$ and
${a}_{ij}^{(2)}$ state that the density, $n(\vec r, t)$, the
mean velocity, $\vec u(\vec r, t)$, and the temperature, $T(\vec r, t)$,
of the distribution function $f$  are uniquely determined by the
corresponding moments of the zero order approximation, $f^{(0)}$,
the Maxwellian distribution function. Consequently  ${a}_{i_n}^{(n)}$ vanish 
for $n>2$ when  the VDF  is strictly an isotropic and
Maxwellian VDF,  $f=f^{(0)}$. 
In Grad's class of kinetic solution defined by (\ref{Grad_expansion})
the higher order coefficients ($n > 2$)  don't contribute to the five first 
order order moments: $n(r,t)$, $\vec u(r,t)$, and $T(r,t)$.
Indeed,  these first order moments are fully determined by $f^{(0)}$, the isotropic, 
displaced Maxwellian VDF which is a priori
adopted for the zero-order approximation of $f(v,r,t)$.  
This imposes a serious restriction and limitation
to Grad's method.  This restriction is similar to that adopted in the
Chapman--Enskog method and described by  equations (\ref{eq:Hilbert_expansion})
and (\ref{n_Enskog})--(\ref{T_Enskog}). 
%In other words both the
%Chapman--Enskog solution (\ref{eq:Hilbert_expansion}) and
%the Grad's solution (\ref{Grad_expansion}) correspond ot a restricted
%or truncated ensemble of velocity distribution functions restricting 
%their variety to VDFs that are everywhere  and always close to the
%local  Maxwellian VDF, $f^{(0)}\left( \vec v, \vec r, t \right)$ with
%the same $n$, $\vec u$ and $T$.

%This constitutes a basic and essential limitation of the Grad 
%and Chapman-Enskog methods/approximations,
%in the cases of collisionless gases/plasmas where the actual/observed 
%VDFs fail to be close to Maxwellian ones, as in the distant solar wind.

Grad's expansion (\ref{Grad_expansion}) is inserted into the Boltzmann equation
(\ref{eq:boltzmann}), and after equating 
the coefficients of $H_{i_n}^{(n)}$ one obtains an infinite series of
differential equations for the linear combinations of the moments
of $f$. This unlimited series of partial differential equations defines
the hierarchy of moments equations which leads to an alternative
approximation of the hydrodynamic transport equations.
We show here the equations corresponding to the second order
terms, $n=2$:
\begin{eqnarray}
\pderiv{a_{ij}^{(2)}}{t} + u_r \pderiv{a_{ij}^{(2)}}{x_r} +
a_{ir}^{(2)} \pderiv{u_{j}}{x_r} + a_{rj}^{(2)} \pderiv{u_{i}}{x_r} +
\left( a_{ij}^{(2)} +\delta_{ij} \right)\frac{1}{kT}\frac{D(kT)}{Dt}
+ \sqrt{kT}\pderiv{a_{ijr}^{(3)}}{x_r} + & & \nonumber \\
\frac{\sqrt{kT}}{n}
a_{ijr}^{(3)} \pderiv{n}{x_r} + \frac{3}{2kT} a_{ijr}^{(3)}
\pderiv{kT}{x_r} + \pderiv{u_i}{x_j} + \pderiv{u_j}{x_i}  &=&
J_{ij}^{(2)} \label{second_order_Grad}
\end{eqnarray}
where tensor notation and  summation convention are used. Remember
that $D/Dt\equiv(\partial/\partial t)+u_r(\partial/\partial x_r) $ 
is the convective time derivative.  
When the force between particles is specified, the contribution of the collision term, denoted
$J_{ij}^{(2)}$ in (\ref{second_order_Grad}), is given by 
an infinite quadratic form of all the Hermite
coefficients (see \citeauthor{Grad1958}, \citeyear{Grad1958},
eqs. 28.14-28.17). 

%The third order ($n=3$) equations are contains complex expression
%and are not given here but can by found in  
%\citeauthor{Grad1958} ( \citeyear{Grad1958}, eqs. 28.18-28.21).
%General equations for order $n>3$ can be found in
%\citeauthor{Jancel+Kahan} ( \citeyear{Jancel+Kahan}, pg. 425-431).
%%they include formulas that are not reproduced here.
 
An important property of the Grad's expansion is evidenced by
equation (\ref{second_order_Grad}): the differential equation for
the moment $a^{(n)}$ involves spatial derivatives of the 
higher order moment $a^{(n+1)}$.
Nevertheless, the corresponding collision term, $J_{ij}^{(2)}$, is given
by sums involving the infinite sequence of  Hermite coefficients of all
orders, $n=0,\dots, \infty$. 
In order to obtain tractable solutions for the collision term, the expansion
(\ref{Grad_expansion}) is truncated and replaced by a    finite sum of 
the first $n$ Hermite polynomials.
Thus the Grad solution of Boltzmann equation assumes that
all the moments, or Hermite coefficients, of order $(n+1)$ and
higher are set equal to zero:
\begin{equation}
\label{Grad_cut_off}
 a_{i_{n+1}}^{(n+1)} = 0
\end{equation}
This ad-hoc truncation is not based on
physics principles other than convenience to limit the mathematical
complexity of the system of differential equations to be solved.

The left hand side of \textit{only one} equation of the Grad's moment 
 hierarchy is affected by the truncation  (\ref{Grad_cut_off}), namely
the equation corresponding to the highest considered order, $n$.
Nevertheless, the collision integral at the right hand side 
of \textit{all}  equations of the hierarchy, 
is altered by the truncation (\ref{Grad_cut_off}). 
A total number of $M$ differential equations results by truncating the
Grad's hierarchy at the (Hermitian) order $(n)$, therefore
 the corresponding Grad expansion is also called the
 $M$-moments method. In section \ref{sec:fluid-models}  
we discussed  the application of the $5$-,
$8$-, and $16$-moment equations to modeling the solar wind.

When the Grad expansion is limited to a few Hermite
coefficients the solution can be interpreted as a generalization
of the usual multi-fluid hydrodynamics. When the expansion is more elaborate and
an increased number of coefficients (or moments) are considered then
Grad's method provides a microscopic estimation of the
distribution function described by the subclass of VDF determined
by (\ref{Grad_expansion}).

The convergence of the Chapman-Enskog and Grad expansions of the velocity 
distribution functions  (\ref{eq:Hilbert_expansion}) and
(\ref{Grad_expansion}) has never been proven mathematically,
not even in the case of gases for which $\epsilon$ is much smaller
than unity,
i.e. for Knudsen numbers much smaller than unity.  A fortiori, the convergence 
of these expansions is not even valid in collisionless gases and plasmas, as for instance in 
coronal and stellar ion-exospheres.
Therefore, it would be presumptuous to consider that the higher the order of the Chapman-Enskog 
or Grad expansion is the better will be hydrodynamic modeling of the system,
and applicable to larger heliocentric distances.

A relevant example may be found in the field of classical hydrodynamics.
In the 50's and 60's the propagation and attenuation of sound waves 
has been investigated in laboratory experiments.  
The wave number and the wave attenuation constant, 
at a given frequency, were measured experimentally \citep{Greenspan1956,Toba1968}. 
These propagation and attenuation constants  were also determined 
theoretically using different hydrodynamical approximations 
of the transport equations and of the moment equations 
\citep{Truesdell1955,SirovichThurber65}
All these studies confirm that the theoretical predictions of these wave propagation constants 
(phase velocity or wave number, and attenuation constant) 
fail to fit the measured values, when the 
ratio between mean free path of the particles and the wave length becomes 
of the order and larger than unity: i.e. when the Knudsen number of the gas 
is of the order of 1 or larger.
%There is no reason to assume that it should be different/better in the exobase transition 
%region of stellar amospheres and planetary ionospheres, 
%where the Knudsen number becomes also larger than unity.

\begin{acknowledgements}
{We wish to thank Eugene N. Parker   and Bernard Shizgal
for their comments and editing of an early version of this
paper. We are grateful to Nicole Meyer-Vernet,
Viviane Pierrard, Herv\'{e} Lamy,
 for their collaboration during the revision
of the manuscript. We appreciate the detailed report of the
referee whose critical comments helped us to improve the revised text.
M. Echim thanks the staff of the Belgian Institute for
Space Aeronomy for long-term and efficient support, and acknowledges
financial support from the Belgian Solar Terrestrial Center of Excellence (STCE)
in Brussels, as well as from the European Space Agency
(PECS project KEEV, contract 98049/2007) and the Romanian Agency 
for Scientific Research - ANCS (SAFIR project, 
contract 98-081/2007, PARTNERIATE-PN2).

The authors are also grateful to
Abdalah Barakat,  Pierre-Louis Blelly, Sunny Tam,
Arkadi Usmanov, Giovani Lapenta, David Bergmans,
for valuable discussions and genuine exchange of
ideas on the occasion of a topical solar and polar 
wind meeting organized during the
STIMM-2 workshop in Sinaia, Romania, June 2007, and
a subsequent solar wind workshop held in Brussels, June 2009.}
\end{acknowledgements}

\newpage
\begin{sidewaystable}
\definecolor{gray}{rgb}{0.7,0.7,0.7}
\renewcommand{\baselinestretch}{0.9}
\normalsize
\renewcommand{\arraystretch}{1.}
\caption{Slow  (or quiet) solar wind observations and model at 1~AU and at a coronal reference level, 
$r_0$.  \label{tab:table1}}
\begin{center}
%\begin{tabular}{ccccccccccccccccc} \hline
%\begin{tabular*}{0.95\textwidth}{@{\extracolsep{\fill}}cccccccccccccccccc}
%\begin{tabular*}{19.95cm}{|p{1.3cm}| p{0.64cm}| p{0.64cm}| p{0.64cm}| p{0.64cm}| p{0.64cm}|
%    p{0.6cm}| p{0.6cm}| p{0.6cm}| p{0.6cm}| p{0.64cm}| p{0.64cm}| p{0.64cm}|
%    p{0.64cm}| p{0.64cm}| p{0.6cm}| p{0.55cm}| @{\extracolsep{\fill}}
%    p{1.2cm}| }
\begin{tabular}{|p{1.3cm}| p{0.64cm}| p{0.64cm}| p{0.6cm}| p{0.6cm}| p{0.6cm}|
    p{0.6cm}| p{0.6cm}| p{0.6cm}| p{0.6cm}| p{0.64cm}| p{0.64cm}| p{0.64cm}|
    p{0.64cm}| p{0.64cm}| p{0.6cm}| p{0.55cm}|  p{1.8cm}| }
\hline
%\toprule
%%%%%%%%%%%%%%%%%%%%%%%%%%%%%%%%%%%%%%%%%%%%%%%%%%%%%%%%%%%%%%%%%%%%%%%%%%%%%%%%%%%%%%%%%%%
%%                                                                                       %%
%%          HEADER TABEL 1                                                               %%
%%                                                                                       %%
%%%%%%%%%%%%%%%%%%%%%%%%%%%%%%%%%%%%%%%%%%%%%%%%%%%%%%%%%%%%%%%%%%%%%%%%%%%%%%%%%%%%%%%%%%%
\rowcolor{gray}Value $@$ 1 A.U. & $\{0\}$ COND \& HYST & $\{10\}$ \mbox{1F-H} &  $\{1\}$ \textbf{Slow} SW Obs.  &  
    $\{22\}$ \textbf{Slow} SW Obs. &  $\{7\}$  \mbox{ E }  &
    $\{2\}$   \mbox{ E }  &  $\{3\}$ \mbox{ E } & 
    $\{4\}$ Hb &  $\{5\}$ Hb  &  $\{6\}$ Hb  & $\{11\}$ 1F-H &  
    $\{12\}$ 1F-H  & $\{13\}$ 2F-H  & $\{14\}$ 2F-H  & $\{15\}$ 2F-H & 
    $\{24\}$ 2F-H \\
\hline
%%%%%%%%%%%%%%%%%%%%%%%%%%%%%%%%%%%%%%%%%%%%%%%%%%%%%%%%%%%%%%%%%%%%%%%%%%%%%%%%%%%%%%%%%%%
%%%%%%%%%%%%%%%%%%%%%%%%%%%%%%%%%%%%%%%%%%%%%%%%%%%%%%%%%%%%%%%%%%%%%%%%%%%%%%%%%%%%%%%%%%%
%%                                                                                       %%
%%          BODY  TABEL 1                                                                %%
%%                                                                                       %%
%%%%%%%%%%%%%%%%%%%%%%%%%%%%%%%%%%%%%%%%%%%%%%%%%%%%%%%%%%%%%%%%%%%%%%%%%%%%%%%%%%%%%%%%%%%
u [km/s] & 0 & 500 $550$  & \multicolumn{1}{|
  >{\columncolor[rgb]{0.9,0.9,0.0}}r|}{$320\pm 20$}
  &\multicolumn{1}{| >{\columncolor[rgb]{0.9,0.9,0.0}}r|}{$392\pm 20$} 
  &  20 & 307  & 320  & 288  & 323  & 313 & 352  & 165  & 250  & 256
& 320  &  (100-600) \\
%\midrule
\hline
$n_p$ [$cm^{-3}$] & 342  & 4 $(24)$  & \multicolumn{1}{|
  >{\columncolor[rgb]{0.9,0.9,0.0}}r|}{$8.7\pm 4.6$}  & 
 \multicolumn{1}{|  >{\columncolor[rgb]{0.9,0.9,0.0}}r|}{$5.55$}  &
 370  & $12.9$  & $7.18$  & $12$  &  & &$6.75$  & $8.5$  & 15  &
 $6.33$  & 13  & (27)  \\
%\midrule
\hline
\mbox{$<T_e>$} [$10^4$ K] & $21.9$ & 100 $(122)$  & \multicolumn{1}{|
  >{\columncolor[rgb]{0.9,0.9,0.0}}r|}{$14\pm 5$}  & \multicolumn{1}{|
  >{\columncolor[rgb]{0.9,0.9,0.0}}r|}{}   & 11  & $11.5$
& $11.7$  & $4.6$  &   & 60  & 28  & 9   & 34  & 16  & 34  & {\tiny$(<40)$}  \\
%\midrule
\hline
\mbox{$<T_p>$} [$10^4$ K] & $21.9$  & 100 $122$  & \multicolumn{1}{|
  >{\columncolor[rgb]{0.9,0.9,0.0}}r|}{$4.4\pm 1.8$}   &
\multicolumn{1}{|
  >{\columncolor[rgb]{0.9,0.9,0.0}}r|}{$8.0$}   & 11  & $4.49$  & $4.79$  & $6.7$  & $1.0$  & $0.41$  & $28$  &
$9$ & $0.44$  & $4.3$  & $3.7$  & $(<50)$  \\
%\midrule
\hline
$T_{e||}$/$T_{e\bot}$ & 1  & 1  &  \multicolumn{1}{|  >{\columncolor[rgb]{0.9,0.9,0.0}}r|}{$1.1\div 1.2$}   &
 \multicolumn{1}{|  >{\columncolor[rgb]{0.9,0.9,0.0}}r|}{}   &  & $3.04$  &  $3.05$  &
 $1$  &  $1$  &  $1$  &  $1$  &  $1.75$  &  $1$  &  $1$  &  $1$  &  $1$  \\
%\midrule
\hline
$T_{p||}$/$T_{p\bot}$  & 1 & 1  & \multicolumn{1}{|
  >{\columncolor[rgb]{0.9,0.9,0.0}}r|}{$2.0\pm 1$} 
   & \multicolumn{1}{| >{\columncolor[rgb]{0.9,0.9,0.0}}r|}{}   &  & $160$
& $0.20$  & $900$  & $50$  & $11$  & $1$   & $1.75$  & $1$  & $1$  & $1$  &  1 \\
%\midrule
\hline
Q [erg/$cm^2$/s] & $0.049$  &   & \multicolumn{1}{|
  >{\columncolor[rgb]{0.9,0.9,0.0}}r|}{$0.24$}   &
\multicolumn{1}{| >{\columncolor[rgb]{0.9,0.9,0.0}}r|}{}  &  &  & $1.32$
$0.20$  &   &   &   &  &  &  &  &  &   $(\ll 8)$ \\
%\bottomrule
\hline
\hline
\rowcolor{gray} Value $@$ $r_0$ &  $\{0\}$ Chapm. 1957 &  $\{10\}$ Parker 1958  &
 $\{1\}$ Hundh. 1972  &  $\{22\}$ Ebert et al. 2009   &
   $\{7\}$ Chb 1960  & $\{2\}$ LSc 1972   &  $\{3\}$ LSb 1971   &
 $\{4\}$ Jock 1970  &  $\{5\}$ Holw. 1970   &
  $\{6\}$ Chen 1972 & $\{11\}$ N$\&$S 1963 &  $\{12\}$ WLC 1966 &
 $\{13\}$  H$\&$S 1968 & $\{14\}$ C$\&$H 1971 & $\{15\}$ H$\&$B 1971
 & $\{24\}$ H$\&$L 1995  \\
%\midrule
\hline
 $r_0$ [$R_S$] & 1  & 1  & \multicolumn{1}{| >{\columncolor[rgb]{0.9,0.9,0.0}}r|}{}   &
\multicolumn{1}{|  >{\columncolor[rgb]{0.9,0.9,0.0}}r|}{}   &
$2.5$  & $6.5$  & $6.6$  & 15  & 15  &   & 1  & 1  & 1  & 1  & 2  & 1  \\
%\midrule
\hline
$n_0$ [$10^4$ $cm^{-3}$]  & 20000  &  20000 & \multicolumn{1}{| >{\columncolor[rgb]{0.9,0.9,0.0}}r|}{}   &
\multicolumn{1}{|  >{\columncolor[rgb]{0.9,0.9,0.0}}r|}{}  & 100  &
$3.1$  & $3.1$  & 90  &   &   & 20000  & 28000  & 3000  & 2300  & 150
& $10^{10}$  \\
%\midrule
\hline
 $T_{e0}$ [$10^6$ K]  & $1.0$  & 1  &  \multicolumn{1}{| >{\columncolor[rgb]{0.9,0.9,0.0}}r|}{}   &
\multicolumn{1}{|  >{\columncolor[rgb]{0.9,0.9,0.0}}r|}{}   &  2 &
 $1.52$  & $1.52$  &  $1.32$   & 1  & &  2  & $1.5$  & 2  & $1.67$  &
 $1.2$    & $7.10^{-3}$  \\
%\midrule
\hline
 $T_{p0}$ [$10^6$ K]  & 1  &  1 &  \multicolumn{1}{| >{\columncolor[rgb]{0.9,0.9,0.0}}r|}{}   &
\multicolumn{1}{|  >{\columncolor[rgb]{0.9,0.9,0.0}}r|}{}   & 2  &
 $0.984$  & $0.984$  & $1.32$  & $0.1$  &   & 2  & $1.5$  & 2  &
 $1.67$  & $1.2$  & $7.10^{-3}$ \\
%\midrule
\hline
 $u_0$ [km/s] & 0  &  &  \multicolumn{1}{| >{\columncolor[rgb]{0.9,0.9,0.0}}r|}{}   &
\multicolumn{1}{|  >{\columncolor[rgb]{0.9,0.9,0.0}}r|}{}    &  & 0  &
 14 & 0  & 175  & &  &  &  &  &  & $5.10^{-6}$ \\
%\midrule
\hline
$\kappa$  &   &  & \multicolumn{1}{| >{\columncolor[rgb]{0.9,0.9,0.0}}r|}{}   &
\multicolumn{1}{|  >{\columncolor[rgb]{0.9,0.9,0.0}}r|}{}   & $\infty$
&  $\infty$  &  $\infty$  &  $\infty$  &  $\infty$  &  $\infty$  &  &  &  &  &  &  \\
%\bottomrule
\hline
\hline
%\end{tabular*}
\end{tabular}
\label{tableEXOS1}
\end{center}
%\end{table}
The different models are identified by numbers in the top row:
$\{0\}$ Conductive hydrostatic model \citep{Chapman57}; $\{1\}$
  observed quiet solar wind at 1~AU \citep{Hundhausen+etal70,
    Hundhausen72}; $\{2\}$ Exospheric model-c
  \citep{LemaireScherer72}; $\{3\}$ Exospheric model-b
  \citep{LemaireScherer71}; $\{ 4\}$ hybrid/semikinetic model
  \citep{Jockers70}; $\{ 5\}$ hybrid/semikinetic model
  \citep{Hollweg70};   $\{ 6\}$ hybrid/semikinetic model
  with spiral IMF and polytropic electron temperature
  distribution\citep{Chen+al72}; $\{ 7\}$  Exospheric Solar Breeze
  model with Pannekoek-Rosseland electric field \citep{Chamberlain60};
   $\{ 8\}$ Exospheric model \citep{Jenssen};  $\{ 9\}$ Exospheric
  model  \citep{BrandtCassinelli66}; $\{10\}$ Isothermal single-fluid
  Euler hydrodynamic model \citep{Parker58, Parker63};
  $\{11\}$ Conductive single-fluid hydrodynamic model
  \citep{NobleScarf63};  $\{12\}$  Conductive and viscous 
  single-fluid hydrodynamic model \citep{WhangLiuChang66};
  $\{13\}$  Conductive two-fluid hydrodynamic model
  \citep{SturrockHartle66};  $\{14\}$ Two-fluid hydrodynamic 
   model with reduced thermal conductivity and enhanced
   non-collisional coupling \citep{CupermanHarten71}; 
   $\{15\}$ Conductive two-fluid hydrodynamic model 
   with adjusted heating of SW protons \citep{HartleBarnes70};
   $\{22\}$ Observed slow solar wind from ULYSSES at 1AU and  $0^0$
   latitude \citep{Ebert+al};  $\{24\}$ Chromospheric conductive 
   two-fluid hydrodynamic with isotropic temperatures and
   adjusted heating; $r < 25R_S$ \citep{Hansteen+Leer}.
\end{sidewaystable}
%\normalsize
\renewcommand{\baselinestretch}{2}
\normalsize
%\end{landscape}

\newpage
\begin{sidewaystable}
\definecolor{gray}{rgb}{0.7,0.7,0.7}
\renewcommand{\baselinestretch}{0.9}
\normalsize
\renewcommand{\arraystretch}{1.}
\caption{Fast solar wind observations and model at 1~AU and at a
  coronal reference level, $r_0$.\label{tab:table2}}
\begin{center}
%\begin{tabular}{ccccccccccccccccc} \hline
%\begin{tabular*}{0.95\textwidth}{@{\extracolsep{\fill}}cccccccccccccccccc}
\begin{tabular*}{17.95cm}{|p{1.3cm}| p{0.7cm}| p{0.7cm}| p{0.7cm}| p{0.7cm}| p{0.7cm}|
    p{0.7cm}| p{0.7cm}| p{0.95cm}| p{0.7cm}| p{0.75cm}| p{1.35cm}| p{0.95cm}|
    p{0.95cm}|} 
\hline
%\toprule
%%%%%%%%%%%%%%%%%%%%%%%%%%%%%%%%%%%%%%%%%%%%%%%%%%%%%%%%%%%%%%%%%%%%%%%%%%%%%%%%%%%%%%%%%%%
%%                                                                                       %%
%%          HEADER TABEL 2                                                               %%
%%                                                                                       %%
%%%%%%%%%%%%%%%%%%%%%%%%%%%%%%%%%%%%%%%%%%%%%%%%%%%%%%%%%%%%%%%%%%%%%%%%%%%%%%%%%%%%%%%%%%%
\rowcolor{gray}Value $@$ 1 A.U. & $\{20\}$ \textbf{Fast} SW Obs. &
    $\{21\}$ Obs. SOHO  &  $\{22\}$ Obs. ULYSSES  &  
    $\{10\}$ \mbox{1F-H} &  $\{16\}$  \mbox{ E }  &
    $\{17\}$  \mbox{2F-H}  &  $\{18\}$ \mbox{3F-H } & 
    $\{19\}$ \mbox{2F-HCH} &  $\{23\}$ 3HB  &  $\{25\}$ \mbox{2F-16M}  & $\{26\}$ \mbox{2F-16M}-BIM &  
    $\{27\}$ 2F-16M &  $\{28\}$ 2F-16M  \\
\hline
%%%%%%%%%%%%%%%%%%%%%%%%%%%%%%%%%%%%%%%%%%%%%%%%%%%%%%%%%%%%%%%%%%%%%%%%%%%%%%%%%%%%%%%%%%%
%%%%%%%%%%%%%%%%%%%%%%%%%%%%%%%%%%%%%%%%%%%%%%%%%%%%%%%%%%%%%%%%%%%%%%%%%%%%%%%%%%%%%%%%%%%
%%                                                                                       %%
%%          BODY  TABEL 2                                                                %%
%%                                                                                       %%
%%%%%%%%%%%%%%%%%%%%%%%%%%%%%%%%%%%%%%%%%%%%%%%%%%%%%%%%%%%%%%%%%%%%%%%%%%%%%%%%%%%%%%%%%%%
u [km/s]  & \multicolumn{1}{|
  >{\columncolor[rgb]{0.9,0.9,0.0}}r|}{667}  & \multicolumn{1}{|
  >{\columncolor[rgb]{0.9,0.9,0.0}}r|}{}
  &\multicolumn{1}{| >{\columncolor[rgb]{0.9,0.9,0.0}}r|}{745} 
  &  500 $500$ & 667  & 710  & 716  & 764  & 593 & 520  & 750 & 1000
& 650 \\
%\midrule
\hline
$n_p$ [$cm^{-3}$]   &  \multicolumn{1}{|
  >{\columncolor[rgb]{0.9,0.9,0.0}}r|}{3}   & \multicolumn{1}{|
  >{\columncolor[rgb]{0.9,0.9,0.0}}r|}{}  & 
 \multicolumn{1}{|  >{\columncolor[rgb]{0.9,0.9,0.0}}r|}{$2.12$}  &
 4 (24)  & $2.7$  & $0.01$  & $2.9$  &  $2.11$ & $6.83$ & $<300$  & 2
 & 2 & 3 \\
%\midrule
\hline
\mbox{$<T_e>$} [$10^4$ K] & \multicolumn{1}{|
  >{\columncolor[rgb]{0.9,0.9,0.0}}r|}{28} &
   \multicolumn{1}{|  >{\columncolor[rgb]{0.9,0.9,0.0}}r|}{}  & \multicolumn{1}{|
  >{\columncolor[rgb]{0.9,0.9,0.0}}r|}{}   & 100 (122)  & $1.22$
& 24  & 37  & $25.8$   & $3.09$  & $<28$  & 10   & 2 & 4   \\
%\midrule
\hline
\mbox{$<T_p>$} [$10^4$ K] & \multicolumn{1}{|
  >{\columncolor[rgb]{0.9,0.9,0.0}}r|}{13}   &
 \multicolumn{1}{|  >{\columncolor[rgb]{0.9,0.9,0.0}}r|}{}   &
\multicolumn{1}{|  >{\columncolor[rgb]{0.9,0.9,0.0}}r|}{$24.6$}   &
100 $122$  & $134$  & $62$  & $0.2$  & $26.5$  & $53.0$  & $<200$  &
$40$ & $20$  & 20 \\
%\midrule
\hline
$T_{e||}$/$T_{e\bot}$ & \multicolumn{1}{|  >{\columncolor[rgb]{0.9,0.9,0.0}}r|}{$1.2$} 
  &  \multicolumn{1}{|  >{\columncolor[rgb]{0.9,0.9,0.0}}r|}{}   &
 \multicolumn{1}{|  >{\columncolor[rgb]{0.9,0.9,0.0}}r|}{}   & 1  & $4.4$  &  $1$  &
 $1$  &  $1$  &  $1.43$  &  $>1.3$  &  $1$  &  $1.3$ & 2  \\
%\midrule
\hline
$T_{p||}$/$T_{p\bot}$  &  \multicolumn{1}{|  >{\columncolor[rgb]{0.9,0.9,0.0}}r|}{$1.2$}
   & \multicolumn{1}{|  >{\columncolor[rgb]{0.9,0.9,0.0}}r|}{} 
   & \multicolumn{1}{| >{\columncolor[rgb]{0.9,0.9,0.0}}r|}{}   & 1  &
 $46$  & 1  & 1  & 1  & 147   & $>8$  & 4 &125 &300  \\
%\midrule
\hline
Q [erg/$cm^2$/s] &  \multicolumn{1}{|
  >{\columncolor[rgb]{0.9,0.9,0.0}}r|}{}   &
 \multicolumn{1}{|  >{\columncolor[rgb]{0.9,0.9,0.0}}r|}{}   &
\multicolumn{1}{| >{\columncolor[rgb]{0.9,0.9,0.0}}r|}{}  &  &  &
 &   &   & $0.015$  & $<0.9$  &  & &  \\
%\bottomrule
\hline
\hline
\rowcolor{gray} Value $@$ $r_0$ &  $\{20\}$ Maks 1995 &  $\{21\}$
M$\&$B 1999  &  $\{22\}$ Eber$\&$al 2000  &  $\{10\}$ Parker 1958   &
   $\{16\}$ MPL 1997  & $\{17\}$ K$\&$al 2004   &  $\{18\}$ LSb E$\&$H
1995 &  $\{19\}$ H$\&$al 1995  &  $\{23\}$ T$\&$C 1999  &
  $\{25\}$ O$\&$L 1999 & $\{26\}$ \mbox{Li 1999} &  $\{27\}$
\mbox{L-S}; \mbox{L$\&$H 2001}
&  $\{28\}$ J;L-S$\&$L 2006 \\
%\midrule
\hline
 $r_0$ [$R_S$]   & \multicolumn{1}{|
 >{\columncolor[rgb]{0.9,0.9,0.0}}r|}{} 
 & \multicolumn{1}{| >{\columncolor[rgb]{0.9,0.9,0.0}}r|}{1}   &
\multicolumn{1}{|  >{\columncolor[rgb]{0.9,0.9,0.0}}r|}{}   &
$1$  & $6.4$  & 1  & 1  & 1  & 1  & 1  & 1  & 1 & 1  \\
%\midrule
\hline
$n_0$ [$10^4$ $cm^{-3}$]   &   \multicolumn{1}{|
  >{\columncolor[rgb]{0.9,0.9,0.0}}r|}{} 
 & \multicolumn{1}{| >{\columncolor[rgb]{0.9,0.9,0.0}}r|}{20000}   &
\multicolumn{1}{|  >{\columncolor[rgb]{0.9,0.9,0.0}}r|}{}  & 20000  &
$3.2$  & $20$  & 70000  & 50000   & 200  & 6000  & 28000  & $9.10^9$ & $9.10^9$  \\
%\midrule
\hline
 $T_{e0}$ [$10^6$ K]    &  \multicolumn{1}{|
 >{\columncolor[rgb]{0.9,0.9,0.0}}r|}{} 
  &  \multicolumn{1}{| >{\columncolor[rgb]{0.9,0.9,0.0}}r|}{}   &
\multicolumn{1}{|  >{\columncolor[rgb]{0.9,0.9,0.0}}r|}{}   &  1 &
 $1.52$  & $1.16$  &  $0.5$   & $0.2$  & $2.03$  &  $0.5$  & $0.6$  &
 $7.10^{-3}$ &  $7.10^{-3}$  \\
%\midrule
\hline
 $T_{p0}$ [$10^6$ K]   &  \multicolumn{1}{|
 >{\columncolor[rgb]{0.9,0.9,0.0}}r|}{} 
  &  \multicolumn{1}{| >{\columncolor[rgb]{0.9,0.9,0.0}}r|}{$0.75$}   &
\multicolumn{1}{|  >{\columncolor[rgb]{0.9,0.9,0.0}}r|}{}   & 1  &
 $1.0$  & $1.16$  & $0.5$  & $0.2$  &  $2.05$  & $0.5$  & $0.6$  &
 $7.10^{-3}$ & $7.10^{-3}$ \\
%\midrule
\hline
 $u_0$ [km/s]   &\multicolumn{1}{| >{\columncolor[rgb]{0.9,0.9,0.0}}r|}{}  &  \multicolumn{1}{| >{\columncolor[rgb]{0.9,0.9,0.0}}r|}{}   &
\multicolumn{1}{|  >{\columncolor[rgb]{0.9,0.9,0.0}}r|}{}    &  &   &
 $0.8$ &   &   & & 10  & $2.4$  & $4.10^{-6}$ &$5.10^{-6}$   \\
%\midrule
\hline
$\kappa$  & \multicolumn{1}{| >{\columncolor[rgb]{0.9,0.9,0.0}}r|}{}  & \multicolumn{1}{| >{\columncolor[rgb]{0.9,0.9,0.0}}r|}{}   &
\multicolumn{1}{|  >{\columncolor[rgb]{0.9,0.9,0.0}}r|}{}   & 
&  2  &    &    &    &    &  &  &  & \\
%\bottomrule
\hline
\hline
\end{tabular*}
\label{tableEXOS2}
\end{center}
%\end{table}
The different models are identified by numbers in the top:
$\{16\}$ Lorentzian exospheric model \citep{Maksimovic97b};
 $\{17\}$ Conductive two-fluid hydrodynamic model with 
Alfv\'{e}n wave heating and accelerating the wind ions \citep{Kim2004};
 $\{18\}$ Conductive three-fluid hydrodynamic model with adjusted
heating \citep{EsserHabbal95}; 
$\{19\}$ Conductive two-fluid hydrodynamic model with adiabatic 
cooling and heating, and acceleration by Alfv\'{e}n waves
\citep{Habbal95};  $\{20\}$ Observed fast solar wind by Helios-1/2
\citep{Maksimovic95}; $\{21\}$ Coronal observations by SOHO 
\citep{MasonBochsler1999};   $\{ 22\}$ Observed fast solar wind at 1AU $\&$ $60^\circ$
lat. ; ULYSSES \citep{Ebert+al}; 
$\{23\}$  Hybrid three-component model including heating by 
wave-particle interactions \citep{Tam+Chang};
$\{25\}$ Gyrotropic 16-moment transport equations with anisotropic
temperatures; no in-situ heating; $r < 30~R_S$ \citep{Olsen+Leer1999}; 
 $\{26\}$ Gyrotropic 16-moment transport equations with anisotropic
proton  temperature, isotropic electron temperature; start at 
transition region \citep{XLi1999}; 
$\{27\}$ Chromospheric 16-moment transport equations with anisotropic 
temperatures with simplified collision terms and proton heated by 
turbulent cascade of Alfv\'{e}n waves  \citep{Lie-Svendsen+Leer+Hansteen2001};
$\{28\}$ Improved gyrotropic 16-moment  transport equations with 
better treatment of heat conduction with proton heating  \citep{Janse+Lie-Svendsen+Leer}.
\end{sidewaystable}
\renewcommand{\baselinestretch}{2}
\normalsize
%\end{landscape}
%_____________________________________________________________
%Table 2
%Comparison between measurements and \textit{kinetic models} of the solar 
%wind of the number density, bulk velocity, parallel and 
%perpendicular temperatures, temperature anisotropies, energy flux 
%and heat conduction flux. During quiet solar wind conditions, 
%the observations taken from Hundhausen (1968) at 1~AU For the high 
%speed solar wind, the observations are made by Helios-1:2 (Maksimovic,
%1995). The different models are identified by Numbers in the top row
%referring to the list of models listed below.%

%Table 2 + legend
%_____________________________________________________________________

%\bibliographystyle{spbasic}
%\bibliography{kinetic_SW_JFL_revOLS,kinetic_theory_revOLS,fluid_revOLS,coronal_heating}          %tranreg
\end{document}